\documentclass[11pt]{article}
\pdfoutput=1
\usepackage{jheppub}

\usepackage{amsmath} 
\usepackage{multirow} 
\usepackage{amsfonts}
\usepackage{amssymb,graphics,psfrag}
\usepackage{array,epsfig,multirow,graphicx,mathtools}
\usepackage{hyperref}
\newcommand{\be}{\begin{equation}}
\newcommand{\ee}{\end{equation}}
\newcommand{\bea}{\begin{eqnarray}}
\newcommand{\eea}{\end{eqnarray}}
  
\def\lam{{\lambda}}

\title{Exact partition functions of Higgsed 5d $T_N$ theories}

\author[a]{Hirotaka Hayashi,}
\author[a, b]{Gianluca Zoccarato}

\affiliation[a]{Instituto de F\'{\i}sica Te\'orica UAM/CSIC, Cantoblanco, 28049 Madrid, Spain}
\affiliation[b]{Departamento de F\'{\i}sica Te\'orica, Universidad Aut\'onoma de Madrid, 28049 Madrid, Spain}

\emailAdd{h.hayashi@csic.es}
\emailAdd{gianluca.zoccarato@csic.es}

\abstract{We present a general prescription by which we can systematically compute exact partition functions of five-dimensional supersymmetric theories which arise in Higgs branches of the $T_N$ theory. The theories may be realised by webs of 5-branes whose dual geometries are non-toric. We have checked our method by calculating the partition functions of the theories realised in various Higgs branches of the $T_3$ theory. A particularly interesting example is the $E_8$ theory which can be obtained by Higgsing the $T_6$ theory. We explicitly compute the partition function of the $E_8$ theory and find agreement with the field theory result as well as enhancement of the global symmetry to $E_8$.}

\begin{document}

\makeatletter
\let\old@fpheader\@fpheader
\renewcommand{\@fpheader}{\old@fpheader\hfill
IFT-UAM/CSIC-14-089}
\makeatother

\maketitle


%
%

\section{Introduction}

String theory is a good candidate for the fundamental theory describing all the forces and matter found in nature, but its applications are not limited to this and
in particular it has proven to be a powerful tool to understand exact results of field theories. For example using the topological vertex method \cite{Iqbal:2002we, 
Aganagic:2003db} or its refinement \cite{Awata:2005fa, Iqbal:2007ii} it is possible to compute five or four-dimensional  Nekrasov partition functions of $U(N)$ gauge theories \cite{Iqbal:2003ix, Iqbal:2003zz, Eguchi:2003sj, Hollowood:2003cv, Taki:2007dh, Awata:2008ed}. 
%
%
There are two possible ways to realise five-dimensional gauge theories that allow the use of the topological vertex to compute the Nekrasov partition function:
either M-theory on toric Calabi--Yau threefolds or webs of $(p,q)$ 5-branes in type IIB string theory. The relation between the two is actually quite simple: the dual of the toric 
diagram
of the Calabi--Yau threefold used as a M-theory background coincides with the web of $(p,q)$ 5-branes that realises the same low energy effective field theory \cite{Leung:1997tw}\footnote{ In this case the five-dimensional gauge theories are obtained by compactifying the worldvolume theories of 5-branes on segments \cite{Aharony:1997ju, Aharony:1997bh}. }.
While the M-theory and 5-branes web perspectives are completely equivalent, the latter offers some advantages, like for instance the fact that the flavour symmetry of the gauge theory becomes manifest after introducing 7-branes \cite{DeWolfe:1999hj, Yamada:1999xr}.

The class of five-dimensional theories that can be obtained using webs of $(p,q)$ 5-branes is not limited to gauge theories however and in particular it is possible to get
more general theories such as the 5d $T_N$ theory \cite{Benini:2009gi} whose four-dimensional versions were originally introduced in \cite{Gaiotto:2009we}. This is particularly important because $T_N$ theories lack a Lagrangian description and so it is not possible to use localisation techniques 
 to compute their partition functions \cite{Nekrasov:2002qd, Nekrasov:2003rj}. But, since the web diagrams that realise these theories are toric\footnote{Strictly speaking, toric should be used for geometries in the dual description. The statement that a web diagram is toric means that the web diagram is dual to a toric Calabi--Yau threefold.},
the computation of their Nekrasov partition function is still possible using the refined topological vertex formalism.

Recently there has been a progress in the computation of the refined topological vertex. In particular it has turned out that the refined topological vertex computation itself automatically contains some factors which are contributions from particles that are decoupled from the theory realised by webs of $(p, q)$ 5-branes \cite{Bergman:2013ala, Bao:2013pwa, Hayashi:2013qwa, Bergman:2013aca}. Using the web diagram it is quite simple to identify these factors as they originate from strings between parallel external legs and only after stripping them off we obtain the partition functions of the theories realised by $(p, q)$ 5-brane webs. For example, the refined topological vertex computation from the web diagram which realises an $SU(2)$ gauge theory with $N_f \leq 4$ flavours yields the $U(2)$ Nekrasov partition functions with $N_f \leq 4$ flavours \cite{Iqbal:2003ix, Iqbal:2003zz, Eguchi:2003sj, Hollowood:2003cv, Taki:2007dh, Awata:2008ed}. It is only after removing the decoupled factors that the partition function becomes that of a $SU(2) \cong Sp(1)$ gauge theory with the same number of flavours \cite{Bao:2013pwa, Hayashi:2013qwa, Bergman:2013aca}. This procedure
for the identification of decoupled factors is general and can be applied to more general theories like the $T_N$ theories and allows to compute their partition function
\cite{Bao:2013pwa, Hayashi:2013qwa}. The importance of the removal of the contributions of decoupled factors has been checked in other important cases: without its 
removal in fact enhancement of the global symmetry of some gauge theories at their superconformal fixed points is not possible \cite{Taki:2013vka, Bergman:2013aca, Zafrir:2014ywa} and checks of dualities at the level of partition function would fail  \cite{Taki:2014pba}.

Let us also mention that the explanation of a similar decoupled factor has been given from the in the context of ADHM quantum mechanics in \cite{Hwang:2014uwa}. For example, $Sp(N)$ gauge theory with $N_f \leq 7$ fundamental and $1$ antisymmetric hypermultiplets can be realised on $N$ D4-branes close to $N_f$ D8-branes and an O8-plane. Gauge instantons on the worldvolume of the D4-branes are given by D0-branes and their contribution to the partition function can be computed
using ADHM quantum mechanics.
 However this computation also receives contributions from D0-D8-O8 bound states and in order to get the correct partition function of the $Sp(N)$ gauge theory 
 with $N_f \leq 7$ fundamental and $1$ antisymmetric hypermultiplets it is necessary to remove these contributions.

It is possible to use the refined topological vertex formalism to compute the partition function of five-dimensional $Sp(1)$ gauge theories with $N_f \leq 5$ flavours\footnote{In the following we will sometimes call a five-dimensional $Sp(1)$ gauge theory with $N_f$ flavours also as $E_{N_f+1}$ theory. This is at the superconformal
point the $SO(2N_f)$ global symmetry of the theory is known to enhance to $E_{N_f+1}$ \cite{Seiberg:1996bd}. In other words, a mass deformation of the $E_{N_f+1}$ theory becomes the five-dimensional $Sp(1)$ gauge theory with $N_f$ flavours. }
for the web diagram realising these theories is toric. This is not possible when the number of flavours is larger than 5 because the web diagram realising these theories 
is no longer toric\footnote{A vertex formalism of unrefined topological string amplitudes which can be applied to certain non-toric geometries has been developed in 
\cite{Diaconescu:2005ik,Diaconescu:2005mv}.}. It is still possible to compute the partition function of five-dimensional $Sp(1)$ gauge theory with $N_f=6,7$ flavours
knowing that these theories can be obtained as an infrared theory in a Higgs branch of some $T_N$ theories \cite{Benini:2009gi}, to be more precise the case of
$N_f=6$ flavours is in the Higgs branch of $T_4$ theory and the case of $N_f=7$ flavours is in the Higgs branch of $T_6$ theory. In order to have some flat directions in the
Higgs branch and thus to give the possibility to some scalars in the hypermultiplets to have a non-zero vacuum expectation value it is necessary to tune
some of the parameters defining the theory (and in some cases some of the Coulomb branch moduli as well). From the perspective of the web of $(p,q)$ 5-branes the effect
of this tuning is to put some of the external 5-branes on the same 7-brane. This procedure has already been successfully applied to the case of $N_f=6$ flavours 
in \cite{Hayashi:2013qwa} and the computation agrees with the one performed using the perspective of the D4--D8--O8 system \cite{Hwang:2014uwa}.
%
%
%
%
The tuning which is necessary to realise five-dimensional $Sp(1)$ gauge theory with $N_f=6$ flavours is however a very simple one, namely in the $T_4$ diagram the
result of the tuning is to put only external D5-branes on the same D7-brane, in more general cases it will be necessary to put other possible external 5-branes in $T_N$
theories, namely NS5-branes and (1,1) 5-branes,
on the same 7-brane as well. So in order to explore in full generality the Higgs branch of $T_N$ theories it is necessary to find a prescription for putting more general
configurations of external $(p,q)$ 5-branes on the same 7-brane. This is of particular interest because the tuning which
realises five-dimensional $Sp(1)$ gauge theory with $N_f=7$ flavours in the Higgs branch of $T_6$ theory involves both D5-branes and NS5-branes.
%
%

The aim of this paper is to find a general procedure of computing the partition functions of IR theories in Higgs branches of the $T_N$ theory realised by non-toric web diagrams. For that we will first find how to tune the parameters of the theory to put external NS5-branes together on one $(0, 1)$ 7-brane in the computation of the partition
function. Originally the tuning for putting D5-branes on the same D7-brane is found by looking for a simple pole in the superconformal index whose residue is interpreted
as the superconformal index of the IR theory in the Higgs branch \cite{Gaiotto:2012uq, Gaiotto:2012xa}. However while in this case the pole is simple to locate for its position 
depends on some flavour fugacities the same will not happen if we want to put some NS5-branes on the same $(0, 1)$ 7-brane. In this case the position of the pole will depend
on an instanton fugacity. This is a difficult problem from a field theoretic point of view because the partition function obtained using localisation techniques is usually
written as a series in the instanton fugacities of the theory. This issue can be solved using the refined topological vertex. A change in the preferred direction in fact allows 
to easily identify the location of the pole, and since the refined topological vertex is conjectured to be invariant under the choice of the preferred direction \cite{Iqbal:2007ii, Awata:2009yc} this change of preferred direction will not affect the resulting partition function. This procedure will allow us to find the tuning for 
putting NS5-branes and $(1,1)$ 5-branes on the same 7-branes\footnote{In the following we will choose a particular S-dual frame, and sometimes refer to D5-branes, NS5-branes and $(1,1)$ 5-branes as horizontal,
vertical and diagonal branes respectively.}.
We will verify the validity of the procedure for both NS5-branes and $(1,1)$ 5-branes applying it to the $T_3$ web diagram, explaining also in detail how to identify 
from the web diagram the contributions of some decoupled singlet hypermultiplets that are present in the Higgs vacuum.

After establishing the tuning for putting all possible external 5-branes together, we will apply the method to the computation of the partition function of the $E_8$ theory which arises an infrared theory in the Higgs branch of the $T_6$ theory. The partition function should agree with the partition function of the $Sp(1)$ gauge theory with 7 fundamental and 1 antisymmetric hypermultiplets obtained in \cite{Hwang:2014uwa}. Although the two computations are done in a completely different way we will find
complete agreement between the two results and  this constitutes a very non-trivial check of the claim that the $E_8$ theory arises as an infrared theory in the Higgs branch of the $T_6$ theory \cite{Benini:2009gi}. Furthermore, the resulting partition function obtained by our method is written by summations of Young diagrams and therefore
it is possible to compute systematically higher order terms of the instanton fugacity.

The organisation of the paper is as follows. In section \ref{sec:Higgs.vertical}, we find a prescription of tuning associated to putting parallel external NS5-branes together on one 7-brane and we will exemplify the prescription by applying it to two theories in the Higgs branches of the $T_3$ theory. In section \ref{sec:E8}, we compute the partition function of the $E_8$ theory whose web diagram involves the tuning for putting the parallel external vertical legs together as well as putting parallel external horizontal legs together. We first describe a general procedure to obtain the partition function of an IR theory in a Higgs branch of the $T_N$ theory, and then apply the various steps to the computation of the partition function of the $E_8$ theory. Some technical details regarding the computation are relegated to appendix \ref{sec:Cartan}. Appendix \ref{sec:defs} collects definitions of the 5d partition functions and the 5d superconformal index used in this paper. In appendix \ref{sec:Higgs.diagonal}, we find a prescription of tuning associated to putting parallel external $(1,1)$ 5-branes together on one 7-brane.

\section{Tuning for coincident NS5-branes}
\label{sec:Higgs.vertical}

It is possible to explore the Higgs branch of five-dimensional $T_N$ theories using webs of  $(p, q)$ 5-branes.
In order to achieve this we will consider the case when the semi-infinite $(p, q)$ 5-branes end on an orthogonal spacetime filling $(p,q)$ 7-brane at a finite distance.
 After putting all the semi-infinite $(p, q)$ 5-branes on $(p, q)$ 7-branes, the global symmetry of the theory is realised on the $(p, q)$ 7-branes and the Higgs branch of the 
 theory opens up when several parallel external 5-branes are put on the same 7-brane. In this situation pieces of 5-branes suspended between the 7-branes can be
 moved in directions off the plane of the web and the positions of the 5-branes suspended between the 7-branes together with part of the gauge field on the 5-branes
 give a parametrisation of the Higgs branch of the theory. Since some of the 7-branes become effectively decoupled when the energy scale is much lower than the
 vacuum expectation value of the hypermultiplets deep in the infrared the theory will have a reduced global symmetry and will therefore be a different class $\mathcal{S}$ theory.
 Moreover moving in the Higgs branch may also affect the dimension of the Coulomb branch due to the s-rule \cite{Hanany:1996ie} and its generalisation 
 \cite{Benini:2009gi}.
%
%

\begin{figure}[t]
\begin{center}
\includegraphics[width=60mm]{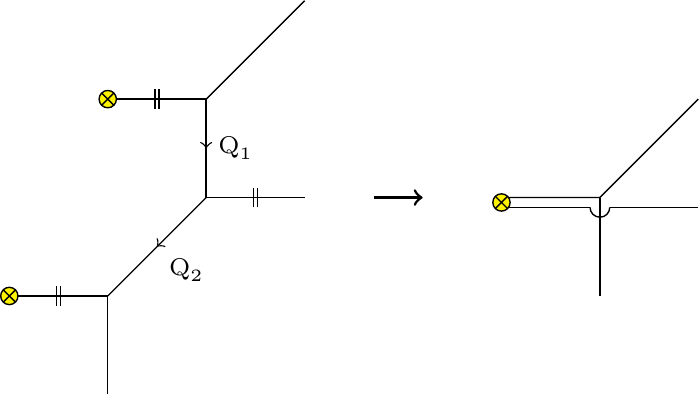}
\end{center}
\caption{The left figure shows a $(p, q)$ 5-brane web that appears as a part of the web diagram of the $T_N$ theory. The process of going to the right figure represents putting parallel horizontal external 5-branes on one 7-brane. $||$ denotes the choice of the preferred direction in the computation of the refined topological vertex. $\otimes$ represents a 7-brane.}
\label{fig:Higgs1}
\end{figure}

Putting parallel external 5-branes on one 7-brane can be achieved by tuning some parameters of the theory realised from a $(p, q)$ 5-brane web and this tuning
can be directly applied to the computation of refined topological string partition functions or five-dimensional Nekrasov partition functions \cite{Hayashi:2013qwa}
(see appendix \ref{sec:defs} for the definitions and the relations between the quantities.). Namely, after inserting the tuning into the partition function of some UV theory such as the $T_N$ theory we obtain the partition function of the low energy theory in the Higgs branch of the UV theory. Let us consider putting two parallel horizontal external 5-branes on a single 7-brane as in figure \ref{fig:Higgs1}. This can be achieved by shrinking the length of the internal 5-branes. 
In the dual M-theory picture \cite{Leung:1997tw}, the length between the 5-branes is related to the K\"ahler parameter of the corresponding two-cycle in a Calabi-Yau threefold. We denote the exponential of the K\"ahler parameters corresponding to the lengths between the 5-branes in figure \ref{fig:Higgs1} by $Q_1$ and $Q_2$ as in \eqref{kahler}. Ref.~\cite{Hayashi:2013qwa} have found that the tuning conditions for putting the parallel external D5-branes together depicted in figure \ref{fig:Higgs1} is achieved by
\be
Q_1 =  Q_2 = \left(\frac{q}{t}\right)^{\frac{1}{2}}, \label{Higgs.horizontal1}
\ee
or
\be
Q_1 =  Q_2 = \left(\frac{t}{q}\right)^{\frac{1}{2}} \label{Higgs.horizontal2}
\ee
where $q$ and $t$ are related to the $\Omega$--deformation parameters as $q = e^{-i\epsilon_2}, t=e^{i\epsilon_1}$ in the comparison with the five-dimensional Nekrasov
partition function.
In fact, both tunings \eqref{Higgs.horizontal1} and \eqref{Higgs.horizontal2} give the same result for the examples studied in \cite{Hayashi:2013qwa}.

One can understand the reason why the two tunings \eqref{Higgs.horizontal1} and \eqref{Higgs.horizontal2} give the same result in the following way. Let us first parameterise $Q_1$ and $Q_2$ by chemical potentials associated with a gauge symmetry and a global symmetry of the five-dimensional theory living on the 5-brane web in figure \ref{fig:Higgs1}.
 In the dual picture in type IIA string theory the two external D5-branes ending on D7-branes
 in figure \ref{fig:Higgs1} can be thought  as flavour branes and the one internal D5-brane in figure \ref{fig:Higgs1} 
can be thought as a colour brane. In the dual picture 
there are also strings between the flavour branes and the colour brane, which yield 
hypermultiplets that have a gauge charge as well as a flavour charge. The mass parameters for the hypermultiplets are related to the lengths between the flavour branes and the colour brane. 
Motivated by this picture we parametrise $Q_1$ and $Q_2$ as
\be
Q_1 = e^{i(\tilde{\nu}_1 - \nu)}, \qquad Q_2 = e^{i(\nu - \tilde{\nu}_2)}. \label{Higgs.para}
\ee
When one computes the refined topological string partition function by using \eqref{Higgs.para} and identifies it with the 5d Nekrasov partition function, it turns out that $\tilde{\nu}_1$ is the classical mass parameter for the hypermultiplet originating from a string between the upper flavour brane and the colour brane, and $\tilde{\nu}_2$ is the classical mass parameter for the 
hypermultiplet originating from a string between the colour brane and the lower flavour brane. 
On the other hand, $\nu$ is the Coulomb branch modulus in the theory. 
Note that we chose the orientation of the 5-branes from top to down as in figure \ref{fig:Higgs1} and we defined positive sign when an arrow of the orientation goes away from the flavour branes or the colour brane. 
Then, the exchange between the chemical potentials $\tilde{\nu}_1$ and $\tilde{\nu}_2$ may be a part of the flavour symmetry $U(2)$ associated with the two 
hypermultiplets. 
Therefore, the refined topological string partition function of the theory computed from the web diagram is invariant under the exchange between the chemical potentials $\tilde{\nu}_1$ and $\tilde{\nu}_2$. If the partition function has the symmetry, the conditions \eqref{Higgs.horizontal1} and \eqref{Higgs.horizontal2} yield the same answer after the tuning. In the case of the $T_N$ theory the exchange between the chemical potentials $\tilde{\nu}_1$ and $\tilde{\nu}_2$ is a part of the flavour symmetry of $SU(N) \subset SU(N) \times SU(N) \times SU(N)$, and hence we can use either \eqref{Higgs.horizontal1} or \eqref{Higgs.horizontal2}\footnote{The asymmetry under the exchange between $\tilde{\nu}_1$ and $\tilde{\nu}_2$ may arise in the contributions of singlet hypermultiplets in the Higgs vacuum as observed in \cite{Hayashi:2013qwa}. The partition function after decoupling the factors of the singlet hypermultiplets is invariant under the exchange. }.

One possible way to understand why either \eqref{Higgs.horizontal1} or \eqref{Higgs.horizontal2} give the correct tuning is looking at the superconformal index of the
theory.
In four-dimension, the index of a class $\mathcal{S}$ theory may be computed as a residue of the superconformal index of a UV theory which leads to the class $\mathcal{S}$ theory in the far infrared \cite{Gaiotto:2012uq, Gaiotto:2012xa}. One may apply the same method to the five-dimensional superconformal index
(see appendix \ref{sec:defs} for its definition), and the superconformal indices studied in \cite{Hayashi:2013qwa} indeed have a simple pole at
\be
Q_1Q_2 = \frac{q}{t}, \label{pole1}
\ee
and also another simple pole at
\be
Q_1Q_2 = \frac{t}{q}. \label{pole2}
\ee
Eq.~\eqref{pole1} and \eqref{pole2} is consistent with \eqref{Higgs.horizontal1} and \eqref{Higgs.horizontal2} respectively. Due to the choice of the preferred directions in figure \ref{fig:Higgs1}, the simple poles are associated with the flavour fugacity $e^{i(\tilde{\nu}_1 - \tilde{\nu}_2)}$. Hence we can easily identify the location of the pole \eqref{pole1} by looking at the perturbative part of the superconformal index.


Note also that shrinking the length of an internal 5-brane does not imply $Q = 1$ in the refined topological vertex computation although that is the case for the unrefined topological vertex. The refined version of the geometric transition suggests $Q=\left(\frac{q}{t}\right)^{\frac{1}{2}}$ or $Q= \left(\frac{t}{q}\right)^{\frac{1}{2}}$ \cite{Dimofte:2010tz, Taki:2010bj, Aganagic:2011sg, Aganagic:2012hs}. The two different results are associated with an overall normalisation ambiguity of the partition function of the refined Chern--Simons theory. By combining this result with \eqref{pole1} or \eqref{pole2}, we can obtain the conclusion \eqref{Higgs.horizontal1} or \eqref{Higgs.horizontal2}.

\begin{figure}[t]
\begin{center}
\includegraphics[width=60mm]{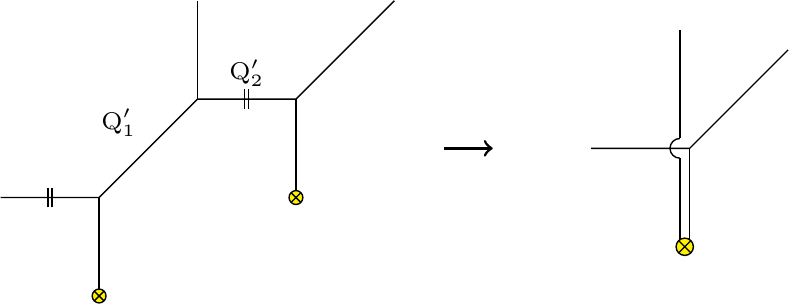}\qquad\qquad
\includegraphics[width=60mm]{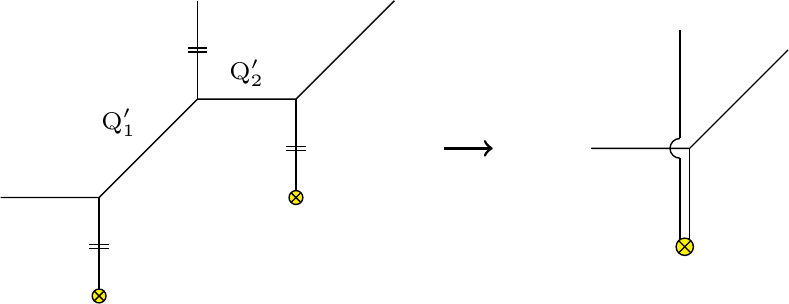}
\end{center}
\caption{Left: The process of putting the parallel vertical external 5-branes on one 7-brane with the particular choice of the preferred direction correlated with the one in figure \ref{fig:Higgs1}. Right: The same process as the left figure but with a difference choice of the preferred directions.}
\label{fig:Higgs2}
\end{figure}

Let us then consider the case of putting two parallel vertical external 5-branes on a single 7-brane as in the left figure of figure \ref{fig:Higgs2}. 
In principle, one may also compute the superconformal index and find a simple pole corresponding to the tuning in the left figure of figure \ref{fig:Higgs2}. However 
in this case the exact location of the pole depends on an instanton fugacity and therefore it is technically challenging to identify it.
This is because the refined topological vertex computation yields the expression expanded by a fugacity assigned along the preferred direction, which is related to an instanton fugacity in the corresponding 5d Nekrasov partition function. For example, when we apply the refined topological vertex computation to the left figure of figure \ref{fig:Higgs2}, we obtain an expression which is expanded by $Q_2^{\prime}$. However, one can circumvent the problem with a different choice of the preferred direction. The refined topological vertex computation is conjectured to be independent of the choice of the preferred direction \cite{Iqbal:2007ii, Awata:2009yc}, namely the partition function will not depend on the choice of the preferred direction.
From the 5-brane web picture, this is related to the S-duality in type IIB string theory.
 Therefore in order to consider the tuning corresponding to the left figure of figure \ref{fig:Higgs2} we can use a different choice of the preferred direction like the one in the right figure of figure \ref{fig:Higgs2}. Then we can sum up the 
expansions associated with both $Q_1^{\prime}$ and $Q_2^{\prime}$, and we can find a location of the poles. In fact, 
the structure of the right figure of figure \ref{fig:Higgs2} is essentially the same as that of the web diagram in figure \ref{fig:Higgs1}, and we can use the result of the tuning for putting the parallel external horizontal 5-branes together on one 7-brane. Therefore the tuning prescription associated with figure \ref{fig:Higgs2} may be given by
\be
Q_1^{\prime} = Q_2^{\prime} = \left(\frac{q}{t}\right)^{\frac{1}{2}}. \label{Higgs.vertical1}
\ee
or 
\be
Q_1^{\prime} = Q_2^{\prime} = \left(\frac{t}{q}\right)^{\frac{1}{2}}. \label{Higgs.vertical2}
\ee
The two tunings \eqref{Higgs.vertical1} and \eqref{Higgs.vertical2} should give the same answer as that was the case for the tunings \eqref{Higgs.horizontal1} and \eqref{Higgs.horizontal2}. For the later computation, we will use \eqref{Higgs.horizontal1} for the tuning associated with putting parallel external D5-branes together on one 7-brane and \eqref{Higgs.vertical1} for the tuning associated with putting the parallel external NS5-branes together on one 7-brane. 
The physical interpretation of the poles associated to \eqref{Higgs.vertical1} or \eqref{Higgs.vertical2} is given in section \ref{sec:pole}.

\subsection{$T_3$ theory revisited}
\label{subs:T3}

We will first exemplify the validity of the tuning \eqref{Higgs.vertical1} by applying it to the two parallel vertical legs of the $T_3$ theory. The infrared theory in the Higgs branch is a free theory with nine hypermultiplets\footnote{The partition function of the free theory in the Higgs branch of the $T_3$ theory by putting two parallel horizontal external 5-branes on one 7-brane has been already obtained in \cite{Hayashi:2013qwa}. }. For that, we will first review the partition function of the $T_3$ theory.

\begin{figure}[t]
\begin{center}
\includegraphics[width=60mm]{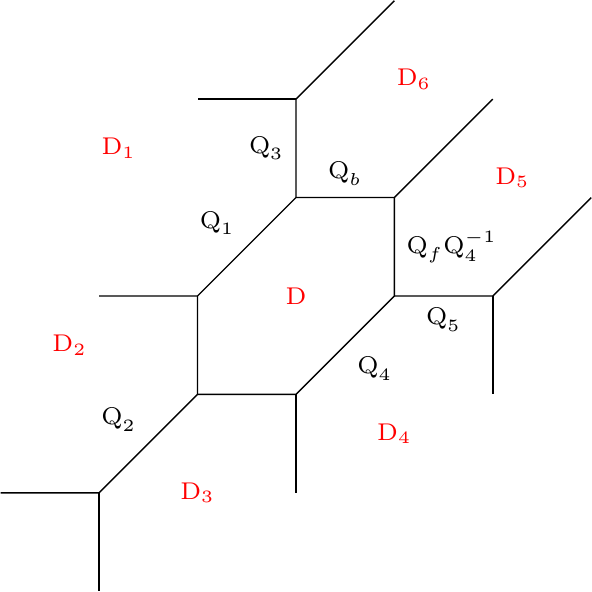}
\end{center}
\caption{The web diagram for the $T_3$ theory. $Q_i, (i=1, 2, 3, 4, 5)$ and $Q_b, Q_f$ parameterise the lengths of the corresponding internal 5-branes. }
\label{fig:T3}
\end{figure}

The web diagram for the $T_3$ theory is depicted in figure \ref{fig:T3}. 
The web diagram in fact can be interpreted as the dual toric diagram of a toric Calabi--Yau threefold \cite{Leung:1997tw} and in the dual picture the five-dimensional theory is obtained by an M-theory compactification on the Calabi--Yau threefold. From this perspective the finite length internal 5-branes are compact two-cycles in the Calabi--Yau threefold and particles in the five-dimensional theory may be understood as M2-branes wrapping two-cycles. 
The faces of the diagram corresponds to divisors in the geometry, and in particular if the face is compact the corresponding divisor will be compact as well (and similarly
 non-compact faces correspond to non-compact divisors). This is important because divisors will give symmetries in the low energy effective action by reducing
the M-theory 3-form on the Poincar\'e dual 2-form, and the corresponding symmetry
will be a gauge symmetry if the corresponding divisor is compact and it will be a global symmetry if the divisor is non-compact.
The rank of the group associated with the symmetry is the number of the divisors. In figure \ref{fig:T3}, we depict a compact divisor by $D$ and six non-compact divisors by $D_a, (a=1, \cdots, 6)$. $D$ is associated with the Cartan generator of the gauge group $U(1) \subset SU(2)$ and $D_a, (a=1, \cdots, 6)$ are associated with the Cartan generators of the global symmetry group of the $T_3$ theory, namely $SU(3) \times SU(3) \times SU(3) \subset E_6$ which is explicitly realised in the web diagram in figure \ref{fig:T3}. The intersection number between the divisor and a two-cycle gives a charge of the particle under the corresponding symmetry. Since the Calabi--Yau threefold picture and the 5-brane web picture are dual to each other, we will use both terminology interchangeably.

By using the picture of the dual Calabi--Yau threefold, one can compute the exact partition function of the five-dimensional $T_3$ theory from the refined topological vertex. The partition function of the $T_3$ theory has been obtained in \cite{Bao:2013pwa, Hayashi:2013qwa} 
\bea
Z_{T_3} &=& Z_0 \cdot Z_{\text{inst}}\cdot Z_{dec}^{-1},\label{T3}
\eea
\bea
Z_0 &=& \prod_{i,j=1}^{\infty}\Big[\frac{\prod_{a=1,4}(1-e^{-i\lambda+im_a}q^{i-\frac{1}{2}}t^{j-\frac{1}{2}})(1-e^{-i\lambda-im_a}q^{i-\frac{1}{2}}t^{j-\frac{1}{2}})}{(1-q^{i}t^{j-1})^{\frac{1}{2}}(1-q^{i-1}t^{j})^{\frac{1}{2}}(1-e^{-2i\lambda}q^it^{j-1})(1-e^{-2i\lambda}q^{i-1}t^j)}\nonumber\\
&&(1-e^{i\lambda + im_2}q^{i-\frac{1}{2}}t^{j-\frac{1}{2}})(1-e^{-i\lambda + im_2}q^{i-\frac{1}{2}}t^{j-\frac{1}{2}})(1-e^{i\lambda - im_3}q^{i-\frac{1}{2}}t^{j-\frac{1}{2}})(1-e^{-i\lambda-im_3}q^{i-\frac{1}{2}}t^{j-\frac{1}{2}})\Big],\nonumber \\ \label{T3-1} 
\eea
\bea
Z_{inst}&=&\sum_{\nu_1, \nu_2, \mu_5}u_2^{|\nu_1|+|\nu_2|}u_1^{|\mu_5|}\Big[\prod_{\alpha=1}^2\prod_{s \in \nu_{\alpha}}\frac{\left(\prod_{a=1}^32i\sin\frac{E_{\alpha\emptyset}-m_a+i\gamma_1}{2}\right)(2i\sin\frac{E_{\alpha 5} - m_4 +i\gamma_1}{2})}{\prod_{\beta=1}^2(2i)^2\sin\frac{E_{\alpha\beta}{2}}{2}\sin\frac{E_{\alpha\beta+2i\gamma_1}}{2}}\nonumber\\
&&\prod_{s\in\mu_5}\frac{\prod_{\alpha=1}^22i\sin\frac{E_{5\alpha +m_4 + i\gamma_1}}{2}}{(2i)^2\sin\frac{E_{55}}{2}\sin\frac{E_{55}+2i\gamma_1}{2}}\Big], \label{T3-2}
\eea
\bea
Z_{dec}^{-1}&=&\prod_{i,j=1}^{\infty}\Big[(1-u_1e^{im_4}q^{i}t^{j-1})(1-u_2e^{-\frac{i}{2}(m_1+m_2+m_3+m_4)}q^{i}t^{j-1})(1-u_1u_2e^{-\frac{i}{2}(m_1+m_2+m_3-m_4)}q^{i}t^{j-1})\nonumber\\
&&(1-u_1e^{-im_4}q^{i-1}t^{j})(1-u_2e^{\frac{i}{2}(m_1+m_2+m_3+m_4)}q^{i-1}t^{j})(1-u_1u_2e^{\frac{i}{2}(m_1+m_2+m_3-m_4)}q^{i-1}t^{j})\Big], \nonumber\\ \label{T3-3}
\eea
where the notation follows the ones in \cite{Hayashi:2013qwa}, namely
\bea
&&q = e^{-\gamma_1 + \gamma_2},\quad t = e^{\gamma_1 + \gamma_2}, \nonumber \\
&&E_{\alpha\beta} = \lambda_{\alpha} - \lambda_{\beta} + i(\gamma_1 + \gamma_2)l_{\nu_{\alpha}}(s) - i(\gamma_1 - \gamma_2)(a_{\nu_{\beta}}(s)+1),
\eea
and $l_{\nu}(i, j) = \nu_i - j, a_{\nu}(i, j) = \nu_j^t - i$. $\gamma_1, \gamma_2$ are related to the $\Omega$--deformation parameters as $i\epsilon_1 = \gamma_1 + \gamma_2, i\epsilon_2 = \gamma_1 - \gamma_2$. $\lambda$ is a Coulomb branch modulus and we set $\lambda_{\emptyset} = \lambda_5 = 0$ and $\lambda_1 = -\lambda_2 = \lambda$. $m_i, (i=1, \cdots, 5)$ are the masses for the $5$ fundamental hypermultiplets. The relations between the K\"ahler parameters and the parameters appearing in \eqref{T3-1}--\eqref{T3-3} are
\bea
&&Q_bQ_1^{\frac{1}{2}}Q_2^{\frac{1}{2}}Q_3^{\frac{1}{2}}Q_4^{-\frac{1}{2}} = u_2,  \quad Q_f = e^{-2i\lambda}, \quad Q_5 = e^{i\lambda}u_1,\\
&&Q_1 = e^{-i\lambda + im_1}, Q_2 = e^{i\lambda + im_2}, Q_3 = e^{i\lambda - im_3}, Q_4 = e^{-i\lambda -im_4}
\eea
The convention of the computation by the refined topological vertex used here is summarised in \cite{Hayashi:2013qwa}.

The partition function \eqref{T3} has been shown to be equal to the partition function of the $Sp(1) $ gauge theory with $5$ flavours under the reparameterisation $u_1 = e^{-im_5}$ and
\be
u = u_2e^{-\frac{i}{2}m_5} \label{shift.T3}
\ee
in \cite{Hayashi:2013qwa} up to the $3$--instanton order, where $u$ is now the instanton fugacity of the $Sp(1)$ gauge theory.

In order to reproduce the partition function of the $T_3$ theory it is important to subtract $Z_{dec}$ in \eqref{T3-3} which contains the contributions of particles which are decoupled from the $T_3$ theory. The contribution is nicely encoded in the web diagram, and it is associated with the contribution of strings between the parallel external legs \cite{Bao:2013pwa, Hayashi:2013qwa, Bergman:2013aca}. We will call this factor as a decoupled factor\footnote{The same factor was called as a non-full spin content in \cite{ Bao:2013pwa} or a $U(1)$ factor in \cite{ Hayashi:2013qwa}. The decoupled factor which we need to subtract  from the index computation of the ADHM quantum mechanics was called $Z_{\text{string}}$ indicating extra string theory states in \cite{Hwang:2014uwa}.}. Only after subtracting the decoupled factor the refined topological vertex computation yield the correct partition function of the $T_3$ theory.

\begin{figure}[t]
\begin{center}
\includegraphics[width=60mm]{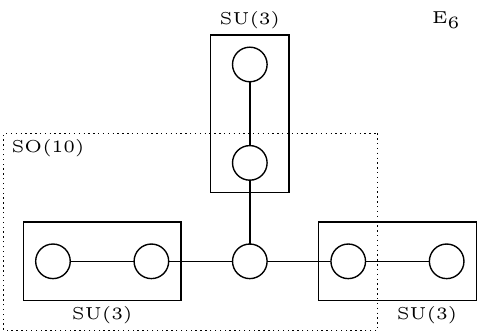}
\end{center}
\caption{The Dynkin diagram of the affine $E_6$ Lie algebra. The nodes in the dotted line represent the Dynkin diagram of $SO(10)$. The nodes in the solid lines denote the Dynkin diagram of $SU(3) \times SU(3) \times SU(3)$.}
\label{fig:E6}
\end{figure}

We can also understand the reason of the shift of the instanton fugacity \eqref{shift.T3} from the global symmetry enhancement to $E_6$. The $Sp(1)$ gauge theory with $5$ flavours perturbatively has an $SO(10) \times U(1)$ global symmetry where the $U(1)$ is the global symmetry associated with the instanton current. The global symmetry is enhanced to $E_6$ at the superconformal fixed point. On the other hand, from the web diagram of figure \ref{fig:T3}, the $SU(3) \times SU(3) \times SU(3)$ global symmetry is manifestly seen. The relation between the Lie algebras is depicted in figure \ref{fig:E6}. 
The Cartan generators of $SU(3) \times SU(3) \times SU(3)$ are associated with the non-compact divisors $D_{a}, a=1, \cdots, 6$. The simple roots of $SU(3) \times SU(3) \times SU(3)$ correspond to the two-cycles parametrised by
\bea
\{Q_1Q_3, Q_2Q_fQ_1^{-1}\} &=& \{e^{i(m_1 - m_3)}, e^{i(m_2 - m_1)}\},  \label{SU(3)1}\\
 \{Q_4Q_5, Q_bQ_1Q_4^{-1}\} &=& \{e^{-i(m_4 + m_5)}, u_2e^{\frac{i}{2}(m_1 + m_2 + m_3 + m_4)}\},\label{SU(3)2}\\
\{Q_fQ_4^{-1}Q_5, Q_bQ_3\} &=& \{e^{i(m_4 - m_5)}, u_2e^{-\frac{i}{2}(m_1 + m_2 + m_3 + m_4)}\}. \label{SU(3)3}
\eea
The charges of particles realised by M2-branes wrapping the two-cycles can be extracted by regarding \eqref{SU(3)1}--\eqref{SU(3)3} as the fugacity $e^{-i\sum_iH_im_i}$ where $H_i, (i=1, \cdots, 5)$ are charges under the Cartan generators of $SO(10)$. Since $SU(3) \times SU(3) \times SU(3) \subset E_6$, the roots of $SU(3) \times SU(3) \times SU(3)$ can be also understood as the roots of $E_6$ which are 
\be
\pm e_i \pm e_j \qquad (1 \leq i \neq j \leq 5), \label{E6.roots1}
\ee 
and 
\be
\frac{1}{2}(\pm e_1 \pm e_2 \pm e_3 \pm e_4 \pm e_5 \pm \sqrt{3}e_6). \label{E6.roots2}
\ee
with the number of the minus signs even. $e_i, (i=1, \cdots 6)$ are the orthonormal bases of $\mathbb{R}^6$. 
In order to match the charges of the particles from M2-branes wrapping the two-cycles \eqref{SU(3)1}--\eqref{SU(3)3} with the charges of the roots \eqref{E6.roots1} and \eqref{E6.roots2}, one has to shift the instanton fugacity $u_2 = ue^{\frac{i}{2}m_5}$. Then we can also see that the particles of M2-branes wrapping the two-cycles \eqref{SU(3)1}--\eqref{SU(3)3} have vectors of charges which are roots or spinor weights of $SO(10)$.

\subsection{Higgsed $T_3$ theory I}
\label{sec:Higgs1}

\begin{figure}[t]
\begin{center}
\includegraphics[width=60mm]{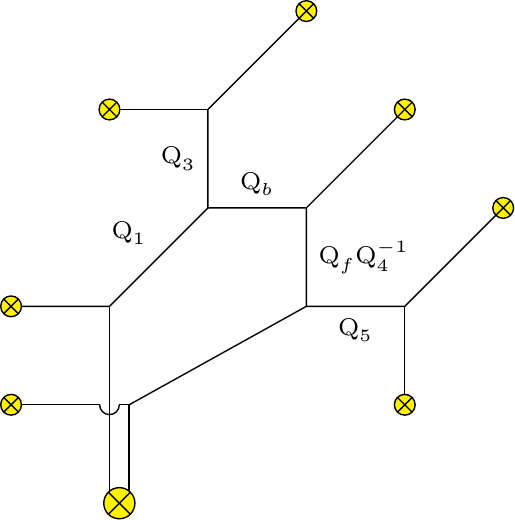}\qquad\qquad
\includegraphics[width=60mm]{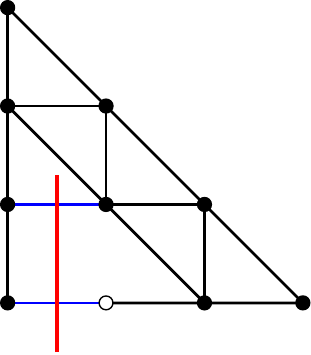}
\end{center}
\caption{Left: The web diagram of the first kind of the Higgsed $T_3$ theory. Right: The dot diagram corresponding to the web on the left. The red line shows the new external leg. }
\label{fig:T3Higgs1}
\end{figure}

Let us then consider a Higgs branch arising by putting two parallel vertical external 5-brane on one 7-brane. We will call the web diagram as the Higgsed $T_3$ web diagram and the infrared theory realised by the diagram as the Higgsed $T_3$ theory. There are two ways to do that, and we first consider putting the two leftmost parallel vertical external 5-branes together as in figure \ref{fig:T3Higgs1}.
We use the tuning \eqref{Higgs.vertical1}, and in this case it corresponds to the tuning of the K\"ahler parameters 
\be
Q_2 = \left(\frac{q}{t}\right)^{\frac{1}{2}}, \quad  Q_bQ_1Q_4^{-1} = \left(\frac{q}{t}\right)^{\frac{1}{2}}. \label{Higgs1}
\ee

By inserting the conditions \eqref{Higgs1}, the partition function of the low energy theory arising in the Higgs branch of the $T_3$ theory becomes\footnote{To get the partition function \eqref{Higgs1-part}, we erase $m_2$, $u_2$ by using the equations \eqref{Higgs1}. We can make a choice of erasing other parameters by using \eqref{Higgs1}, which does not affect any physics.} 
\bea
Z_{T_{\mathcal{IR}}} &=& Z_0 \cdot Z_{inst} \cdot Z^{-1}_{dec}, \label{Higgs1-part}
\eea
\bea
Z_0 &=& \prod_{i,j=1}^{\infty}\Big[\frac{\prod_{a=1,4}(1-e^{-i\lambda + im_a}q^{i-\frac{1}{2}}t^{j-\frac{1}{2}})(1-e^{-i\lambda - im_a}q^{i-\frac{1}{2}}t^{j-\frac{1}{2}})}{(1-q^{i}t^{j-1})^{-\frac{1}{2}}(1-q^{i-1}t^{j})^{\frac{1}{2}}(1-e^{-2i\lambda}q^{i-1}t^j)}\nonumber\\
&&\times(1-e^{i\lambda - im_3}q^{i-\frac{1}{2}}t^{j-\frac{1}{2}})(1-e^{-i\lambda - im_3}q^{i-\frac{1}{2}}t^{j-\frac{1}{2}})\Big],
\eea
\bea
Z_{inst} &=& \sum_{\nu_1, \mu_5}\left(e^{\frac{i}{2}\lambda - \frac{i}{2}(m_1 + m_3 + m_4)}\left(\frac{q}{t}\right)^{\frac{3}{4}}\right)^{|\nu_1|} u_1^{|\mu_5|}\Big[\prod_{s\in\nu_1}\frac{\prod_{a=1,3}\left(2i\sin\frac{E_{1\emptyset}-m_a + i\gamma_1}{2}\right)(2i\sin\frac{E_{15}-m_4+i\gamma_1}{2})}{(2i)^2\sin\frac{E_{11}}{2}\sin\frac{E_{11}+2i\gamma_1}{2}(2i\sin\frac{E_{1\emptyset}+\lambda+2i\gamma_1}{2})}\nonumber\\
&&\prod_{s \in \mu_5}\frac{(2i\sin\frac{E_{51}+m_4+i\gamma_1}{2})(2i\sin\frac{E_{5\emptyset} + \lambda + m_4 + i\gamma_1}{2})}{(2i)^2\sin\frac{E_{55}}{2}\sin\frac{E_{55}+2i\gamma_1}{2}}\Big], \label{Higgs1-inst}
\eea
\bea
Z^{-1}_{dec} &=& \prod_{i,j=1}^{\infty}\Big[(1-u_1e^{-im_4}q^{i-1}t^{j})(1-q^it^{j-1})(1-u_1e^{-im_4}q^it^{j-1})\nonumber\\
&&(1-u_1e^{im_4}q^it^{j-1})(1-e^{i\lambda - i(m_1+m_3 + m_4)}q^{i+\frac{1}{2}}t^{j-\frac{3}{2}})(1-u_1e^{i\lambda - i(m_1+m_3)}q^{i+\frac{1}{2}}t^{j-\frac{3}{2}})\Big],\nonumber\\
 \label{Higgs1-decoupled}
\eea
where $T_{\mathcal{IR}}$ represents the low energy theory which arises in the Higgs branch of the $T_3$ theory. Note that the Young diagram summation of $\nu_2$ disappears due to the first tuning of \eqref{Higgs1}\footnote{When we use the tuning \eqref{Higgs.vertical2}, the simplification of the disappearance of the Young diagram summation of $\nu_2$ does not happen.}. After the first tuning of \eqref{Higgs1}, the factor $\sin\left(\frac{E_{2\emptyset}+\lambda}{2}\right)$ appears. This term  always contains zero in the product of the Young diagram $\nu_2$ and therefore the Young diagram summation of $\nu_2$ vanishes.

In fact, the instanton partition function \eqref{Higgs1-inst} can be written by the product of the Plethystic exponentials 
\be
\label{eq:strange}
\begin{split}
Z_{inst}  &= \prod_{i,j=1}^{\infty}\Big[\frac{(1-Q_1Q_3Q_bq^{i-\frac{1}{2}}t^{j-\frac{1}{2}})(1-Q_1Q_3Q_4^{-1}Q_5Q_fQ_bq^{i-\frac{1}{2}}t^{j-\frac{1}{2}})(1-Q_bq^{i-\frac{1}{2}}t^{j-\frac{1}{2}})}{(1-u_1e^{-im_4}q^{i-1}t^{j})(1-u_1e^{-im_4}q^it^{j-1})(1-u_1e^{im_4}q^it^{j-1})}\\
&\times\frac{(1-Q_4^{-1}Q_5Q_bQ_fq^{i-\frac{1}{2}}t^{j-\frac{1}{2}})(1-Q_5q^{i-\frac{1}{2}}t^{j-\frac{1}{2}})(1-Q_3Q_4^{-1}Q_bQ_fq^{i-\frac{1}{2}}t^{j-\frac{1}{2}})}{(1-e^{i\lambda - i(m_1+m_3 + m_4)}q^{i+\frac{1}{2}}t^{j-\frac{3}{2}})(1-u_1e^{i\lambda - i(m_1+m_3)}q^{i+\frac{1}{2}}t^{j-\frac{3}{2}})(1-e^{-i\lambda-im_1}q^{i-\frac{1}{2}}t^{j-\frac{1}{2}})}\\
&\times\frac{(1-Q_4Q_5q^{i-\frac{1}{2}}t^{j-\frac{1}{2}})(1-e^{-2i\lambda}q^{i-1}t^{j})}{(1-e^{-i\lambda-im_4}q^{i-\frac{1}{2}}t^{j-\frac{1}{2}})(1-e^{-i\lambda-im_3}q^{i-\frac{1}{2}}t^{j-\frac{1}{2}})}\Big], 
\end{split}\ee
Checking the equality (\ref{eq:strange}) is not straightforward for in the instanton partition function \eqref{Higgs1-inst} the original instanton fugacity $u_2$ is replaced with other parameters  $\lambda, m_1, m_3, m_4$ by which the original instanton partition function is not expanded. However, we can still check the equality \label{strange} by carefully choosing expansion parameters. To do this, we need to choose expansion parameters so that we can use the expression \eqref{Higgs1-inst} truncated at some finite order of $|\nu_1|$. We first rewrite the equations on both sides of \eqref{Higgs1-inst} by $Q_1, Q_3, Q_4, Q_f$. In fact, at the zeroth order of $u_1$, both Eq.~\eqref{Higgs1-inst} and the right--hand side of (\ref{eq:strange}) can be expanded by $Q_f$ and $Q_4$ and there are no poles with respect to $Q_f$ and $Q_4$. Furthermore, if we expand \eqref{Higgs1-inst} until $k=|\nu_1|$, the expression \eqref{Higgs1-inst} is exact until $\mathcal{O}(Q_f^aQ_4^b)$ with $a + b = k$. therefore, we can check the equality (\ref{eq:strange}) by truncating the Young diagram summation of $\nu_1$ at finite order. We have checked the equality (\ref{eq:strange}) until $k=3$ order. As for the equality of the order $\mathcal{O}(u_1^l)$, the negative power of $Q_f$ and $Q_4$ appears. However, when one factors out $Q_f^{-\frac{l}{2}}Q_4^{-l}$ at each order of $\mathcal{O}(u_1^l)$, the expression of \eqref{Higgs1-inst} is exact 
until $\mathcal{O}(Q_f^aQ_4^b)$ with $a + b = k$ if we include the Young diagram summation $\nu_1$ until $|\nu_1|  = k$. Therefore, we can include the expansion until $|\nu_1| = k$, and check the equality (\ref{eq:strange}). We have checked it up to $(l, k) = (2, 2)$.

The equality (\ref{eq:strange}) enables us to write \eqref{Higgs1-part} by the product of Plethystic exponentials
\bea
Z_{T_{\mathcal{IR}}} 
&=& \prod_{i,j=1}^{\infty}\Big[(1-Q_1Q_3Q_bq^{i-\frac{1}{2}}t^{j-\frac{1}{2}})(1-Q_1Q_3Q_4^{-1}Q_5Q_bQ_fq^{i-\frac{1}{2}}t^{j-\frac{1}{2}})(1-Q_bq^{i-\frac{1}{2}}t^{j-\frac{1}{2}})\nonumber\\
&&\times(1-Q_4^{-1}Q_5Q_bQ_fq^{i-\frac{1}{2}}t^{j-\frac{1}{2}})(1-Q_5q^{i-\frac{1}{2}}t^{j-\frac{1}{2}})(1-Q_3Q_4^{-1}Q_bQ_fq^{i-\frac{1}{2}}t^{j-\frac{1}{2}})\nonumber\\
&&\times(1-e^{-i\lambda+im_1}q^{i-\frac{1}{2}}t^{j-\frac{1}{2}})(1-e^{i\lambda-im_3}q^{i-\frac{1}{2}}t^{j-\frac{1}{2}})(1-e^{-i\lambda+im_4}q^{i-\frac{1}{2}}t^{j-\frac{1}{2}})\Big]\nonumber\\
&&\times\Big[(1-q^it^{j-1})^{\frac{3}{2}}(1-q^{i-1}t^{j})^{-\frac{1}{2}}(1-Q_4Q_5q^{i-\frac{1}{2}}t^{j-\frac{1}{2}})\Big], \label{9free1}
\eea
where the factors in the first big bracket in \eqref{9free1} correspond to nine hypermultiplets of the infrared theory in the Higgs branch of the $T_3$ theory. On the other hand, the factors in the last big bracket in \eqref{9free1} correspond to singlet hypermultiplets as a result of the Higgsing.

The physical meaning of the partition function \eqref{9free1} becomes more clear when one use the parameters associated with the unbroken global symmetry $SU(3) \times SU(3) \times U(1)$. Originally, the generators of the global symmetry are $D_a, (a=1, \cdots, 6)$ and we define the parameters $\mu_a, (a = 1, \cdots, 6)$ as
\be
t_{SU(3) \times SU(3) \times SU(3)} = -i\left(\mu_1D_1 + \mu_2D_2 + \mu_1^{\prime}D_3 + \mu_2^{\prime}D_4 + \tilde{\mu}_1 D_5 + \tilde{\mu}_2 D_6\right).
\ee
Due to the tuning \eqref{Higgs1}, the generators of the unbroken flavour symmetry in the Higgsed vacuum is determined such that $Q_2$ and $Q_bQ_1Q_4^{-1}$ do not have any charge under the unbroken global symmetry. Then the generators of the unbroken global symmetry after the Higgsing can be chosen as
\be
t_{SU(3) \times SU(3) \times U(1)} =-i\left(\mu_1D_1 + \mu_2(D_2 + D) + \tilde{\mu}_1 D_5 + \tilde{\mu}_2 D_6 + \mu (D_3 + 2D_4 + D)\right). \label{Cartan.Higgs1}
\ee
By using the generators \eqref{Cartan.Higgs1}, the chemical potentials assigned to the two-cycles in figure \ref{fig:T3Higgs1} are then
\bea
&&Q_1 = e^{i(-\nu_1-\tilde{\nu}_3 - \mu)}, \quad Q_3 = e^{i(\nu_2 + \tilde{\nu}_3 + \mu)}, \quad Q_4 = e^{i(\nu_3 + \tilde{\nu}_1 - 2\mu)}, \quad Q_5 = e^{i(-\nu_3 - \tilde{\nu_1} - \mu)}, \nonumber\\
&&Q_b=e^{i(-\nu_2-\tilde{\nu}_2-\mu)}, \quad Q_f = e^{i(2\nu_3 - \tilde{\nu}_3 - \mu)}, \label{charge1}
\eea
where we used \eqref{kahler} and the divisor \eqref{Cartan.Higgs1} is Poincar\'e dual to the K\"ahler form. $\nu_i, (i=1, 2, 3)$ and $\tilde{\nu}_i, (i=1, 2, 3)$ are defined as
\bea
\nu_1 = \mu_1, \quad \nu_2 = -\mu_1 + \mu_2, \quad \nu_3 = -\mu_2,\label{charge.rewrite1}\\
\tilde{\nu}_1 = \tilde{\mu}_1, \quad \tilde{\nu}_2 = -\tilde{\mu}_1 + \tilde{\mu}_2, \quad \tilde{\nu}_3 = -\tilde{\mu}_2, \label{charge.rewrite2}
\eea
which satisfy $\sum_{i=1}^3\nu_i = \sum_{i=1}^3\tilde{\nu}_i = 0$.  

By using the parameterisation \eqref{charge1}, the partition function \eqref{9free1} becomes 
\bea
Z_{T_{\mathcal{IR}}} &=& \prod_{i,j=1}^{\infty}\Big[(1-e^{i(-\nu_1-\tilde{\nu}_2-\mu)}q^{i-\frac{1}{2}}t^{j-\frac{1}{2}})(1-e^{i(-\nu_1-\tilde{\nu}_1-\mu)}q^{i-\frac{1}{2}}t^{j-\frac{1}{2}})(1-e^{i(-\nu_2-\tilde{\nu}_2-\mu)}q^{i-\frac{1}{2}}t^{j-\frac{1}{2}})\nonumber\\
&&\times(1-e^{i(-\nu_2-\tilde{\nu}_1-\mu)}q^{i-\frac{1}{2}}t^{j-\frac{1}{2}})(1-e^{i(-\nu_3-\tilde{\nu}_1-\mu)}q^{i-\frac{1}{2}}t^{j-\frac{1}{2}})(1-e^{i(\nu_3+\tilde{\nu}_3+\mu)}q^{i-\frac{1}{2}}t^{j-\frac{1}{2}})\nonumber\\
&&\times(1-e^{i(-\nu_1-\tilde{\nu}_3-\mu)}q^{i-\frac{1}{2}}t^{j-\frac{1}{2}})(1-e^{i(\nu_2+\tilde{\nu}_3+\mu)}q^{i-\frac{1}{2}}t^{j-\frac{1}{2}})(1-e^{i(-\nu_3-\tilde{\nu}_1-\mu)}q^{i-\frac{1}{2}}t^{j-\frac{1}{2}})\Big]\nonumber\\
&&\times\Big[(1-q^it^{j-1})^{\frac{3}{2}}(1-q^{i-1}t^{j})^{-\frac{1}{2}}(1-e^{i(-3\mu)}q^{i-\frac{1}{2}}t^{j-\frac{1}{2}})\Big] \label{10hyper1-2}
\eea
We can explicitly see that \eqref{10hyper1-2} is the partition function of the free $9$ hypermultiplets associated with the global symmetry $SU(3) \times SU(3) \times U(1)$ up to singlet hypermultiplet contributions in the last big bracket, and this is in agreement with the field theory expectations.

Let us comment on how the singlet hypermultiplets in the second big bracket in \eqref{10hyper2} arise from the web digram in figure \ref{fig:T3Higgs1}. In order to understand their origin from the web, we rewrite the contributions as
\be
Z_{extra} = \left\{(1-q^it^{j-1})^{-\frac{1}{2}}(1-q^{i-1}t^{j})^{-\frac{1}{2}}\right\}\left\{(1-q^it^{j-1})^{2}\right\} \left\{(1-e^{i(-3\mu)}q^{i-\frac{1}{2}}t^{j-\frac{1}{2}})\right\}. \label{T3Higgs1.singlet}
\ee
The two factors in the first curly bracket in \eqref{T3Higgs1.singlet} simply come from the Cartan parts of the original $T_3$ theory. The two factors in the second curly bracket can be thought of as the contributions from M2-branes wrapping the two-cycle with the K\"ahler parameter $Q_2$ and $Q_bQ_1Q_4^{-1}$ respectively. This is essentially the same situation as the case of putting two horizontal external 5-branes together discussed in \cite{Hayashi:2013qwa}. At the computational level, one of the two factors in the second curly bracket comes from one of the decoupled factor in \eqref{Higgs1-decoupled}. This is because the instanton summation from the refined topological vertex automatically contains the decoupled factor $Z_{dec}$. Therefore, the singlet hypermultiplet contribution from the M2-brane wrapping the two-cycle with the K\"ahler parameter $Q_bQ_1Q_4^{-1}$  is automatically canceled in the instanton summation of $\nu_1$ when one does not take into account the decoupled factor. Then, the factor $(1-q^it^{j-1})$ in \eqref{Higgs1-decoupled} recovers the the contribution from the M2-brane wrapping $Q_bQ_1Q_4^{-1}$ which was canceled in the computation of the instanton summation of $\nu_1$.

There is also another factor of a singlet hypermultiplet which depends on the parameters associated with the flavour symmetry of the theory in \eqref{T3Higgs1.singlet}. The singlet hypermultiplet which is in the third curly bracket in \eqref{T3Higgs1.singlet} may be inferred from the web diagram of figure \ref{fig:T3Higgs1}. Since it is a contribution of a singlet which is decoupled from the infrared theory in the Higgs branch of the $T_3$ theory, it is associated with the contribution from new parallel external legs which only appear after the Higgsing as considered in \cite{Hayashi:2013qwa}. This is analogous to the decoupled factor \eqref{T3-3} before the Higgsing, which is the contribution from the parallel external legs in the $T_3$ web diagram. After the Higgsing of the first kind, an internal line becomes an external line. The new external line can be easily identified from the dot diagram depicted in the right figure of figure \ref{fig:T3Higgs1}. The dot diagram was introduced in \cite{Benini:2009gi}, and it is the dual diagram of the web diagram corresponding to a theory in a Higgs branch. The dual of the usual web diagram is a toric diagram with all the dots are denoted by black dots. The dot diagram introduces a white dot which implies the 5-branes which are separated by the white dot are on top of each other. Then, if an external line of the web diagram after a tuning crosses a line which is not on boundaries of the dot diagram, then the external line corresponds to a new external leg. For the current example, the new external leg is depicted in red color in the dot diagram of figure \ref{fig:T3Higgs1}. Then we have new parallel external legs whose distance is parameterised by $Q_4Q_5$. Therefore, M2-branes wrapping the two-cycle whose K\"ahler parameter is $Q_4Q_5$ gives a singlet hypermultiplet contribution. The contribution is nothing but the very last factor in \eqref{T3Higgs1.singlet}. Note also that in this case, the factors in the second curly bracket in \eqref{T3Higgs1.singlet} may be regarded as the contributions from new parallel external legs where the parallel external legs are on top of each other in figure \ref{fig:T3Higgs1}.

\subsection{Higgsed $T_3$ theory II}
\label{sec:Higgs2}

\begin{figure}[t]
\begin{center}
\includegraphics[width=60mm]{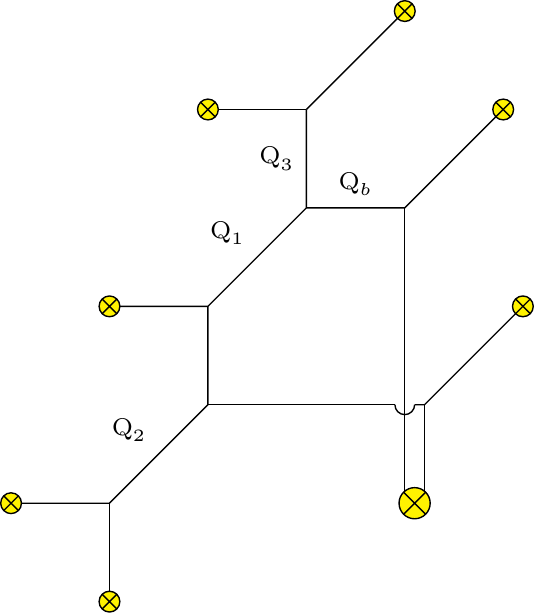}\qquad\qquad
\includegraphics[width=60mm]{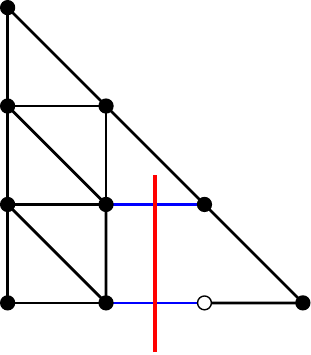}
\end{center}
\caption{Left: The web diagram of the second kind of the Higgsed $T_3$ theory. Right: The dot diagram of the web diagram on the left. The red line shows the new external leg.}
\label{fig:T3Higgs2}
\end{figure}

In this section we consider a different Higgs branch realised by putting the two rightmost parallel vertical legs together, corresponding to the figure \ref{fig:T3Higgs2}. 
By applying \eqref{Higgs.vertical1} again, the Higgs branch can be achieved by choosing the following tuning of the K\"ahler parameters
\be
Q_4 = \left(\frac{q}{t}\right)^{\frac{1}{2}}, \quad Q_5 = \left(\frac{q}{t}\right)^{\frac{1}{2}}. \label{Higgs2}
\ee
The partition function of the infrared theory in this Higgs branch of the $T_3$ theory becomes
\bea
Z_{T_\mathcal{IR}} &=&Z_0 \cdot Z_{inst}\cdot Z_{dec}^{-1}, \label{Higgs2-part}
\eea
\bea
Z_0 &=& \prod_{i,j=1}^{\infty}\Big[\frac{(1-e^{-i\lambda+im_1}q^{i-\frac{1}{2}}t^{j-\frac{1}{2}})(1-e^{-i\lambda-im_1}q^{i-\frac{1}{2}}t^{j-\frac{1}{2}})}{(1-q^{i}t^{j-1})^{-\frac{1}{2}}(1-q^{i-1}t^{j})^{\frac{1}{2}}(1-e^{-2i\lambda}q^it^{j-1})}\nonumber\\
&&(1-e^{i\lambda + im_2}q^{i-\frac{1}{2}}t^{j-\frac{1}{2}})(1-e^{-i\lambda + im_2}q^{i-\frac{1}{2}}t^{j-\frac{1}{2}})(1-e^{i\lambda - im_3}q^{i-\frac{1}{2}}t^{j-\frac{1}{2}})(1-e^{-i\lambda-im_3}q^{i-\frac{1}{2}}t^{j-\frac{1}{2}})\Big], \nonumber 
\eea
\bea
Z_{inst}&=&\sum_{\nu_1, \nu_2, \mu_5}u_2^{|\nu_1|+|\nu_2|}\left(e^{-i\lambda}\left(\frac{q}{t}\right)^{\frac{1}{2}}\right)^{|\mu_5|}\Big[\prod_{\alpha=1}^2\prod_{s \in \nu_{\alpha}}\frac{\left(\prod_{a=1}^32i\sin\frac{E_{\alpha\emptyset}-m_a+i\gamma_1}{2}\right)(2i\sin\frac{E_{\alpha 5} +\lambda  +2i\gamma_1}{2})}{\prod_{\beta=1}^2(2i)^2\sin\frac{E_{\alpha\beta}{2}}{2}\sin\frac{E_{\alpha\beta+2i\gamma_1}}{2}}\nonumber\\
&&\prod_{s\in\mu_5}\frac{\prod_{\alpha=1}^22i\sin\frac{E_{5\alpha}-\lambda}{2}}{(2i)^2\sin\frac{E_{55}}{2}\sin\frac{E_{55}+2i\gamma_1}{2}}\Big], \label{Higgs2-inst}
\eea
\bea
Z_{dec}^{-1}&=&\prod_{i,j=1}^{\infty}\Big[(1-e^{-2i\lambda}q^{i}t^{j-1})(1-u_2e^{\frac{i}{2}\lambda-\frac{i}{2}(m_1+m_2+m_3)}q^{i+\frac{1}{4}}t^{j-\frac{5}{4}})(1-u_2e^{-\frac{3i}{2}\lambda-\frac{i}{2}(m_1+m_2+m_3)}q^{i+\frac{1}{4}}t^{j-\frac{5}{4}})\nonumber\\
&&(1-q^{i}t^{j-1})(1-u_2e^{-\frac{i}{2}\lambda+\frac{i}{2}(m_1+m_2+m_3)}q^{i-\frac{5}{4}}t^{j+\frac{1}{4}})(1-u_2e^{-\frac{i}{2}\lambda+\frac{i}{2}(m_1+m_2+m_3)}q^{i-\frac{1}{4}}t^{j+\frac{3}{4}})\Big], \nonumber\\
\eea

The tuning of $Q_4$ by \eqref{Higgs2} simplifies the Young diagram summation of $\mu_5$. The instanton partition function \eqref{Higgs2-inst} can be non-zero if $\mu_{5, i} \leq \nu_{2, i}$ for all $i$. Here $\nu_i$ implies an i--th row of the Young diagram $\nu$. Namely, the summation of $\mu_5$ vanishes if any i--th row of $\mu_5$ is not greater than the i--th row of $\nu_2$. This is due to the term $\sin\left(\frac{E_{52} - \lambda}{2}\right)$ in \eqref{Higgs2-inst}. If $\mu_{5, 1} > \nu_{2, 1}$, then the function $\sin\left(\frac{E_{52} - \lambda}{2}\right)$ at $(1, |\mu_{5,1}|) \in \mu_5$ gives zero. If we then assume $\mu_{5, 1} \leq \nu_{2, 1}$ and $\mu_{5, 2} > \nu_{2, 2}$, the function $\sin\left(\frac{E_{52} - \lambda}{2}\right)$ at $(2, |\mu_{5, 2}|) \in \mu_5$ yields zero. In this way, the term $\sin\left(\frac{E_{52} - \lambda}{2}\right)$ gives zero unless $\mu_{5, i} \leq \nu_{2, i}$ for all $i$. Therefore, until the order $\mathcal{O}(u_1^{|\nu_1| + |\nu_2|})$ with $|\nu_1| + |\nu_2| = k$, the expansion by $u_2$ is exact when one includes the expansion regarding $\mu_5$ until $|\mu_5| = |\nu_2| \leq k$. Note also that there are non-zero contributions from $|\mu_5| \neq |\nu_2|$ although the two-cycles associated with the Young diagrams $\mu_5$ and $\nu_2$ are connected with each other in the web diagram.

The instanton partition function \eqref{Higgs2-inst} again can be written as the product of Plethystic exponentials
\bea
Z_{inst} &=& \prod_{i,j=1}^{\infty}\Big[\frac{(1-Q_bq^{i-\frac{1}{2}}t^{j-\frac{1}{2}})(1-Q_1Q_3Q_bq^{i-\frac{1}{2}}t^{j-\frac{1}{2}})}{(1-u_2e^{\frac{i}{2}\lambda-\frac{i}{2}(m_1+m_2+m_3)}q^{i+\frac{1}{4}}t^{j-\frac{5}{4}})(1-u_2e^{-\frac{3i}{2}\lambda-\frac{i}{2}(m_1+m_2+m_3)}q^{i+\frac{1}{4}}t^{j-\frac{5}{4}})}\nonumber\\
&&\times\frac{(1-Q_2Q_3Q_bQ_fq^{i-\frac{1}{2}}t^{j-\frac{1}{2}})(1-Q_1Q_2Q_bq^{i-\frac{1}{2}}t^{j-\frac{1}{2}})}{(1-u_2e^{-\frac{i}{2}\lambda+\frac{i}{2}(m_1+m_2+m_3)}q^{i-\frac{5}{4}}t^{j+\frac{1}{4}})(1-u_2e^{-\frac{i}{2}\lambda+\frac{i}{2}(m_1+m_2+m_3)}q^{i-\frac{1}{4}}t^{j+\frac{3}{4}})}\Big], \nonumber \\\label{Higgs2-PE}
\eea
The equality of \eqref{Higgs2-PE} has been checked up to $\mathcal{O}(u_2^2)$. Then, the partition function \eqref{Higgs2-part} becomes
\bea
Z_{T_{\mathcal{IR}}} &=& \prod_{i,j=1}^{\infty}\Big[(1-Q_bq^{i-\frac{1}{2}}t^{j-\frac{1}{2}})(1-Q_1Q_3Q_bq^{i-\frac{1}{2}}t^{j-\frac{1}{2}})(1-Q_2Q_3Q_bQ_fq^{i-\frac{1}{2}}t^{j-\frac{1}{2}})\nonumber\\
&&\times(1-e^{-i\lambda+im_1}q^{i-\frac{1}{2}}t^{j-\frac{1}{2}})(1-e^{-i\lambda-im_1}q^{i-\frac{1}{2}}t^{j-\frac{1}{2}})(1-e^{i\lambda + im_2}q^{i-\frac{1}{2}}t^{j-\frac{1}{2}})\nonumber\\
&&\times(1-e^{-i\lambda + im_2}q^{i-\frac{1}{2}}t^{j-\frac{1}{2}})(1-e^{i\lambda - im_3}q^{i-\frac{1}{2}}t^{j-\frac{1}{2}})(1-e^{-i\lambda-im_3}q^{i-\frac{1}{2}}t^{j-\frac{1}{2}})\Big]\nonumber\\
&&\times\Big[(1-q^{i}t^{j-1})^{\frac{3}{2}}(1-q^{i-1}t^{j})^{-\frac{1}{2}}(1-Q_1Q_2Q_bq^{i-\frac{1}{2}}t^{j-\frac{1}{2}})\Big]. \label{10hyper2}
\eea
As with the case of \eqref{9free1}, the factors in the first big bracket stand for the nine free hypermultiplets and the factors in the last big bracket represent the singlet hypermultiplet contributions.

One can again rewrite the partition function \eqref{10hyper2} by the parameters associated with the global symmetry $SU(3) \times SU(3) \times U(1)$. The generators of the unbroken global symmetry can be found by requiring that $Q_2$ and $Q_bQ_1Q_4^{-1}$ have no charge under the unbroken global symmetry in the Higgs branch. Then the generators are
\be
t_{SU(3)\times SU(3) \times U(1)} = -i\left(\mu_1D_1 + \mu_2 D_2 + \tilde{\mu}_1(D_5 + D) + \tilde{\mu}_2 D_6 + \mu(2D_3 + D_4 + D)\right).
\ee
The parameterisation of the two-cycles with finite size in figure \ref{fig:T3Higgs2} is 
\bea
&&Q_1 = e^{i(\nu_2+\tilde{\nu}_2-\mu)}, \quad Q_2 = e^{i(\nu_3 + \tilde{\nu}_1 - \mu)}, \quad Q_3 = e^{i(-\nu_1 - \tilde{\nu}_2 + \mu)}, \nonumber\\
&&Q_b = e^{i(\nu_1 + \tilde{\nu}_3 - \mu)}, \quad Q_f = e^{i(-\tilde{\nu}_1 + \tilde{\nu}_2)}, \label{charge2}
\eea
where we again use \eqref{charge.rewrite1} and \eqref{charge.rewrite2}. By using the parameters \eqref{charge2}, the partition function \eqref{10hyper2} can be written 
\bea
Z_{T_{\mathcal{IR}}} &=& \prod_{i,j=1}^{\infty}\Big[(1-e^{i(\nu_1+\tilde{\nu}_3-\mu)}q^{i-\frac{1}{2}}t^{j-\frac{1}{2}})(1-e^{i(\nu_2+\tilde{\nu}_3-\mu)}q^{i-\frac{1}{2}}t^{j-\frac{1}{2}})(1-e^{i(\nu_3+\tilde{\nu}_3-\mu)}q^{i-\frac{1}{2}}t^{j-\frac{1}{2}})\nonumber\\
&&\times(1-e^{i(\nu_2+\tilde{\nu}_2-\mu)}q^{i-\frac{1}{2}}t^{j-\frac{1}{2}})(1-e^{i(-\nu_2-\tilde{\nu}_1+\mu)}q^{i-\frac{1}{2}}t^{j-\frac{1}{2}})(1-e^{i(\nu_3+\tilde{\nu}_1-\mu)}q^{i-\frac{1}{2}}t^{j-\frac{1}{2}})\nonumber\\
&&\times(1-e^{i(\nu_3+\tilde{\nu}_2-\mu)}q^{i-\frac{1}{2}}t^{j-\frac{1}{2}})(1-e^{i(-\nu_1+\tilde{\nu}_2+\mu)}q^{i-\frac{1}{2}}t^{j-\frac{1}{2}})(1-e^{i(-\nu_1-\tilde{\nu}_1+\mu)}q^{i-\frac{1}{2}}t^{j-\frac{1}{2}})\Big]\nonumber\\
&&\times\Big[(1-q^{i}t^{j-1})^{\frac{3}{2}}(1-q^{i-1}t^{j})^{-\frac{1}{2}}(1-e^{-3i\mu}q^{i-\frac{1}{2}}t^{j-\frac{1}{2}})\Big]. \label{10hyper2-2}
\eea
Therefore, we can explicitly see that the partition function describes the free $9$ hypermultiplets associated with the global symmetry $SU(3) \times SU(3) \times U(1)$ plus singlet hypermultiplets in this case also.  

We can again understand the origin of the singlet hypermultiplets from the web diagram of figure \ref{fig:T3Higgs2} as in section \ref{sec:Higgs2}. The singlet hypermultiplets contribution which only depend on the $\Omega$--deformation parameters come from the Cartan parts of the original $T_3$ theory and also M2-branes wrapping the two-cycle with the K\"ahler parameter $Q_4$ or $Q_5$. The total contributions explain the factors $(1-q^{i}t^{j-1})^{\frac{3}{2}}(1-q^{i-1}t^{j})^{-\frac{1}{2}}$ in \eqref{10hyper2-2}. Also, the contribution of the very last factor of \eqref{10hyper2-2} comes from the new parallel external legs after the Higgsing. The new external leg in the dot diagram is depicted in red color in the right figure of figure \ref{fig:T3Higgs2}, which corresponds to the two-cycle with the K\"ahler parameter $Q_fQ_4^{-1}$. Then the distance between the new parallel external legs is parameterised by $Q_1Q_2Q_b$. Hence the singlet hypermultiplet from M2-branes wrapping the two-cycle yields the contribution which is nothing but the very last factor in \eqref{10hyper2-2}.

\subsection{Physical interpretation of poles}
\label{sec:pole}

The five-dimensional  superconformal index \eqref{5d.index} may have many poles in the flavour fugacities and the residue of the poles have a physical meaning. In order to understand the essence of the physical meaning, we follow the argument of \cite{Gaiotto:2012uq,Gaiotto:2012xa} and write an index of a theory $\mathcal{T}$  schematically as
\be
I(a, b) = \text{Tr}(-1)^Fa^fb^g, \label{index.schematic}
\ee
where $f, g$ are flavour charges of two flavour symmetries whose fugacities are $a$ and $b$ respectively. Let us suppose that the index \eqref{index.schematic} has a pole like
\be
I(a, b) = \frac{\tilde{I}(a, b)}{1 - a^{f_{\mathcal{O}}}b^{g_{\mathcal{O}}}}. \label{index1}
\ee
where $f_{\mathcal{O}}$ and ${g_{\mathcal{O}}}$ are the flavour charges of a bosonic operator $\mathcal{O}$. The index \eqref{index1} diverges when $ a^{f_{\mathcal{O}}}b^{g_{\mathcal{O}}} = 1$. The divergence arises due to a bosonic zero--mode of the operator $\mathcal{O}$, and arbitrary high powers of the operator $\mathcal{O}$ contribute to the index. The residue of the index \eqref{index1} at the pole $a^{f_{\mathcal{O}}}b^{g_{\mathcal{O}}} = 1$ is then given by
\be
\tilde{I}(b^{-\frac{g_{\mathcal{O}}}{f_\mathcal{O}}}, b) = \text{Tr}(-1)^Fb^{g^{\prime}}, \label{index2}
\ee
where $g^{\prime}$ is 
\be
g^{\prime} = g - \frac{g_{\mathcal{O}}}{f_{\mathcal{O}}}f. \label{shift}
\ee
 The shift of the charge \eqref{shift} in the residue \eqref{index2} indicates that the operator $\mathcal{O}$ gets a vacuum expectation value and only one flavour symmetry whose charge is give by \eqref{shift} is left unbroken in the vacuum. Therefore, the residue \eqref{index2} should correspond to an index of an IR theory $\mathcal{T}_{IR}$ that is realised at the end point of the RG flow of the UV theory $\mathcal{T}$ induced by the vacuum expectation value of the operator $\mathcal{O}$. The residue \eqref{index2} typically contains contributions of free hypermultiplets. The genuine index of the IR theory $\mathcal{T}_{IR}$ is obtained after removing the contributions.

This technique was applied in \cite{Hayashi:2013qwa} to obtain tuning conditions \eqref{Higgs.horizontal1} or \eqref{Higgs.horizontal2} for yielding a 5d partition function of an IR theory which is realised in the far infrared limit in a Higgs branch of a UV theory. If we consider a UV theory $\mathcal{T}$ whose web diagram realisation contains a diagram in figure \ref{fig:Higgs1}, the superconformal index \eqref{5d.index} may have poles \cite{Hayashi:2013qwa}
\be
I(\gamma_1, \gamma_2, m_i, u) = \frac{\tilde{I}(\gamma_1, \gamma_2, m_i, u)}{(1 - Q_1Q_2e^{-2\gamma_1})(1 - Q_1^{-1}Q_2^{-1}e^{-2\gamma_1})}. \label{index3}
\ee
assuming $e^{-\gamma_1}\ll 1$. The index \eqref{index3} has a pole at $Q_1Q_2e^{-2\gamma_1} = 1$, which corresponds to \eqref{Higgs.horizontal2}. Therefore the operator associated to the divergence has charges $(j_r, j_l) = (0, 0)$, $j_R = 1$ and also the flavour charge $-1, +1$ associated to the fugacity $e^{-i\nu_1}, e^{-i\nu_2}$ respectively. This is nothing but a part of a mesonic operator in the adjoint representation of the $U(2)$. Therefore, the residue of \eqref{index3} evaluated at the pole $Q_1Q_2e^{-2\gamma_1}  = 1$ should correspond to an index of an IR theory $\mathcal{T}_{IR}$ at the end point of the RG flow triggered by the vacuum expectation value of the mesonic operator from the UV theory $\mathcal{T}$. This result agrees with the 5-brane web picture. The mesonic operator is associated to a string connecting the upper external horizontal 5-brane with the lower external horizontal 5-brane, In order to open up the Higgs branch, we tune parameters such that the two external 5-branes are put together. This means that the meson becomes massless and gives rise to a flat direction. Hence, stripping off the piece of the 5-brane between the two 7-branes correspond to giving the vacuum expectation value for the mesonic operator. Therefore, giving the vacuum expectation value for the mesonic operator exactly corresponds to moving to the Higgs branch we are considering. The index has another pole at  $Q_1^{-1}Q_2^{-1}e^{-2\gamma_1} = 1$, which corresponds to \eqref{Higgs.horizontal1}. This pole is associated to another part of the mesonic operator with $(j_r, j_l) = (0, 0)$, $j_R = 1$ and also the flavour charge $+1, -1$ associated to the fugacity $e^{-i\nu_1}, e^{-i\nu_2}$ respectively. The vacuum expectation value of the mesonic operator yields the same IR theory $\mathcal{T}_{IR}$.

The physical interpretation of the pole in the case of putting two external vertical 5-branes together in figure \ref{fig:Higgs2} is essentially the S-dual version of that in the case of figure \ref{fig:Higgs1}. By using the partition function computed by the refined topological vertex with the vertical lines chosen as the preferred directions, we can see that the superconformal index has simple poles 
\be
I(\gamma_1, \gamma_2, m_i, u) = \frac{\tilde{I}(\gamma_1, \gamma_2, m_i, u)}{(1 - Q_1^{\prime}Q_2^{\prime}e^{-2\gamma_1})(1 - Q_1^{\prime -1}Q_2^{\prime -1}e^{-2\gamma_1})}. \label{index4}
\ee
The important difference from the poles in \eqref{index3} is that the fugacity $Q_2^{\prime}$ may contain an instanton fugacity. The operator corresponds to the divergence has charges $(j_r, j_l) = (0, 0)$ and $j_R = 1$ but now it also carries the instanton number. From the brane picture, the operator is associated to a string between the external vertical 5-branes.

In the explicit example of the Higgsed $T_3$ theory in section \ref{sec:Higgs1}, we used the pole located at 
\be
Q_b^{-1}Q_1^{-1}Q_2^{-1}Q_4e^{-2\gamma_1}=  u^{-1}e^{\frac{-i}{2}\left(m_1+m_2+m_3+m_4+m_5\right)}e^{-2\gamma_1} = 1.
\ee
Therefore the operator responsible for the divergence has charges $(j_r, j_l) = (0, 0)$ and $j_R = 1$ and its charges form a weight of the Weyl spinor representation of $SO(10)$ with positive chirality\footnote{Note that we introduced the fugacity by $e^{-i\sum_i H_im_i}$ in \eqref{5d.index}.}, and carries the instanton number $-1$. On the other hand, for the example of the Higgsed $T_3$ theory in section \ref{sec:Higgs2}, we used the pole located at 
\be
Q_4^{-1}Q_5^{-1}e^{-2\gamma_1} = e^{i(m_4 + m_5)}e^{-2\gamma_1} = 1. 
\ee
The pole is associated to the perturbative operator that has charges $(j_r, j_l) = (0, 0)$ and $j_R = 1$, and has a vector of charges which is a root of $SO(10)$.

\section{The partition function of the $E_8$ theory}
\label{sec:E8}

In this section we will apply the tuning discussed in section \ref{sec:Higgs.vertical} to the diagram of $T_6$ theory to realise the $E_8$ theory which we know can
be realised in Higgs branch of $T_6$ theory deep in the infrared \cite{Benini:2009gi}. In section \ref{sec:Higgs.vertical}, we have seen how to obtain the partition functions of the free theory in the Higgs branch of the $T_3$ theory. We propose here the general procedure to obtain a partition function of an infrared theory realised in a Higgs branch by putting several 5-branes on a 7-brane. 

\begin{enumerate}
\item We first compute the partition function of a UV theory by the refined topological vertex method. It is important to remove the decoupled factors which are associated with the parallel external legs.

\item In order to put two or more external 5-branes on the same 7-brane we impose a condition \eqref{Higgs.horizontal1} or \eqref{Higgs.horizontal2} in the case of horizontal 5-branes, or \eqref{Higgs.vertical1} or \eqref{Higgs.vertical2} in the case of vertical 5-branes\footnote{For the tuning of putting the parallel diagonal external 5-branes together, we can use the condition \eqref{Higgs.diagonal1} or \eqref{Higgs.diagonal2}.}. Moreover if necessary we also tune some of the K\"ahler parameters of 
the internal two-cycles of the diagram. Whether this is necessary or not it is determined by consistency constraints of the geometry, and quite interestingly this is equivalent
to the propagation of the generalised s-rule inside the diagram \cite{Benini:2009gi}.

\item We parameterise the lengths of internal 5-branes or the K\"ahler parameters of compact two-cycles by the chemical potentials associated with unbroken gauge symmetries and those of unbroken global symmetries. The unbroken symmetries can be determined by requiring that the tuned two-cycles have no charge under the unbroken symmetries in the Higgs vacuum. Linear combinations of the Cartan generators of the unbroken global symmetries are associated with masses and instanton fugacities in the perturbative regime. 

\item After inserting the tuning conditions as well as the new parameterisation, we almost obtain the partition function of the low energy theory in the Higgs branch of the UV theory. However, there can be still some contributions from singlet hypermultiplets. We need to remove such contributions. The singlet hypermultiplet factor which depends on some parameters associated with flavour symmetries in the theory may be inferred from the web diagram. The contribution of such a singlet hypermultiplet is associated with strings between new parallel external 5-branes which only appear after moving to the Higgs branch. Note that such a singlet hypermultiplet contribution can depend on an instanton fugacity. The other singlet hypermultiplet factor which only depends on the $\Omega$--deformation parameters appears in the perturbative part, namely the zero-th order of the instanton fugacities. Once we obtain the perturbative part, we can identify those contributions. 

\item After eliminating the singlet hypermultiplet contributions, we finally obtain the partition function of the infrared theory in the Higgs branch.
\end{enumerate}

In this section, we will obtain the partition function of the $E_8$ theory by applying this procedure to the $T_6$ diagram.

\subsection{$T_6$ partition function}
\begin{figure}[t]
\begin{center}
\includegraphics[width=120mm]{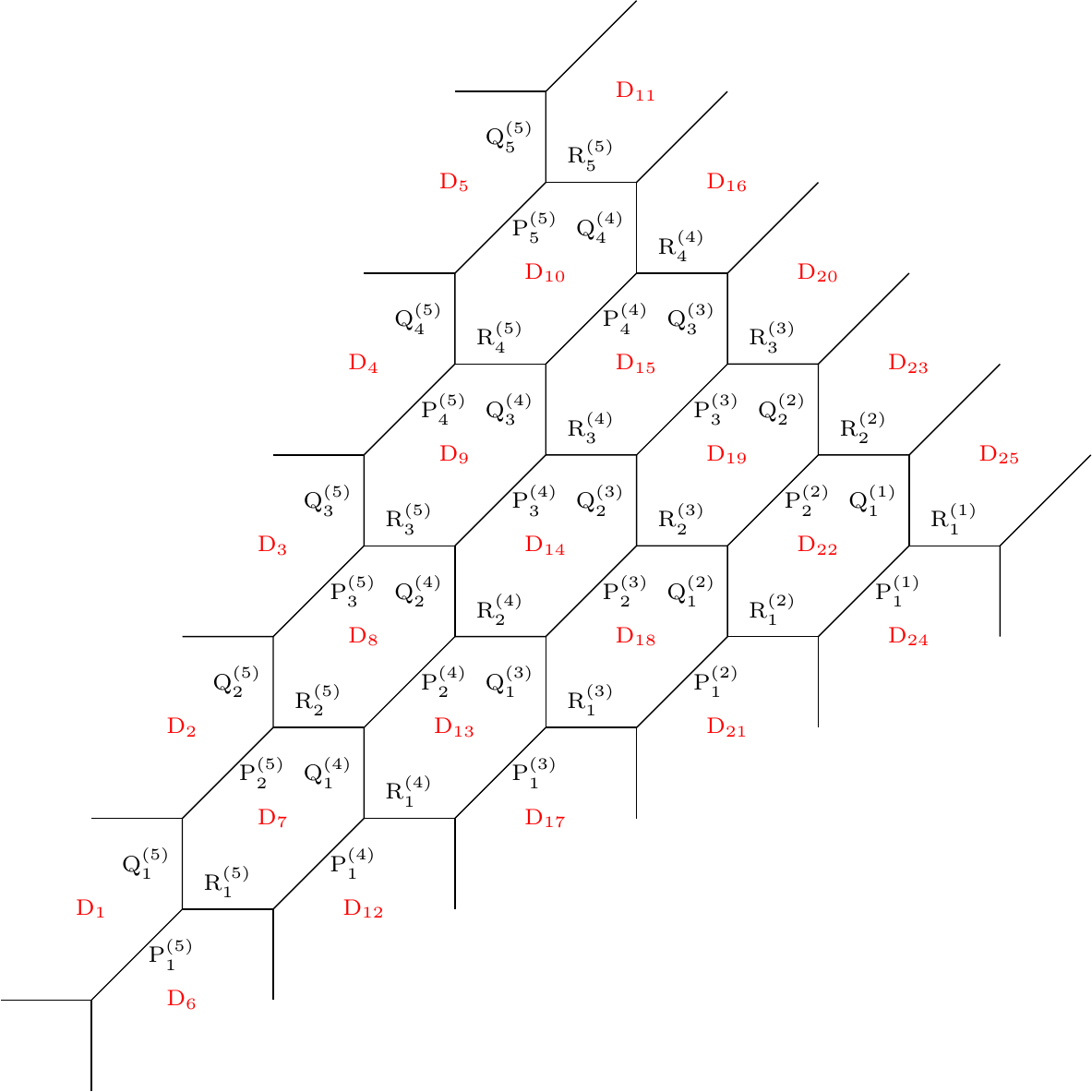}
\end{center}
\caption{The web diagram for the $T_6$ theory. }
\label{fig:T6}
\end{figure}
In this section we review the partition function of the $T_6$ theory. This theory can be obtained by compactifying M-theory on the blow-up of  $\mathbb{C}^3/(\mathbb{Z}_6\times \mathbb{Z}_6)$ whose toric diagram we show in 
figure \ref{fig:T6}. In the figure we also show how the fugacities $P^{(n)}_k$, $Q^{(n)}_k$ and $R^{(n)}_k$ are associated to the two cycles present in the geometry. Note that the geometry imposes some conditions on these fugacities
\be\label{cons}
Q_k^{(n)}P_k^{(n)} = Q_k^{(n+1)}P_{k+1}^{(n+1)}\,, \quad R_k^{(n+1)} Q_k^{(n+1)} = R_{k+1}^{(n+1)} Q_k^{(n)}\,,
\ee
so that the actual number of K\"ahler parameters is 25.
The partition function of this theory was computed in \cite{Bao:2013pwa, Hayashi:2013qwa} and here we simply quote the result
\be
 Z_{T_6} = (M(t,q) M(q,t))^{5}\, Z_0 \,Z_{inst}\, Z_{dec}^{-1}\,,
\ee
\be
M(t,q) = \prod_{i,j=1}^{\infty} (1-q^{i} t^{j-1})^{-1}\,,
\ee
\be\label{Z:inst}\begin{split}
Z_0 &=\prod_{i,j=1}^\infty \left\{\frac{\left[\prod_{a \leq b}(1-e^{-i \lambda_{5;b}+i \tilde m_a}q^{i-\frac{1}{2}}t^{j-\frac{1}{2}})\prod_{b < a}(1-e^{i \lambda_{5;b}-i \tilde m_a}q^{i-\frac{1}{2}}t^{j-\frac{1}{2}})\right]}{\prod_{n=1}^5 \prod_{a<b}(1-e^{i \lambda_{n;a}-i\lambda_{n,b}} q^i t^{j-1})(1-e^{i \lambda_{n;a}-i\lambda_{n,b}} q^{i-1} t^{j})}\right\}\\
&\times \prod_{n=2}^5 \prod_{a \leq b}(1- e^{i \lambda_{n;a}-i \lambda_{n-1;b}+i \hat m_n}q^{i - \frac{1}{2}}t^{j-\frac{1}{2}})(1- e^{i \lambda_{n-1;b}-i \lambda_{n;a}-i \hat m_n}q^{i - \frac{1}{2}}t^{j-\frac{1}{2}})\,,
\end{split}\ee
\be\begin{split}
Z_{inst} &= \sum_{\vec Y_1 ,\dots, \vec Y_5}\left\{\prod_{n=1}^4 u_n^{|\vec Y_n|}\prod_{\alpha=1}^n\prod_{s\in Y_{n,\alpha}}\frac{\left[\prod_{\beta=1}^{n+1}2i \sin \frac{E_{\alpha \beta} - \hat m_{n+1} +i \gamma_1}{2}\right]\left[\prod_{\beta=1}^{n-1}2i \sin \frac{E_{\alpha \beta} + \hat m_{n} +i \gamma_1}{2}\right]}{\prod_{\beta=1}^n(2i)^2 \sin \frac{E_{\alpha \beta}}{2} \sin \frac{E_{\alpha \beta}+2i \gamma_1}{2}} \right\} \\
&\times\left\{u_5^{|\vec Y_5|}\prod_{\alpha=1}^5\prod_{s\in Y_{5,\alpha}}\frac{\left[\prod_{\kappa=1}^{6}2i \sin \frac{E_{\alpha \emptyset} - \tilde m_{\kappa} +i \gamma_1}{2}\right]\left[\prod_{\beta=1}^{4}2i \sin \frac{E_{\alpha \beta} + \hat m_{5} +i \gamma_1}{2}\right]}{\prod_{\beta=1}^5(2i)^2 \sin \frac{E_{\alpha \beta}}{2} \sin \frac{E_{\alpha \beta}+2i \gamma_1}{2}}\right\}\,,
\end{split}\ee
\be
Z_{dec}^{-1} = \prod_{i,j=1}^\infty \prod_{1\leq a < b \leq 6}\left(1-\bigg(\prod_{n=a}^{b-1}R_1^{(n)}P_1^{(n)}\bigg)q^{i-1}t^j\right)\left(1-\bigg(\prod_{n=a}^{b-1}R_n^{(n)}Q_n^{(n)}\bigg)q^{i}t^{j-1}\right)\,.
\ee
In writing the partition function we have used the Coulomb branch moduli $\lambda_{n;k}$ with $1\leq k\leq n=2,\dots,5$ defined by
\be
P_k^{(n-1)} Q_k^{(n-1)}= \exp(-i \lambda_{n;k+1}+i \lambda_{n;k})\,,
\ee 
and subject to the condition $\sum_{k=1}^n \lambda_{n;k}=0$. Moreover the parameters $\hat m_n$ with $n=2,\dots 5$ are defined by
\be
P_k^{(n-1)} = \exp( i \lambda_{n;k}-i \lambda_{n-1;k}+i\hat  m_n)\,,
\ee
and the parameters $\tilde m_k$ with $k= 1, \dots 6$ by
\be
P^{(5)}_k Q^{(5)}_k= \exp(-i \tilde m_{k+1}+i \tilde m_k)\,, \quad P_k^{(5)} = \exp(i \tilde m_k -i \lambda_{5;k})\,.
\ee
Finally the parameters $u_k$ with $k=1,\dots,5$ are defined as
\be
u_k= \sqrt{R_1^{(k)}P_1^{(k)}R_k^{(k)} Q_k^{(k)}}\,.
\ee


\subsection{The $E_8$ theory from $T_6$ theory}

\begin{figure}[t]
\begin{center}
\includegraphics[width=150mm]{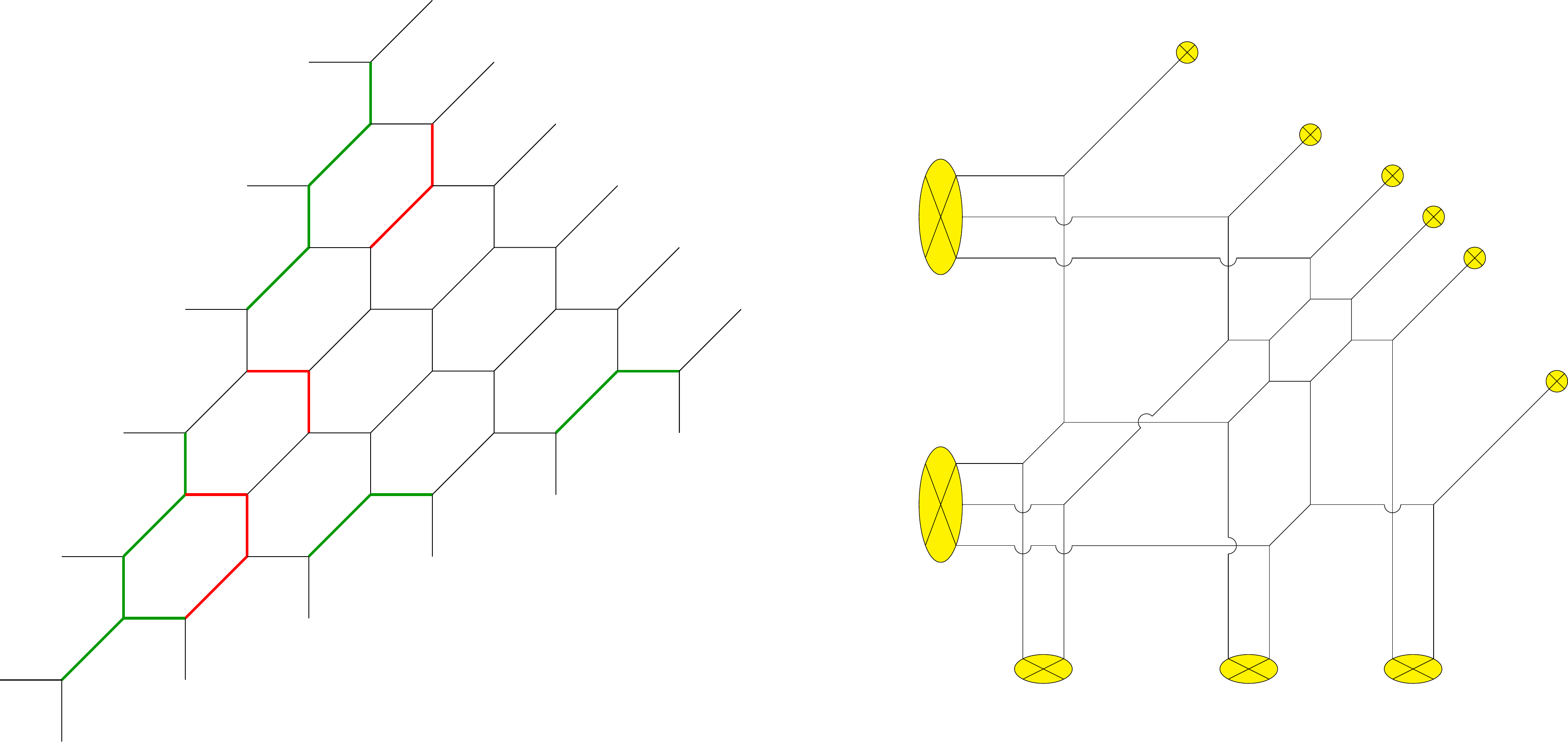}
\end{center}
\caption{Higgsed $T_6$ diagram. On the left the original diagram, in green the curves whose K\"ahler parameters are restricted to engineer the $E_8$ theory and in red the curves whose K\"ahler parameters are restricted because of
the geometric constraint \eqref{cons}. On the right the resulting web diagram after the Higgsing.}
\label{fig:higt6}
\end{figure}

\begin{figure}[!t]
\begin{center}
\includegraphics[width=110mm]{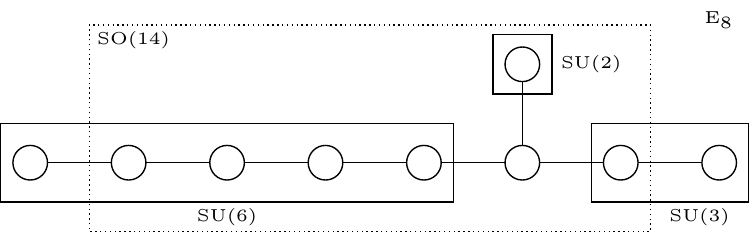}
\end{center}
\caption{The Dynkin diagram of the affine $E_8$ Lie algebra. The nodes in the dotted line represent the Dynkin diagram of $SO(14)$. The nodes in the solid lines denote the Dynkin diagram of $SU(6) \times SU(3) \times SU(2)$.}
\label{fig:e8dynk}
\end{figure}

It was argued in \cite{Benini:2009gi} that it is possible to engineer a theory with an $E_8$ global symmetry in the Higgs branch of the $T_6$ theory and we show in figure \ref{fig:higt6} the web diagram that realises this theory. The resulting theory has a manifest $SU(6)\times SU(3) \times SU(2)$ global symmetry which is believed to enhance to $E_8$ at the superconformal fixed point\footnote{As argued in \cite{Benini:2009gi} the monodromy given by the system of 11 7-branes is conjugate
to the monodromy of the affine $E_8$ configuration. In particular it is possible to collapse 10 of the 11 7-branes to produce a 7-brane with $E_8$ gauge symmetry.}. A similar story happens for a 5d $Sp(1)$ gauge theory with $N_f=7$ 
fundamental flavours whose manifest global $SO(14) \times U(1)$ symmetry enhances to $E_8$ as well at the superconformal point \cite{Seiberg:1996bd, Morrison:1996xf, Douglas:1996xp, Intriligator:1997pq}. The relation between these Lie algebras and their embedding inside the affine $E_8$
Dynkin diagram is shown in figure \ref{fig:e8dynk}.
Furthermore these theories have Coulomb branch and Higgs branch with the same dimensions,
namely $\text{dim}_\mathbb{C}(\mathcal{M}_C)=1$ and $\text{dim}_\mathbb{H}(\mathcal{M}_H)=29$. As we will see later the partition function will have $E_8$ symmetry providing further evidence for the enhancement of
the global symmetry.
In order to achieve this diagram from the web diagram of the $T_6$ theory it is necessary to perform a tuning of the K\"ahler parameters of some of the curves in the diagram in order to group some of the external 5-branes on a single 7-brane. From figure \ref{fig:higt6} we see that we need to group the three upper left legs, the three lower left legs, the two leftmost lower legs, the two central lower legs and the two rightmost lower legs. To group the three upper left legs we need to impose
\be
 Q_5^{(5)}=P_5^{(5)} = Q_4^{(5)} = P_4^{(5)} = \left(\frac{q}{t}\right)^{\frac{1}{2}}\,,
\ee
and to group the three lower left legs the conditions are
\be
P_1^{(5)} =  Q_1^{(5)}=P_2^{(5)} = Q_2^{(5)} = \left(\frac{q}{t}\right)^{\frac{1}{2}}\,,
\ee
by using \eqref{Higgs.horizontal1}. Finally for the leftmost lower legs we impose
\be
P_1^{(5)} = R_1^{(5)}=\left(\frac{q}{t}\right)^{\frac{1}{2}}\,,
\ee
for the central ones we impose
\be
P_1^{(3)} = R_1^{(3)}=\left(\frac{q}{t}\right)^{\frac{1}{2}}\,,
\ee
and finally for the rightmost lower legs we impose
\be
P_1^{(1)} = R_1^{(1)}=\left(\frac{q}{t}\right)^{\frac{1}{2}}\,,
\ee
by using \eqref{Higgs.vertical1}. While these conditions are sufficient to realise the desired pattern for external legs we also need to take into account the geometric constraints of the web diagram \eqref{cons} and in the end some additional K\"ahler parameters will
be restricted. Quite interestingly applying these geometric constraints appears to be equivalent to the propagation of the generalised s-rule presented in \cite{Benini:2009gi}. In the end we will have that the geometric constraints
\eqref{cons} will imply the following conditions on K\"ahler parameters
\be
Q_4^{(4)}=P_4^{(4)} =Q_1^{(4)}=P_1^{(4)} =R_2^{(5)}=R_3^{(5)} = Q_2^{(4)} =\left(\frac{q}{t}\right)^{\frac{1}{2}}\,,
\ee


\subsection{$Sp(1)$ gauge theory parametrisation}
\label{sec:parameterization}

In this section we describe how to define the instanton fugacity of the $Sp(1)$ gauge theory analysing the global  $SU(6) \times SU(3) \times SU(2)$ symmetry inside $E_8$. The first step is to determine the unbroken generators of 
the unbroken flavour symmetry $SU(6) \times SU(3) \times SU(2)$. In the original $T_6$ theory there are 25 generators, 10 of these generators are associated to compact divisors in the geometry and are parameterised by 
the Coulomb branch
moduli while the remaining 15 are associated to non-compact divisors and realise the $SU(6) \times SU(6) \times SU(6)$ flavour symmetry. 

After fixing some K\"ahler parameters to realise the $Sp(1)$
with 7 flavours gauge theory only a reduced number of generators will be unbroken, namely there will be a single Coulomb branch modulus and the generators of the $SU(6) \times SU(3) \times SU(2)$  flavour symmetry. The
unbroken generators are easily identified as the linear combinations of compact and non-compact divisors of the geometry that do not intersect any of curves whose K\"ahler parameter is restricted. This procedure yields as 
expected 9 linearly independent generators which we wish to identify with the generators of $SU(6) \times SU(3) \times SU(2)$  and the generator associated to the Coulomb branch modulus. First we label the divisors in the 
geometry as in figure \ref{fig:T6}. Naively we would associate the generators of the $SU(6)$ part of the flavour symmetry with the non-compact divisors $D_{11}$, $D_{16}$, $D_{20}$, $D_{23}$, and $D_{25}$, the generators
of the $SU(3)$ part of the flavour symmetry with the non-compact divisors $D_{12}$ and $D_{21}$, and the generator of the $SU(2)$ part of the flavour symmetry with the non-compact divisors $D_{3}$ while the generator associated
with the Coulomb branch modulus with $D_{19}$. This allows us to identify one of the generators of $SU(6)$ as the linear combination of unbroken generators that contains $D_{11}$ with coefficient 1 but does not contain any of the 
other flavour generators and the gauge generator. A similar procedure can be applied to the other generators as well allowing the identifications of the generators of the flavour symmetry. For concreteness we list up the Cartan generators for $SU(6) \times SU(3) \times SU(2)$ in appendix \ref{sec:Cartan}.

Let us first define the mass parameters $m_i, (i=1, \cdots, 7)$ as follows,
\be\begin{split}
&Q_3^{(3)} = e^{i\lambda -  im_1}\,, \quad P_3^{(3)} = e^{-i\lambda + im_2}\,, \quad R_3^{(4)} = e^{i\lambda + im_3}\,, \quad P_2^{(3)} = e^{i\lambda + im_4}\,,\\
&Q_1^{(2)} = e^{i\lambda + im_5}\,, \quad P_2^{(2)} = e^{-i\lambda - im_6}\,, \quad R_2^{(2)} = e^{i\lambda - im_7}\,,\quad R^{(3)}_3 = e^{i\tilde{u}-i\lambda}\,.\label{Sp1.para2}
\end{split}\ee
The dependence of the Coulomb branch modulus $\lambda$ is determined by the intersection between the compact divisor $D_{19}$ and two-cycles. The two-cycles in \eqref{Sp1.para2} are the ones which have non-zero intersection number with $D_{19}$. We also introduced $\tilde{u}$ whose linear combination with $m_i, (i=1, \cdots, 7)$ eventually becomes a chemical potential for the instanton fugacity of the $Sp(1)$ gauge theory. By using the parameters in \eqref{Sp1.para2}, we find that the fugacities for particles which have charges equal to the roots of the flavour symmetry are
\be\begin{split}
SU(6)&: \quad \{e^{im_2 -i m_4},e^{-im_2 - im_3},e^{im_1 - i\tilde{u}},e^{-i m_6+im_7 },e^{- i m_5+im_6 }\}\,,\\
SU(3)&: \quad \{ e^{-im_3 -i m_5 -i m_6 - i\tilde{u}},e^{-im_2 -i m_4 +i m_7 -i \tilde{u}}\}\,,\\
SU(2)&:\quad \{e^{ im_1 - i m_2  - i m_4 - i m_5  - i m_6 - i\tilde{u}}\}\,. \label{SU(2)}
\end{split}\ee
The masses $m_i, (i=1, \cdots, 7)$ are the chemical potentials associated to the fugacity $e^{-i\sum_i H_i m_i}$ where $H_i, (i=1, \cdots, 7)$ are the Cartan generators of $SO(14)$.

As in section \ref{subs:T3} we would like the simple roots of  $SU(6)\times SU(3) \times SU(2)$ to be understood as roots of $E_8$. Recalling that the roots of $E_8$ are\footnote{$e_i, (i=1, \cdots, 8)$ are the orthonormal bases of $\mathbb{R}^8$.}
\begin{equation}
\pm(e_i \pm e_j),
\end{equation}
with $i, j = 1, \cdots, 8$ and 
\begin{equation}
\frac{1}{2}(\pm e_1 \pm e_2 \pm e_3 \pm e_4 \pm e_5 \pm e_6 \pm e_7 \pm e_8),
\end{equation}
with an even number of minus signs, we see that the chemical potentials for the particles whose vector of charges is a root of $SU(6)\times SU(3) \times SU(2)$ fit in the $E_8$ root system if we choose
\be
\tilde{u} = \frac{1}{2}m_8 + \frac{1}{2}(m_1 - m_2 - m_3 - m_4 - m_5 - m_6 + m_7)\,.
\ee
Writing the instanton fugacity of the $Sp(1)$ gauge theory as
\be
u = e^{\frac{i}{2} m_8}
\ee
we find that
\be
R_3^{(3)} = u e^{-i\lambda + \frac{i}{2}(m_1 - m_2 - m_3 - m_4 - m_5 - m_6 + m_7)} \equiv u e^{-i \lambda + i f(m)}\,.
\ee
where for later purposes we have the defined a particular linear combination of masses $f(m)$.

In the perturbative regime of the $Sp(1)$ gauge theory with 7 flavours, the mass parameters are associated with the $SO(14)$ flavour symmetry and the instanton current supplies another $U(1)$ symmetry. However, not all the simple roots of $SO(14)$ inside $E_8$ as in figure \ref{fig:e8dynk} are written by $\pm m_i \pm m_j, (i, j = 1, \cdots, 7)$ in \eqref{SU(2)}. This is because we are in a different Weyl chamber of the $E_8$ Cartan subalgebra. If we perform a sequence of Weyl reflections, we can write the mass parameters of the particles whose charges form a root of $SO(14)$ inside $E_8$ as $m_i-m_{i+1}, m_6+m_7, (i=1, \cdots, 6)$.


\subsection{Singlets in the Higgs vacuum}
\begin{figure}[t]
\begin{center}
\includegraphics[width=150mm]{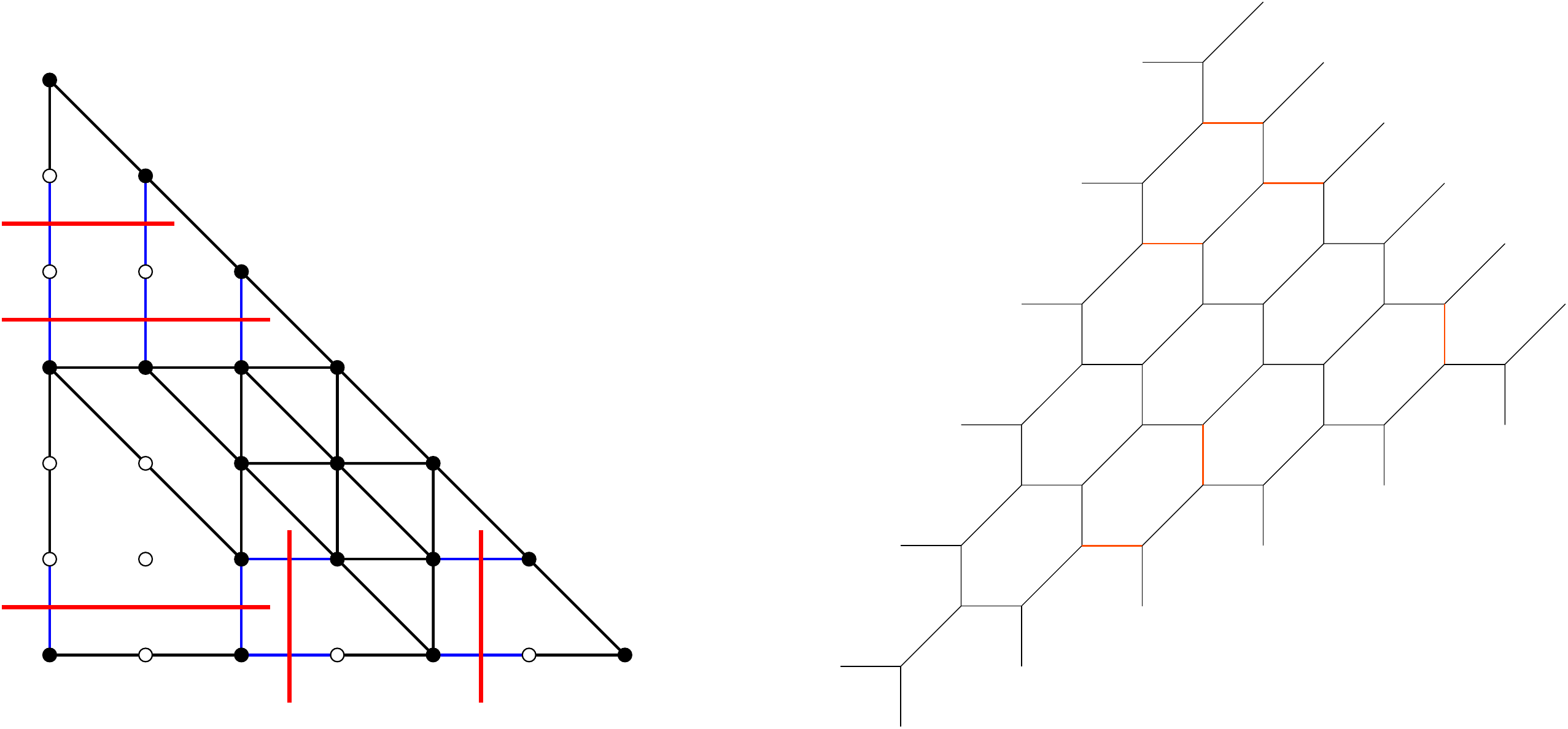}
\end{center}
\caption{Parallel external legs in the Higgsed $T_6$ diagram. On the left the identification via the dot diagram, showing in blue parallel pairs of vertical and horizontal legs that can be connected without crossing diagonal lines and in red the corresponding line.
On the right the original $T_6$ diagram with highlighted in orange the legs that become external after Higgsing.}
\label{fig:ext}
\end{figure}
As already noted in \cite{Hayashi:2013qwa} and explained in section \ref{sec:Higgs.vertical} applying the tuning to the $T_6$ partition function will not give simply the partition function of $Sp(1)$ gauge theory with $N_f=7$ 
fundamental flavours 
as there will be additional contributions coming from singlet hypermultiplets. Therefore the actual partition function of the $E_8$ theory will be
\be
Z_{E_8} = Z_{T_6}^H / Z_{extra}\,,
\ee
where we called $Z_{T_6}^H$ the $T_6$ partition function after tuning the K\"ahler parameters and gathered in $Z_{extra}$ the contributions due to singlet hypermultiplets. In this section, we identify $Z_{extra}$ for the infrared theory in the Higgs branch of the $T_6$ theory corresponding to figure \ref{fig:T6}.

We will start by explaining how to identify
the singlet hypermultiplets factors that only depend on the $\Omega$--deformation parameters. This kind of singlets originate from M2-branes wrapping two cycles and linear combinations of two cycles whose K\"ahler parameter is 
$(q/t)^{\frac{1}{2}}$ and their contributions to the partition function can be understood locally in the diagram. This allows us to split the discussion in six different parts: looking at figure \ref{fig:higt6} we see that the kind of curves
we are interested in appear in the upper left part , in the bottom left part, in the middle top part, in the middle bottom part and in the bottom right part of the diagram. We will now discuss all these contributions separately.
In the upper left part the contribution involves the curves $Q_5^{(5)}$, $P_5^{(5)}$, $Q_4^{(5)}$, $P_4^{(5)}$, 
and the contribution due to singlet hypermultiplets and vector multiplets is
\be
Z_{singl}^{(1)} = \prod_{i,j=1}^{\infty} (1-q^i t^{j-1})^3 (1-q^{i+1}t^{j-2})\,.
\ee
In the bottom left part the contribution is a bit more involved, but being the contribution local we can select a part of the diagram that looks like the higgsed $T_3$ diagram of section \ref{sec:Higgs.vertical}. Being careful not to 
subtract the decoupled factor from parallel diagonal legs that are not external in this case we get the following contribution
\be
Z_{singl}^{(2)} = \prod_{i,j=1}^{\infty} (1-q^i t^{j-1})^6 (1-q^{i+1}t^{j-2})\,.
\ee
In the middle left part we have only the curves $Q_2^{(4)}$ and $R_3^{(5)}$. 
In this case, we need to be careful of subtracting a part of the vector multiplet coming from M2-branes wrapping the two-cycle whose K\"ahler parameter is $Q_2^{(4)}R_3^{(5)}$. Then, the final contribution is simply
\be
Z_{singl}^{(3)} = \prod_{i,j=1}^{\infty} (1-q^i t^{j-1})\,.
\ee
Finally we have the contributions in the middle top part (that involves the curves $Q_4^{(4)}$ and $P_4^{(4)}$),  in the middle bottom part (that involves the curves $P_1^{(3)}$ and $R_1^{(3)}$) and in the bottom right part (that involves the curves $P_1^{(1)}$ and $R_1^{(1)}$). These contributions are identical
and are
\be
Z_{singl}^{(4)}=Z_{singl}^{(5)} = Z_{singl}^{(6)} = \prod_{i,j=1}^{\infty} (1-q^i t^{j-1})^2\,.
\ee
We are thus able to write the contribution to the partition function coming from decoupled hypermultiplets that only depend on the $\Omega$--deformation parameters
\be
Z_{singl} = \prod_{k=1}^6Z_{singl}^{(k)} = \prod_{i,j=1}^{\infty} (1-q^i t^{j-1})^{16} (1-q^{i+1}t^{j-2})^2\,. \label{T6Higgs.singlet1}
\ee

Next we turn to the discussion of decoupled hypermultiplets that depend on the parameters associated with the flavour symmetry. In \cite{Hayashi:2013qwa} this contribution was identified with the perturbative part of the partition function
of hypermultiplets and vector multiplets which come from strings stretching between parallel branes that become external after Higgsing. However while in the examples presented in \cite{Hayashi:2013qwa} the identification of
branes becoming external after Higgsing presented no difficulty in the case of $E_8$ theory this identification is a bit more subtle because of the propagation of the generalised s-rule inside the diagram, and we 
will apply the rule used in  \ref{sec:Higgs1} and \ref{sec:Higgs2} to identify new external legs after Higgsing using the dot diagrams introduced in \cite{Benini:2009gi}. 
We briefly describe the rule here again. We identify a new horizontal external leg with a pair of vertical segments in the dot diagram, one external and one internal, that 
can be connected with a horizontal line without crossing any diagonal line in the dot diagram. 
A similar identification of parallel external legs works for vertical and diagonal legs in the diagram. Using this procedure we can identify which legs are external for the dot diagram of $E_8$ theory, and we show in figure \ref{fig:ext} 
the result. In the result of the computation we need to discard the hypermultiplets that only depend on the $\Omega$--deformation parameters as these have already been included in $Z_{singl}$. Including also the contributions due to the
higgsed Cartan part as well as \eqref{T6Higgs.singlet1} we find that the total contribution is 
\be\begin{split}
Z_{extra} &=(M(q,t)M(t,q))^{\frac{9}{2}} \prod_{i,j=1}^\infty\left(1-q^{i+1} t^{j-2}\right)^2 \left(1-q^i t^{j-1}\right)^{16}\times\\
 &\times(1- u e^{-i m_1+im_2+ im_4+im_5+im_6+i f(m)}q^i t^{j-1})^2 \times\\&\times(1- u e^{-i m_1+im_2+ im_4+im_5+im_6+i f(m)}q^{i-1} t^j)^2\times\\
 &\times(1- u e^{-i m_1+im_2+ im_4+im_5+im_6+i f(m)}q^{i-2} t^{j+1}) \times\\&\times (1- u e^{-i m_1+ im_2+im_4+im_5+im_6+i f(m)}q^{i+1} t^{j-2}) \times\\
&\times(1-u e^{im_2+i m_4+im_7+if(m)} q^{i} t^{j-1})(1-u e^{im_2+i m_4+im_7+if(m)} q^{i-1} t^{j})\times\\
&\times(1-u e^{i m_3+im_5+im_6+if(m)} q^{i-1} t^{j})(1-u e^{i m_3+im_5+im_6+if(m)} q^{i} t^{j-1})\times\\
&\times(1-u^2e^{im_2+i m_3+im_4+im_5+im_6+im_7+2if(m)} q^{i-1} t^{j})\times\\
&\times (1-u^2e^{im_2+i m_3+im_4+im_5+im_6+im_7+2if(m)} q^{i} t^{j-1})\,.
\end{split}\ee

\subsection{The partition function of $Sp(1)$ with $N_f=7$ flavours}

Here we write the resulting partition function of the $E_8$ theory. We recall from the previous section that
\be
Z_{E_8} = Z_{T_6}^H / Z_{extra}\,,
\ee
where $Z_{T_6}^H$ is the $T_6$ partition function after tuning the K\"ahler parameters and $Z_{extra}$ includes the contributions of singlet hypermultiplets. 
Before writing the result some comments are needed regarding the instanton summation in \eqref{Z:inst} as the tuning of some K\"ahler parameters greatly simplifies it. This happens for the same reasons as in sections \ref{sec:Higgs1} and \ref{sec:Higgs2}, namely the tuning of the K\"ahler parameters will imply the appearance of terms of the form $\sin\left(\frac{E_{ \alpha \beta} -\lambda_\alpha+\lambda_\beta}{2}\right)$ giving a zero in the instanton summation
whenever $Y_\alpha > Y_\beta$\footnote{We define this inequality as $Y_{\alpha, i} > Y_{\beta, i}$ for all the rows of $Y_{\alpha}$ and $Y_{\beta}$}. As in some cases the Young diagram $Y_\beta$ is trivial this implies that the only possible diagram contributing to the instanton summation is $Y_\alpha = \emptyset$.
In the end only 8 Young diagram summations will be non-trivial, and we will call the non-trivial Young diagrams as
\be\begin{split}
& R^{(5)}_3 \rightarrow Y_1\quad  R^{(4)}_2 \rightarrow Y_2 \quad R^{(4)}_3 \rightarrow Y_3 \quad R^{(3)}_2 \rightarrow Y_4   \\
& R^{(3)}_3 \rightarrow Y_5 \quad R^{(2)}_1 \rightarrow Y_6 \quad R^{(2)}_2 \rightarrow Y_7 \quad R^{(1)}_1 \rightarrow Y_8\,.
\end{split}
\ee
Moreover the result will vanish if $Y_1 > Y_2$ and $Y_8 > Y_6$.

We write the $T_6$ partition function after tuning the K\"ahler parameters as
\be
Z^H_{T_6} = (M(q,t) M(t,q))^5Z_0^H Z_{inst}^H( Z_{dec}^{/\!/ }Z_{dec}^{||})^{-1}\,,\label{Partition.E8}
\ee
where
\be\begin{split}
Z_{0}^H=& \prod_{i,j=1}^\infty\frac{ (1-q^{i+1} t^{j-2})^2 (1-q^i t^{j-1})^{13}(1- e^{-i \lam + i m_2}q^{i-\frac{1}{2}}t^{j-\frac{1}{2}})}{(1- e^{-2 i \lam}q^i t^{j-1})(1- e^{-2 i \lam}q^{i-1} t^{j})(1-e^{i m_5-i m_6}q^i t^{j-1})(1-e^{im_4-im_2}q^i t^{j-1})}\times \\ 
\times&(1-e^{-i \lam - i m_2}q^{i-\frac{1}{2}}t^{j-\frac{1}{2}})(1-e^{i \lam + i m_4}q^{i-\frac{1}{2}}t^{j-\frac{1}{2}})(1-e^{-i \lam + i m_4}q^{i-\frac{1}{2}}t^{j-\frac{1}{2}})\times \\ 
\times & (1-e^{i \lam -i m_1}q^{i-\frac{1}{2}}t^{j-\frac{1}{2}})(1-e^{-i \lam - i m_1}q^{i-\frac{1}{2}}t^{j-\frac{1}{2}})(1-e^{-i \lam - i m_6}q^{i-\frac{1}{2}}t^{j-\frac{1}{2}})\times \\ 
\times&(1-e^{-i \lam + i m_6}q^{i-\frac{1}{2}}t^{j-\frac{1}{2}}) (1-e^{i\lam +i m_5}q^{i-\frac{1}{2}}t^{j-\frac{1}{2}})(1-e^{-i \lam + i m_5}q^{i-\frac{1}{2}}t^{j-\frac{1}{2}})\\
\times&\frac{ (1-u e^{-i m_1+im_2+ im_4+im_5+im_6+i f(m)}q^{i+1} t^{j-2} )(1- u e^{-i m_1+im_2+ im_4+im_5+im_6+i f(m)}q^i t^{j-1})^2 }{(1-u e^{i \lam+im_2 +i m_4+i m_5+im_6 +i f(m)}q^{i-\frac{3}{2}} t^{j+\frac{1}{2}} )  (1-u e^{-i \lam+im_2 + im_4+im_5+im_6+i f(m)}q^{i-\frac{3}{2}} t^{j+\frac{1}{2}} )}\\
\times & (1-u e^{-i m_1+im_2+ im_4+im_5+im_6+i f(m)}q^{i-1} t^j )^3(1-u e^{i m_4 + i m_5 +im_6 + i f(m)}q^{i-1} t^j) \\
\times&(1-u e^{-i m_1+im_2+ im_4+im_5+im_6+i f(m)}q^{i-2} t^{j+1} )  (1-u e^{im_2+i m_5+im_6 + i f(m)}q^{i-1} t^j )\\
\times&(1- u e^{im_2+i m_4+im_6 + i f(m)}q^{i-1} t^{j})(1- u e^{im_2+i m_4 +im_5+ i f(m)}q^{i-1} t^{j})\,.
\end{split}
\ee
\be\begin{split}
1/Z_{dec}^{/\!/}&= \prod_{i,j=1}^{\infty}(1- e^{im_4-im_2}q^i t^{j-1})(1- e^{im_3+im_4}q^i t^{j-1})(1- e^{im_2+im_3}q^i t^{j-1})\times\\
&\times (1- e^{-im_7+im_6}q^i t^{j-1})(1- e^{im_5-im_7}q^i t^{j-1})(1- e^{im_5-im_6}q^i t^{j-1})\times\\
&\times (1-u e^{-im_1+im_3+im_4+if(m)}q^i t^{j-1})(1-u e^{-im_1+im_3+im_4+im_5-im_7+if(m)}q^i t^{j-1})\times\\
&\times (1-u e^{-im_1+im_2+im_3+if(m)}q^i t^{j-1})(1-u e^{-im_1+im_3+im_4-im_7+if(m)+im_6}q^i t^{j-1})\times\\
&\times (1-u e^{-im_1+im_5-im_7+if(m)}q^i t^{j-1})(1-u e^{-im_1+im_2+im_3-im_7+if(m)+im_6}q^i t^{j-1})\times\\
&\times (1-u e^{-im_1+if(m)}q^i t^{j-1})(1-u e^{-im_1+im_2+im_3+im_5-im_7+if(m)}q^i t^{j-1})\times\\
&\times(1-u e^{-im_1-im_7+if(m)+im_6}q^i t^{j-1})\,,
\end{split}\ee
\be\begin{split}
1/Z_{dec}^{||}&= \prod_{i,j=1}^{\infty}(1- q^{i}t^{j-1})^3(1- u e^{im_2+im_4-im_7+if(m)}q^{i-1} t^{j})(1- u e^{im_2+im_4-im_7+if(m)}q^{i} t^{j-1})\times\\
&\times(1- u e^{im_2+im_3+im_5+im_6+if(m)-im_2}q^{i-1} t^{j})(1- u e^{im_3+im_5+im_6+if(m)}q^{i} t^{j-1})\times\\
&\times(1- u e^{im_3+im_5+im_6+if(m)}q^{i-2} t^{j+1})(1- u^2 e^{im_2+im_3+im_4+im_5+im_6-im_7+2if(m)}q^{i-1} t^{j})\times\\
&\times(1- u e^{im_2+im_4-im_7+if(m)}q^{i-2} t^{j+1})(1- u^2 e^{im_2+im_3+im_4+im_5+im_6-im_7+2if(m)}q^{i} t^{j-1})\times\\
&\times (1- u e^{im_2+im_4-im_7+if(m)}q^{i-1} t^{j})(1- u^2 e^{im_2+im_3+im_4+im_5+im_6-im_7+2if(m)}q^{i-1} t^{j})\times\\
&\times (1- u e^{im_3+im_5+im_6+if(m)}q^{i-1} t^{j})(1- u^2 e^{im_2+im_3+im_4+im_5+im_6-im_7+2if(m)}q^{i-2} t^{j+1})\,,\end{split}\ee
\be
Z^H_{inst} = \sum_{Y_1,\dots , Y_8} u_3^{|Y_4|+|Y_5|}Z_L(Y_4,Y_5) Z_M Z_R(Y_4,Y_5)
\ee
\be\begin{split}
Z_L(Y_4,Y_5) &= \prod_{\alpha=2,3}\left[\prod_{s \in Y_\alpha} \frac{\left(2i \sin \frac{E_{\alpha 4}-m^L_1+i\gamma_1}{2}2i\sin\frac{E_{\alpha\emptyset}-m^L_2+i\gamma_1}{2}2i\sin\frac{E_{\alpha5}-m^L_3+i\gamma_1}{2}\right)(2i\sin\frac{E_{\alpha 1} -\lambda_2  +2i\gamma_1}{2})}{\prod_{\beta=2,3}(2i)^2\sin\frac{E_{\alpha\beta}}{2}\sin\frac{E_{\alpha\beta}+2i\gamma_1}{2}}\right.\\
&\left.\prod_{s\in Y_1}\frac{\prod_{\alpha=2,3}2i\sin\frac{E_{1\alpha}+\lambda_2}{2}}{(2i)^2\sin\frac{E_{11}}{2}\sin\frac{E_{11}+2i\gamma_1}{2}}\right] u_4^{|Y_2|+|Y_3|} u_5^{|Y_1|}\\
&\prod_{\alpha=4,5}\prod_{s \in Y_\alpha}(2i)^2\sin \frac{ E_{\alpha2}+m_1^L+i \gamma_1}{2} \sin \frac{ E_{\alpha 3}+m_1^L+i \gamma_1}{2} \,, 
\end{split}\ee
\be\begin{split}
Z_M &= \prod_{\alpha=4,5}\prod_{s \in Y_\alpha}\frac{2i \sin \frac{E_{\alpha\emptyset} - m_1+i \gamma_1}{2} 
}{2i \sin\frac{ E_{\alpha\emptyset}+i \log u -m_2-m_4-m_5-m_6-f(m)+3i \gamma_1}{2} \prod_{\beta=4,5} (2i)^2 \sin \frac{E_{\alpha\beta}}{2}\sin \frac{E_{\alpha\beta}+2i \gamma_1}{2}}\,,
\end{split}\ee
\be\begin{split}
Z_R(Y_4,Y_5) &= \prod_{\alpha=6,7}\left[\prod_{s \in Y_\alpha} \frac{\left(2i \sin \frac{E_{\alpha 4}-m^R_1+i\gamma_1}{2}2i\sin\frac{E_{\alpha\emptyset}-m^R_2+i\gamma_1}{2}2i\sin\frac{E_{\alpha5}-m^R_3+i\gamma_1}{2}\right)(2i\sin\frac{E_{\alpha 8} -\lambda_6  +2i\gamma_1}{2})}{\prod_{\beta=6,7}(2i)^2\sin\frac{E_{\alpha\beta}}{2}\sin\frac{E_{\alpha\beta}+2i\gamma_1}{2}}\right.\\
&\left.\prod_{s\in Y_8}\frac{\prod_{\alpha=6,7}2i\sin\frac{E_{8\alpha}+\lambda_6}{2}}{(2i)^2\sin\frac{E_{88}}{2}\sin\frac{E_{88}+2i\gamma_1}{2}}\right] u_2^{|Y_6|+|Y_7|}u_1^{|Y_8|} \\
& \prod_{\alpha=4,5}\prod_{s \in Y_\alpha}(2i)^2 \sin \frac{ E_{\alpha6}+m_1^R+i \gamma_1}{2} \sin \frac{ E_{\alpha 7}+m_1^R+i \gamma_1}{2}
\end{split}\ee

where we defined the parameters
\be\begin{split}
&\lambda_2 = - \lambda_3 = -\frac{1}{2}(m_2-m_4)\,, \quad \lambda_6 = - \lambda_7 = -\frac{1}{2}(m_6-m_5)\,, \quad \lam_1 = \lam_8 = \lam_\emptyset = 0\,, \quad \lam_4 = - \lam_5 = -\lam\,,\\
& m^L_1 = m^L_3 = -\frac{1}{2} (m_4+m_2)\,, \quad m_2^L = -i \log u +\frac{1}{2} (m_4+m_2)+m_5+m_6+f(m) - i \gamma_1\,,\\
& m^R_1 = m^R_3 = -\frac{1}{2} (m_5+m_6)\,, \quad m_2^R = -i \log u +\frac{1}{2} (m_5+m_6)+m_2+m_4+f(m) - i \gamma_1\,,
\end{split}\ee

and the instanton fugacities
\be\begin{split}
&u_5= e^{i(m_4-m_2)/2}\,, \quad u_4 = u^{1/2} e^{\gamma_1} e^{i[2m_3+m_5+g(m)]/2}\,, \quad u_3 = u^{1/2} e^{-\gamma_1}e^{i[-m_1+f(m)]/2}\,,\\
&u_2 = u^{1/2} e^{\gamma_1}e^{i[m_4+m_5+g(m)]/2}\,, \quad u_1 = e^{-\gamma_1} e^{i(m_5-m_6)/2}\,.
\end{split}\ee

The $Z_{inst}^H$ part in \eqref{Partition.E8} has a peculiar structure. It is written by gluing $Z_L(Y_4, Y_5)$ and $Z_R(Y_4, Y_5)$ with $Z_M$. This is almost identical to gluing the two $Z_{inst}$ parts of the Higgsed $T_3$ theory in section \ref{sec:Higgs2} with additional bi-fundamental hypermultiplets and $U(2)$ vector multiplet along the two-cycles whose the K\"ahler parameters are $R_3^{(3)}$ and $R_2^{(3)}$. The difference only appears in $Z_M$ where a hypermultiplet contribution from the two-cycle with the K\"ahler parameter $Q_3^{(3)}$ in the numerator, and also the remnant of the $U(3)$ vector multiplet due to the tuning of the two-cycles whose K\"ahler parameters are $P_1^{(3)}$ and $R_1^{(3)}$ in the denominator. 

We would like to extract the perturbative part of the partition function, namely we would like to take the limit $\lim_{u\rightarrow 0} Z_{E_8}$ and see if this correctly reproduces the perturbative part of $Sp(1)$ gauge theory with
$N_f=7$ fundamental flavours. We start by taking the terms in $Z_{inst}^H$ with $Y_4= Y_5 = \emptyset$ because taking these Young diagrams to be non-trivial only adds terms that vanish in the limit $u\rightarrow 0 $. Doing this 
the instanton summation becomes the product of two factors

\be
\left(\sum_{Y_1,Y_2,Y_3}  Z_L(\emptyset,\emptyset)\right)\left(\sum_{Y_6,Y_7,Y_8} Z_R(\emptyset,\emptyset)\right)\,.
\ee
Using the definitions of $Z_L$ and $Z_R$ it is easy to see that $Z_L (\emptyset,\emptyset)$ and $Z_R (\emptyset,\emptyset)$ are simply the instanton part of the Higgsed $T_3$ diagram described in section \ref{sec:Higgs2}.
Knowing the result of the summation it is quite easy to extract from it the perturbative part and the result is

\be
\left. \left(\sum_{Y_1,Y_2,Y_3} Z_L(\emptyset,\emptyset)\right)\right|_{u=0}=\prod_{i,j=1}^\infty \frac{(1- e^{i \lam+i m_3} q^{i-1/2} t^{j-1/2})(1- e^{-i \lam + i m_3} q^{i-1/2} t^{j-1/2})}{(1-e^{im_2+im_3}q^i t^{j-1})(1-e^{im_3+im_4}q^i t^{j-1})}\,,
\ee
\be
\left.\left(\sum_{Y_6,Y_7,Y_8} Z_R(\emptyset,\emptyset)\right)\right|_{u=0} = \prod_{i,j=1}^\infty \frac{(1-e^{i\lam - i m_7}q^{i-1/2}t^{j-1/2})(1-e^{-i \lam-im_7}q^{i-1/2}t^{j-1/2})}{(1-e^{-im_7+im_6}q^i t^{j-1})(1-e^{im_5-im_7}q^i t^{j-1})}\,.
\ee
We are now able to write the partition function as
\be
Z_{E_8} = Z_{pert} Z_{n.p.}\,,
\ee
where
\be\begin{split}
 Z_{pert}&=(M(q,t)M(t,q))^{\frac{1}{2}}\prod_{i,j=1}^\infty\frac{ (1- e^{-i \lam + i m_2}q^{i-\frac{1}{2}}t^{j-\frac{1}{2}})(1-e^{-i \lam - i m_2}q^{i-\frac{1}{2}}t^{j-\frac{1}{2}})}{(1- e^{-2 i \lam}q^i t^{j-1})(1- e^{-2 i \lam}q^{i-1} t^{j})}\times \\ 
&\times(1-e^{i \lam + i m_4}q^{i-\frac{1}{2}}t^{j-\frac{1}{2}})(1-e^{-i \lam + i m_4}q^{i-\frac{1}{2}}t^{j-\frac{1}{2}})(1-e^{i \lam -i m_1}q^{i-\frac{1}{2}}t^{j-\frac{1}{2}})\times \\ 
&\times (1-e^{-i \lam - i m_1}q^{i-\frac{1}{2}}t^{j-\frac{1}{2}})(1-e^{-i \lam - i m_6}q^{i-\frac{1}{2}}t^{j-\frac{1}{2}})(1-e^{-i \lam + i m_6}q^{i-\frac{1}{2}}t^{j-\frac{1}{2}})\times \\ 
&\times (1-e^{i\lam +i m_5}q^{i-\frac{1}{2}}t^{j-\frac{1}{2}})(1-e^{-i \lam + i m_5}q^{i-\frac{1}{2}}t^{j-\frac{1}{2}})(1-e^{i\lam - i m_7}q^{i-\frac{1}{2}}t^{j-\frac{1}{2}})\times\\
&\times(1-e^{-i \lam-im_7}q^{i-\frac{1}{2}}t^{j-\frac{1}{2}})(1- e^{i \lam+i m_3} q^{i-\frac{1}{2}} t^{j-\frac{1}{2}})(1- e^{-i \lam + i m_3} q^{i-\frac{1}{2}} t^{j-\frac{1}{2}})\,,
\end{split}\ee
\be\begin{split}
Z_{n.p.} &= Z_{inst}^H\prod_{i,j=1}^\infty\frac{ (1- u e^{im_2+i m_5+im_6 + i f(m)}q^{i-1} t^j)(1-u e^{i m_4 + i m_5 +i m_6 + i f(m)}q^{i-1} t^j)}{(1- u e^{i \lam +i m_4+i m_5 +i g(m)}q^{i-\frac{3}{2}} t^{j+\frac{1}{2}})(1- u e^{-i \lam + im_4+im_5+i g(m)}q^{i-\frac{3}{2}} t^{j+\frac{1}{2}})}\times\\
&\times(1-u e^{-im_1+if(m)}q^i t^{j-1})(1-u e^{-im_1-im_7+if(m)+im_6}q^i t^{j-1})\times\\
&\times(1-u e^{im_1+im_2+im_3+if(m)}q^i t^{j-1})(1-u e^{-im_1+im_5-im_7+if(m)}q^i t^{j-1})\times\\
   &\times(1- u e^{im_2+i m_4 +im_6+ i f(m)}q^{i-1} t^{j})(1-u e^{-im_1+im_3+im_4+im_5-im_7+if(m)}q^i t^{j-1})\times\\
&\times(1- u e^{im_2+im_4-im_7+if(m)}q^{i-2} t^{j+1})(1-u e^{-im_1+im_2+im_3+im_5-im_7+if(m)}q^i t^{j-1})\times\\
&\times (1- u e^{-i m_1+im_2+ im_4+im_5+im_6+i f(m)}q^{i-1} t^j) (1-u e^{-im_1+im_3+im_4+if(m)}q^i t^{j-1})\times\\
&\times(1- u e^{im_2+i m_4 +im_5+if(m)}q^{i-1} t^{j})  (1-u e^{-im_1+im_3+im_4-im_7+if(m)+im_6}q^i t^{j-1})\times\\
&\times(1- u e^{im_3+im_5+im_6+if(m)}q^{i-2} t^{j+1})(1-u e^{-im_1+im_2+im_3-im_7+if(m)+im_6}q^i t^{j-1})\times\\
&\times(1- u e^{im_3+im_5+im_6+if(m)}q^{i-1} t^{j})(1- u^2 e^{im_2+im_3+im_4+im_5+im_6-im_7+2if(m)}q^{i-1} t^{j})\times\\
&\times(1- u e^{im_2+im_4-im_7+if(m)}q^{i-1} t^{j}) (1- u^2 e^{im_2+im_3+im_4+im_5+im_6-im_7+2if(m)}q^{i-2} t^{j+1})\times\\
& \times \frac{(1-e^{im_2+im_3}q^i t^{j-1})(1-e^{im_3+im_4}q^i t^{j-1})(1-e^{-im_7-im_6}q^i t^{j-1})(1-e^{im_5-im_7}q^i t^{j-1})}{(1- e^{i \lam+i m_3} q^{i-\frac{1}{2}} t^{j-\frac{1}{2}})(1- e^{-i \lam + i m_3} q^{i-\frac{1}{2}} t^{j-\frac{1}{2}})(1-e^{i\lam - i m_7}q^{i-\frac{1}{2}}t^{j-\frac{1}{2}})(1-e^{-i \lam-im_7}q^{i-\frac{1}{2}}t^{j-\frac{1}{2}})}\,.
\end{split}\ee
With this choice we have that $Z_{n.p.}|_{u=0} = 1$.

\subsection{Partition function at 1-instanton level}
Having successfully reproduced the perturbative part of the partition function of $Sp(1)$ with 7 flavours we would like now to discuss the partition function at 1-instanton level. In order to compute it (and also
the partition function at higher instanton level) we will need to take the Young diagrams $Y_4$ and $Y_5$ to be non-trivial and perform the instanton summation for the remaining ones. We have already noticed the equality between the 
instanton part of the Higgsed $T_3$ diagram in section \ref{sec:Higgs2}
and the contributions $Z_L (\emptyset,\emptyset)$ and $Z_R (\emptyset,\emptyset)$, and somehow taking $Y_4$ or $Y_5$ to be non-trivial is related somehow to the instanton part of a Higgsed $T_3$ diagram with non-trivial
representation on an external leg with some additional hypermultiplets. We can consider the following quantity
\be
\tilde Z_L(Y_4,Y_5) \equiv \frac{ \sum_{Y_1,Y_2,Y_3}   Z_L(Y_4,Y_5)}{ \sum_{Y_1,Y_2,Y_3} Z_L(\emptyset,\emptyset)}\,,
\ee
and a similar quantity involving $Z_R(Y_4,Y_5)$.
Knowing the result of the summation for $Z_L(\emptyset,\emptyset)$ if we are able to compute $\tilde Z_L(Y_4,Y_5)$ we automatically have the result of the summation for $Z_L(Y_4,Y_5)$. We have observed that expressing
$\tilde Z_L (Y_4,Y_5)$ as a series in the instanton fugacity $u_4$ the series stops at a finite order. More specifically we expect that at level $k= |Y_4|+|Y_5|$ the series terminates at order $u_4^k$ with higher order terms
vanishing. We have checked this explicitly up to $k=2$ for higher orders of $u_4$. We emphasise that the termination of the series happens separately for each choice of $Y_4$ and $Y_5$ in the external legs, not only for the
sum of all contributions with fixed $k$.
Using this it is possible to compute explicitly the partition function at 1-instanton level and the result matches with field theory one \cite{Kim:2012gu}
\be
Z_{k=1}^{Sp(1)} = \frac{1}{32} \left[\frac{\prod_{a=1}^7 2i \sin \frac{m_a}{2}}{i^2 \sinh \frac{\gamma_1\pm \gamma_2}{2}\sin \frac{i\gamma_1+\lam}{2}}+\frac{\prod_{a=1}^7 2 \cos \frac{m_a}{2}}{ \sinh \frac{\gamma_1\pm \gamma_2}{2}\cos \frac{i\gamma_1+\lam}{2}}\right]\,,
\ee
where we used the notation $\sin (a\pm b) = \sin (a+b) \sin (a-b)$.

\subsection{2-instanton order and the comparison with field theory result}

We would like to understand if $Z_{E_8}$ correctly reproduces the partition function of an $Sp(1)$ gauge theory with 7 fundamental flavours at 2-instanton level, however it is first useful 
to review how the computation of the instanton partition function is performed in field theory. It is possible to engineer 5d $Sp(N)$ gauge theory with $N_f \leq 7$ in string theory on 
the worldvolume of $N$ D4-branes in the proximity of $N_f$ D8-branes and an O8-plane. In this system instantons in the 5d gauge theory are D0-branes and as we will discuss later
the partition function at $k$ instanton level can be computed as a Witten index in the ADHM quantum mechanics on the worldvolume of $k$ D0-branes. 
Note that in this system an additional hypermultiplet in the antisymmetric representation of $Sp(N)$ is present which originates from strings stretching between the $N$ D4-branes and 
the orientifold plane (or the mirror $N$ D4-branes). The presence of the antisymmetric hypermultiplet is important even for the case of $N=1$ where the antisymmetric representation is trivial for it changes the 
instanton calculation providing 
non-perturbative couplings 
due to small instantons. Even the naive expectation that in the final result for $N=1$ the contribution 
due to the antisymmetric representation simply factors out of the partition function is not true for $N_f=7$ as noted in \cite{Kim:2012gu, Hwang:2014uwa} and the computation performed 
without including the antisymmetric representation does not give the correct partition function (for instance the superconformal index does not respect the $E_8$ symmetry).
However it is important to note that the computation will contain the contributions of additional states that are present in the string theory realisation but are not present in the field
theory, states that can be interpreted as due to strings in the system D0-D8-O8, and once the contributions due to these states are canceled the 5d partition function is correctly
reproduced.

The quantity we would like to discuss is a Witten index $Z^k_{QM}$ \eqref{witten} for the ADHM quantum mechanics on the worldvolume of $k$ D0-branes.
Knowing the index $Z_{QM} = \sum_k u^k Z_{QM}^k$ it is possible to compute the instanton part of the 5d partition function as $Z_{inst} = Z_{QM} / Z_{string}$ where $Z_{string}$
contains the contributions of additional states that are present in the string theory realisation but not present in the field theory. We will write its explicit expression later, but first we 
will discuss how to compute $Z^k_{QM}$. The result can be expressed as a contour integral in the space of zero modes given by the holonomies of the gauge field and the scalar in
the vector multiplet in the ADHM quantum mechanics. Since the gauge group $\hat G$ of the ADHM quantum mechanics is compact the holonomies of the vector 
field actually  live in a compact space and the space of zero modes will be the product of $r$ cylinders where $r$ is the rank of $\hat G$. 

For the case of $Sp(N)$ gauge theories 
some additional care is needed for $\hat G= O(k)$ which is not connected. In this case the $k$ instanton index is
\be
Z^k_{QM} = \frac{1}{2} (Z_+^k + Z_-^k)
\ee
where $Z_\pm^k$ is the index for the $O(k)_\pm$ component. The correct definition of the contour of integration is discussed in \cite{Hwang:2014uwa} and here we will simply state the result for the case we are interested in.
The rank of $O(2)_+$ is 1 so that the moduli space is a cylinder and we have that
\be\begin{split}
&Z^{2}_+ = \oint_{\mathcal{C}} [d\phi] Z_{vec}^+ Z_{anti}^+(m)\prod_{i=1}^7 Z_{fund}^+ (m_i)\,,\\
&Z_{vec}^+ = \frac{1}{2^9}\frac{ \sinh \gamma_1}{ \sinh \frac{\pm \gamma_2+\gamma_1}{2} \sinh \frac{\pm 2\phi \pm \gamma_2+\gamma_1}{2} \sinh \frac{\pm \phi \pm i\lam+\gamma_1}{2}}\,,\\
&Z_{anti}^+(m) =\frac{ \sinh \frac{\pm i m- \gamma_2}{2}\sinh \frac{\pm \phi \pm i\lam-i m}{2}}{\sinh \frac{\pm i m-\gamma_1}{2} \sinh \frac{\pm 2\phi \pm i m - \gamma_1}{2}}\,,\\
&Z_{fund}^+(m_i) =2 \sinh \frac{\pm \phi +i m_i}{2}\,,
\end{split}\ee
where $m$ is the mass of the hypermultiplet in the antisymmetric representation. Moreover the measure of integration is simply $[d\phi] = \frac{1}{2\pi} d\phi$. As we see the integrand has simple poles at the zeroes of the hyperbolic sines with the general form
\be
\frac{1}{\sinh \frac{Q \phi + \dots}{2}}\,.
\ee
The contour of integration $\mathcal{C}$ is defined to surround the poles with $Q>0$, or alternatively we can define the contour of integration as the unit circle in the variable $z=e^\phi$ and substitute $t= e^{-\gamma_1}$ 
in $Z_{vec}^+$ and $T = e^{-\gamma_1}$ in $Z_{anti}^+$ and taking $t<1$ and $T>1$. The two procedures are equivalent for the poles with $Q>0$ will lay inside the unit circle in $z$ if $t$ is taken sufficiently small and $T$ 
sufficiently large. In our case the contour $\mathcal{C}$ will surround 10 poles, 6 of which will come from $Z_{vec}^+$ and 4 from $Z_{anti}^+$, we choose not to write the result of the computation here being it quite long.
The situation is much simpler for $Z_-^2$ for the rank of $O(2)_-$ is 0 and no integration is needed. The result is
\be\begin{split}
&Z^{2}_- = Z_{vec}^- Z_{anti}^-(m)\prod_{i=1}^7 Z_{fund}^- (m_i)\,,\\
&Z_{vec}^- = \frac{1}{32}\frac{ \cosh \gamma_1}{ \sinh \frac{\pm \gamma_2+\gamma_1}{2} \sinh ( \pm \gamma_2+\gamma_1) \sinh ( \pm i\lam+\gamma_1)}\,,\\
&Z_{anti}^-(m) =-\frac{ \cosh \frac{\pm im- \gamma_2}{2}\sin(\pm \lam+m)}{\sinh \frac{ im\pm\gamma_1}{2}\sinh ( i m\pm\gamma_1)}\,,\\
&Z_{fund}^-(m_i) =2i  \sin m_i\,.
\end{split}\ee

The only last piece necessary for the computation of the partition function is the factor $Z_{string}$ that as explained before will cancel from $Z_{QM}$ will cancel the contributions due to additional states present in the string theory
realisation of $Sp(1)$ gauge theory. This contribution was computed in \cite{Hwang:2014uwa} and the result for $N_f=7$ is
\be\label{eq:Zstring}
Z_{string} = \text{PE} \big[f_7(x,y , v,w_i,u)\big] \,, 
\ee
where $x= e^{-\gamma_1}$, $y= e^{-\gamma_2}$, $v= e^{-i m}$, $u$ is the instanton fugacity of $Sp(1)$ gauge theory and $w_i = e^{\frac{i}{2} m_i}$ with $i=1,\dots ,7$.  In \eqref{eq:Zstring} we also defined
the Plethystic exponential of a function $f(x)$ as
\be
\text{PE}[f(x)]= \exp \left[\sum_{n=1}^\infty \frac{1}{n} f(x^n)\right]\,.
\ee
Finally in \eqref{eq:Zstring} $f_7$ is
\be
f_7 = \frac{u\, x^2}{(1-x y)(1-x/y)(1-xv)(1-x/v)}\left[\chi(w_i)^{SO(14)}_{\mathbf{64}}+u\chi(w_i)^{SO(14)}_{\mathbf{14}}\right]\,.
\ee

Knowing this it is possible to extract the instanton partition function of $Sp(1)$ with 7 flavours and one anti-symmetric hypermultiplet at instanton level 2 and check whether there is agreement with the result coming from $Z_{E_8}$. While it has not been possible so far to
check agreement between the two expression because of computational difficulties however it has been possible to check that the two expressions agree in the special limit where all but two masses of the fundamental 
hypermultiplets are taken to zero. Moreover expanding the two expressions in the fugacity $x= e^{-\gamma_1}$ we have found complete agreement between the two expression up to order $x^3$.

Another check is to see the perturbative flavour symmetry $SO(14)$ at each instanton level. We have checked that the 2-instanton part we obtained is indeed invariant under the Weyl symmetry of $SO(14)$. This is also a non-trivial evidence that our calculation yields the correct result of the 2-instanton part of the $E_8$ theory. Further check will be discussed in the next section and involves the computation of the superconformal index.

\subsection{Superconformal index of the $E_8$ theory}
Knowledge of the 5d Nekrasov partition function allows us to perform the computation of the superconformal index which will allow us to verify explicitly the non-perturbative enhancement of the flavour symmetry. The
superconformal index of a 5d theory (or equivalently the partition function on $S^1 \times S^4$) is defined as \eqref{5d.index}.
The computation of the superconformal index can be performed using localisation techniques and the result is \cite{Kim:2012gu}\footnote{For the case of $SU(2)$
gauge theories it was noticed in \cite{Iqbal:2012xm} that it is also possible to use directly the refined topological string partition function.}
\be\label{eq:index}
I(\gamma_1,\gamma_2,m_i,u ) = \int [d\lambda]_H\, \text{PE}\left[f_{mat} ( x,y,e^{i\lambda},e^{im_i})+f_{vec} ( x,y,e^{i\lambda})\right] \left|I^{inst}(x,y,e^{i \lam},e^{i m_i},u)\right|^2\,,
\ee
where $f_{mat}$ and $f_{vec}$ take into account the perturbative contributions given by hypermultiplets and vector multiplets and they are
\be
f_{mat}( x,y,e^{i\lambda},e^{i m_i}) = \frac{ x}{(1-xy)(1-x/y)} \sum_{\mathbf{w}\in \mathbf{W}}\sum_{i =1}^{N_f} (e^{-i \mathbf{w} \cdot \lam-i m_i}+e^{i \mathbf{w} \cdot \lam+i m_i})
\ee
\be
f_{vec}( x,y,e^{i\lambda}) = -\frac{ xy + x/y}{(1-xy)(1-x/y)} \sum_{\mathbf{R}} e^{-i \mathbf{R} \cdot \lam}
\ee
where $\mathbf{R}$ is the set of all roots of the Lie algebra of the gauge group and $\mathbf{W}$ is the weight system for the representation of the hypermultiplets. Moreover in \eqref{eq:index} $[d\lam]_H$ denotes the
the Haar measure of the gauge group which for $Sp(N)$ is equal to
\be
[d\lambda]_H = \frac{2^N}{N!} \left[\prod_{i=1}^N \frac{d\lambda_i}{2 \pi}\sin^2 \lambda_i\right]\prod_{i<j}^N\left[2\sin \left(\frac{\lambda_i-\lambda_j}{2}\right)2\sin \left(\frac{\lambda_i+\lambda_j}{2}\right)\right]^2\,,
\ee
and $|I^{inst}(x,y,e^{i \lam},e^{i m_i},u)|^2$ includes the contributions due to instantons and is given by
\be\label{eq:ind-inst}\begin{split}
\left|I^{inst}(x,y,e^{i \lam},e^{i m_i},u)\right|^2& = I^{inst}_{north}(x,y,e^{i \lam},e^{i m_i},u) I^{inst}_{south}(x,y,e^{i \lam},e^{i m_i},u) =  \\
&= \left[\sum_{k=0}^\infty u^{-k} I^{k}(x,y,e^{-i\lam},e^{-i m_i})\right]\left[\sum_{k=0}^\infty u^{k} I^{k}(x,y,e^{i\lam},e^{i m_i})\right]\,.
\end{split}\ee
In \eqref{eq:ind-inst} $I^{inst}_{north}(x,y,e^{i \lam},e^{i m_i},u) $ contains the contributions due to anti-instantons localised at the north pole of $S^4$ and $I^{inst}_{south}(x,y,e^{i \lam},e^{i m_i},u) $
contains the contributions of instantons localised at the south pole of $S^4$. Here $I^k$ agrees with the $k$--instanton part of the 5d Nekrasov partition function $Z^k_{QM}$ computed using ADHM quantum mechanics.


We have been able to compute the superconformal index using $Z_{E_8}$ expanding it in the fugacity $x$ 
up to order $x^3$ and the result is\footnote{$\chi_{\mathbf{2}} (y) = y + 1/y$ is the character of the fundamental representation of $SU(2)$.}
\be\label{eq:indexE8}\begin{split}
I &= 1 + (1+ \chi_{\mathbf{91}}^{SO(14)} + u \,\chi_{\mathbf{64}}^{SO(14)}+u^{-1} \chi_{\mathbf{\overline{64}}}^{SO(14)}+ u^2 \chi_{\mathbf{14}}^{SO(14)}+u^{-2} \chi_{\mathbf{\overline{14}}}^{SO(14)})x^2\\
&+ \chi_{\mathbf{2}}(y)(1+1+ \chi_{\mathbf{91}}^{SO(14)} + u \,\chi_{\mathbf{64}}^{SO(14)}+u^{-1} \chi_{\mathbf{\overline{64}}}^{SO(14)}+ u^2 \chi_{\mathbf{14}}^{SO(14)}+u^{-2} \chi_{\mathbf{\overline{14}}}^{SO(14)})x^3+\dots\\
& = 1+ \chi_{\mathbf{248}}^{E_8}\, x^2 +\chi_{\mathbf{2}}(y) (1+\chi_{\mathbf{248}}^{E_8})x^3+\dots
\end{split}\ee
which is expected from the branching
\be\begin{split}
E_8 &\supset SO(14) \times U(1)\\
\mathbf{248} &\rightarrow \mathbf{1}_0 + \mathbf{91}_0 + \mathbf{64}_1  +\mathbf{\overline{64}}_{-1}  +\mathbf{14}_2  +\mathbf{\overline{14}}_{-2}\,. 
\end{split}\ee
In \eqref{eq:indexE8} we have assumed that contributions with higher instanton number will appear in the superconformal index only with higher powers of $x$. Finally let us mention that we have expanded the partition function
$Z_{E_8}$ at order $x^4$ and found the following contributions to the superconformal index
\be
1+\chi_{\mathbf{3080}}^{SO(14)} + u^2\chi_{\mathbf{\overline{1716}}}^{SO(14)}+u^{-2}\chi_{\mathbf{1716}}^{SO(14)} + \chi_{\mathbf{3}}(y) (1+ \chi_{\mathbf{248}}^{E_8})
\ee
which again is consistent with the results of \cite{Kim:2012gu,Hwang:2014uwa}. However the complete expression at order $x^4$ has not been reproduced because part of the expression involves contributions at 3 and 4 instanton number.
A similar computation has been performed using the field theory result for the Nekrasov partition function \cite{Hwang:2014uwa} and the same result has been obtained. This provides further evidence for the equality of the partition function at instanton
level 2 computed from $Z_{E_8}$ and the field theory result.

\section{Conclusion and discussion}

In this paper, we have obtained the prescription to compute partition functions of five-dimensional class $\mathcal{S}$ theories which are realised as low energy theories in Higgs branches of the $T_N$ theory. Although the web diagrams of the resulting theories are non-toric, one can obtain their exact partition functions by inserting the conditions of the tunings of the parameters in the theories corresponding to putting parallel horizontal external 5-branes together, putting parallel vertical external 5-branes together or putting parallel diagonal 5-branes together. The first type of the tuning was found in \cite{Hayashi:2013qwa}, and we have further extended the result including the latter two tunings. Their validity has been exemplified by applying them to the theories in the corresponding Higgs branches of the $T_3$ theory. The tunings inside the web diagrams are determined by consistency conditions from the geometry. The three types of the tunings are enough for moving to any Higgs branch of the $T_N$ theory.

With this general prescription, we have computed the exact partition function of the $E_8$ theory which arises in the far infrared of a Higgs branch of the $T_6$ theory. In the Higgs vacuum, there are singlet hypermultiplets which are decoupled from the $E_8$ theory. We have determined their contributions, and in particular we propose that the singlet hypermultiplets which depend on the parameters associated with flavour symmetries can be understood as the decoupled factor associated with new parallel external legs of the web diagram in the Higgs branch. Identifying the singlet hypermultiplets is important since their contributions depend on the instanton fugacity of $Sp(1)$. The proposal works perfectly for the examples we have computed. The final expression of the partition function is written by the summation of the eight Young diagrams. 
We observed that the six Young diagrams summations terminate at finite order with the fixed order for 
the other two Young diagrams. The other two Young diagrams are related to the summation with respect to the instanton fugacity. Therefore, we can evaluate the partition function exactly at some order of the instanton fugacity. We have also compared the our result with the partition function of the $Sp(1)$ gauge theory with 7 fundamental and 1 anti-symmetric hypermultiplets obtained in \cite{Hwang:2014uwa}. Although the method we obtained the partition is completely different from the one in \cite{Hwang:2014uwa}, we found the quite non-trivial agreement as expected.

In the computation of the $E_8$ theory, the singlet hypermultiplets contributions which depend on the parameters associated with the flavour symmetry were totally determined by the factors coming from the new parallel external legs in the Higgs branch of the web diagram. However, not all the singlet hypermultiplets contributions which only depend on the $\Omega$--deformation parameters are interpreted in this way. It is interesting to find a method which can determine the total contribution of singlet hypermultiplets in a Higgs branch purely from a web diagram. Practically, the singlet hypermultiplets contributions which only depend on the $\Omega$--deformation parameters are all contained in the perturbative part of the partition function. Therefore, we can identify them easily once we obtain the perturbative part. 

In the computation of a partition function of a theory from a web diagram or a web diagram for a Higgsed $T_N$ theory, we often end up with a partition function with Young diagrams summations related to flavour fugacities. For the partition function of the $T_3$ theory and the $E_7$ theory, we essentially need the exact partition function of the $T_2$ theory where a Young diagram is assigned to each horizontal external legs, with some additional hypermultiplets which are bi-fundamental between the $Sp(1)$ and the flavour symmetry associated to the Young diagram summation of the partition function of the $T_2$ theory. For the partition function of the $E_8$ theory, we need the exact partition function of the Higgsed $T_3$ theory where a Young diagram is assigned to two upper horizontal external legs, with some additional hypermultiplets which are bi-fundamental between the $Sp(1)$ and the flavour symmetry associated to the Young diagram summation of the partition function of the Higgsed $T_3$ theory. Since the Young diagram summation is related to a summation of a flavour fugacity, the summation may terminate at finite order. Indeed we have observed the termination of the summation in the case of the computation of the $E_8$ theory in this paper as well as in the case of the $E_7$ theory and the $E_6$ theory in \cite{Hayashi:2013qwa}. It is interesting to show and explore the origin of the termination of the Young diagrams summations. This computation can be also used for a prediction of the exact partition function of the theory in the Higgs branch of the $T_3$ theory where non-trivial Young diagrams are assigned to two upper horizontal external legs. Since the Young diagram summation of a flavour fugacity often occurs in the computation from a web diagram, other computation using some web diagram may predict exact results for some Young diagram summation in other theories.

Finally, since our prescription can be used for web diagrams of any Higgs branch, it is interesting to compute partition functions of other theories realised by some non-toric diagrams. Particular examples are higher rank $E_{6, 7, 8}$ theories discussed in \cite{Benini:2009gi}.

\section*{Acknowledgments}

We thank Hee-Cheol Kim for helpful discussions and for the participation of the project at the initial stage as well as useful comments on the manuscript. We also thank Joonho Kim, Futoshi Yagi for interesting and fruitful discussions. H.H. would like to thank King's College London, Korea Institute for Advanced Study, Universit\"at Heidelberg and Institut de Physique Th\'eorique at Saclay during a part of the project, and also Mainz Institute for Theoretical Physics during the last stage of the project, for kind hospitality and financial support. H.H. would also like to thank the European COST action ``The String Theory Universe" for a short-term visiting grant. The work of H.H. and G.Z. is supported by the grant FPA2012-32828 from the MINECO, the REA grant agreement PCIG10-GA-2011-304023 from the People Programme of FP7 (Marie Curie Action), the ERC Advanced Grant SPLE under contract ERC-2012-ADG-20120216-320421 and the grant SEV-2012-0249 of the ``Centro de Excelencia Severo Ochoa" Programme. The work of G.Z. is also supported through a grant from ``Campus Excelencia Internacional UAM+CSIC".

\appendix
\section{5d partition function and 5d superconformal index}
\label{sec:defs}

In this appendix, we summarise the definitions of the quantities we consider in this paper.  

\paragraph{Topological string partition function.} The genus $g$ topological string amplitude $\mathcal{F}_g$ is a generating function of the ``number" of maps from a genus $g$ Riemann surface to various two-cycles $\alpha$ in a Calabi--Yau threefold $X$\footnote{In \eqref{GW}, we formally include the contributions of families of two-cycles which involve the integration over the moduli space of the maps and also those of the constant maps which involve the integration over the moduli space of the genus $g$ Riemann surface.},
\be
\mathcal{F}_g = \sum_{\alpha \in H_2(X, \mathbb{Z})}N_{\alpha}^g \; Q_{\alpha}, \label{GW}
\ee
where $Q_{\alpha}$ is the exponential of a K\"ahler parameter associated to the cycle $\alpha$ given by
\be
Q_{\alpha} = e^{-\int_{\alpha}J}, \label{kahler}
\ee
and $J$ is the K\"ahler form of $X$ and $N_{\alpha}^g$ is the genus $g$ Gromov--Witten invariant. We can further define a generating function for the genus $g$ topological string amplitude as 
\be
Z(g_s) = \exp\left(\sum_{g=0}^{\infty}\mathcal{F}_g g_s^{2g-2}\right),
\ee
which is called the topological string partition function.

The spacetime interpretation of the topological string amplitude $\mathcal{F} = \sum_{g=0}^{\infty}\mathcal{F}_g g_s^{2g-2}$ has been given in \cite{Gopakumar:1998ii, Gopakumar:1998jq} by considering an M-theory compactification of the Calabi--Yau threefold $X$. From this viewpoint, the contribution come from M2-branes wrapping various two-cycles $\beta \in H_2(X, \mathbb{Z})$, that give rise to particles in five dimensions with spin $(J_L, J_R)$ with respect to $SU(2)_L \times SU(2)_R \subset SO(4)$ acting 
on $\mathbb{R}^4$. $SU(2)_R$ is identified with the $SU(2)$ Lefschetz action on the moduli space of the deformation of $\beta$ inside $X$. On the other hand, $SU(2)_L$ is identified with the $SU(2)$ Lefschetz action on the moduli space of the flat bundle over $\beta$. The explicit expression of the topological string amplitude in terms of the reformulation of \cite{Gopakumar:1998ii, Gopakumar:1998jq} is
\be
\mathcal{F} =  \sum_{\beta, J_L, k > 0}\sum_{l=-J_L}^{J_L}(-1)^{2J_L}\frac{n_{J_L}^{(\beta)}}{k}\frac{q^{2lk}}{\left(q^{\frac{k}{2}}-q^{-\frac{k}{2}}\right)^2}Q_{\beta}^k, \label{top.unrefined}
\ee
where $q = e^{ig_s}$ and $n_{J_L}^{(\beta)}$ is related to the BPS degeneracy $n_{J_L. J_R}^{\beta}$ of the M2-branes wrapping $\beta$ with spin $(J_L, J_R)$ by $n_{J_L}^{(\beta)} = \sum_{J_R}(-1)^{2J_R}(2J_R + 1)n_{J_L. J_R}^{\beta}$. 

The refine version of the topological string amplitude was also proposed in \cite{Hollowood:2003cv} by introducing another parameter ``$t$"
\be
\mathcal{F}_{ref} =  \sum_{\beta, J_L, J_R, k > 0}\sum_{l=-J_L}^{J_L}\sum_{r=-J_R}^{J_R}(-1)^{2J_L+2J_R}\frac{n_{J_L, J_R}^{(\beta)}}{k}\frac{(tq)^{lk}(t/q)^{rk}}{\left(q^{\frac{k}{2}}-q^{-\frac{k}{2}}\right)\left(t^{\frac{k}{2}}-t^{-\frac{k}{2}}\right)}Q_{\beta}^k\,,
\ee
which reduces to \eqref{top.unrefined} when $t = q$. When $X$ is a non-compact toric Calabi--Yau manifold, the powerful technique of the refined topological vertex formalism in \cite{Iqbal:2007ii} computes 
\be
\tilde{Z}_{ref} = \exp\left(\mathcal{F}_{\text{ref}}\right), \label{ref1}
\ee
up to a prefactor that is the refined version of the constant map contribution given by 
\be
\left(M(t, q)M(q, t)\right)^{\frac{\chi(X)}{4}}, \label{constant.map}
 \ee
where $M(t. q) = \prod_{i, j =1}^{\infty}\left(1 - q^it^{j-1}\right)^{-1}$ and $\chi(X)$ is the Euler characteristic of $X$. {We call \eqref{ref1} as the refined topological string partition function.}

{When $X$ is a non-compact toric Calabi--Yau threefold, the toric diagram is dual to a web of $(p, q)$ 5-branes \cite{Leung:1997tw}. In this dual picture, the same five-dimensional theory is living on the web of $(p, q)$ 5-branes and we will mainly employ this point of view in this article.}

In the refined topological partition function, there can be some contributions from M2-branes wrapping two-cycles that may be moved to infinity and hence are decoupled. From the toric diagram, those contributions are associated to the two-cycles between parallel external legs \cite{Bao:2013pwa, Hayashi:2013qwa, Bergman:2013aca}. We can further define a different refined topological string partition function by
\be
Z_{ref} = \tilde{Z}_{ref}/Z_{dec}, \label{ref2}
\ee
where we denote the decoupled contributions by $Z_{dec}$.

\paragraph{5d Nekrasov partition function.} 

The 5d Nekrasov partition function is the partition function of a 5d theory on $\mathbb{R}^4 \times S^1$ on the so-called $\Omega$--background \cite{Nekrasov:2002qd}.  The $\Omega$--background yields a non-trivial fibration of $\mathbb{R}^4$ over the circle $S^1$. The rotation of the two orthogonal 2--planes is given by the $\Omega$--deformation parameters $\epsilon_1, \epsilon_2$ that act $(z_1, z_2) \in \mathbb{C}^2 \cong \mathbb{R}^4 \rightarrow (e^{i\epsilon_1}z_1, e^{i\epsilon_2}z_2)$. Due to the introduction of the $\Omega$--background, the $k$--instanton partition function reduces to the Witten index $Z_{QM}^k$ of the ADHM quantum mechanics where the ADHM data become the dynamical degrees and the ADHM constraints are the D-term conditions. The Witten index $Z_{QM}^k$ is defined as 
\be
Z^k_{QM} (\epsilon_1,\epsilon_2,\alpha_1,z)= \text{Tr}\left[(-1)^F e^{-\beta \{Q,Q^\dag\}}e^{-i\epsilon_1 (j_1+j_R)}e^{-i\epsilon_2 (j_2+j_R)}e^{-i\lambda_i \Pi_i}e^{-iz F^{\prime}}\right]\,, \label{witten}
\ee
where $F$ is the Fermion number operator, $j_1, j_2$ are the Cartan generators of the $SO(4)$ symmetry rotating two orthogonal 2--planes. $j_R$ is the Cartan generator for the $SU(2)_R$ R--symmetry. $\Pi_i$ are the Cartan generators for the gauge group of the theory, and the chemical potential $\lambda_i$ are the Coulomb branch moduli. Finally, $F^{\prime}$ denotes the Cartan generators for the other flavor symmetries and $z$ is the corresponding chemical potential. $Q$ is a supercharge that commutes with all the fugacities and $Q^{\dagger}$ is its conjugate. In this case, the supercharge $Q$ has spin $(j_1, j_2) = (-\frac{1}{2}, -\frac{1}{2})$ and $j_R = \frac{1}{2}$, and $Q^{\dagger}$ has spin $(j_1, j_2)=(\frac{1}{2}, \frac{1}{2})$ and $j_R=-\frac{1}{2}$. 
Hence, the Witten index is defined so that we count the BPS states that are annihilated by $Q$ and also $Q^{\dagger}$. For simplicity, we use the same symbols for the Cartan generators and the eigenvalues of the states under the Cartans.

The full instanton part of the Nekrasov partition function is given by $Z_{QM} = \sum_{k=0}^{\infty}u^kZ_{QM}^k$ where $u$ is the instanton fugacity. Note that in 5d an instanton is associated with a global $U(1)$ symmetry with the current $j$
\be
j = \ast \text{Tr}(G \wedge G),
\ee
where $G$ is a 5d gauge field strength and $\ast$ is the Hodge star operator in five dimensions. The Nekrasov partition function $\tilde{Z}_{Nekra}$ is obtained by multiplying $Z_{QM}$ by the perturbative part $Z_{0}$
\be
\tilde{Z}_{Nekra} = Z_{0}\cdot Z_{QM},\label{nekra1}
\ee
where the perturbative partition function from a vector multiplet  is 
\bea
Z_{0}^{vm} = &&\prod_{m, n = 1}^{\infty}\Big[(1 - e^{i((n-1)\epsilon_1 - m\epsilon_2)})^r(1 - e^{i(n\epsilon_1 - (m-1)\epsilon_2)})^r\nonumber\\
&&\prod_{{\bold R} \in \text{root}}\left(1 - e^{i{\bold R}\cdot \lambda+i((n-1)\epsilon_1 - m\epsilon_2)}\right)\left(1 - e^{i{\bold R}\cdot \lambda+i(n\epsilon_1 - (m-1)\epsilon_2)}\right)\Big]^{-\frac{1}{2}},
\eea
where $r$ is the rank of the gauge group. The perturbative partition function from a hypermultiplet in a representation is 
\be
Z_{0}^{hm} =\prod_{m, n = 1}^{\infty}\prod_{{\bold W}}\left(1 - e^{i{\bold W}\cdot \lambda - im +i((n-\frac{1}{2})\epsilon_1 - (m-\frac{1}{2})\epsilon_2)}\right),
\ee
where ${\bold W}$ are weights of the representation.

The ADHM quantum mechanics can be also embedded in string theory. In string theory the instanton particle may be realised by D0-branes moving on D4-brane. The presence of D8-branes and an O8-plane can introduce flavours in the 5d theory. It was pointed out in \cite{Hwang:2014uwa} that the string theory embedding of the ADHM quantum mechanics contains extra UV degrees of freedom that make the ADHM quantum mechanics UV complete. These extra UV degrees of freedom do not appear in the 5d quantum field theory, and should be removed from \eqref{nekra1}. Therefore, the correct 5d Nekrasov partition function is 
\be
Z_{Nekra} = \tilde{Z}_{Nekra}/Z_{string}, \label{nekra2}
\ee
where we call $Z_{string}$ as the contributions of the extra UV degrees of freedom.

\paragraph{Relation.} When the Calabi--Yau threefold $X$ is chosen such that the low energy effective field theory of the M-theory compactification on $X$ yields a gauge theory, the refined topological partition function \eqref{ref2} compute the index of the 5d BPS states in the gauge theory. This is essentially the same computation of the Nekrasov partition function and it turns out that 
\be
Z_{ref} = Z_{Nekra}, \label{relation}
\ee
after appropriately redefining the parameters. $q$ and $t$ in \eqref{ref2} are related to the $\Omega$--deformation parameters by $q=e^{-i\epsilon_2}, t= e^{i\epsilon_1}$. The other chemical potentials in the Nekrasov partition function are related to the K\"ahler parameters of two-cycles in the Calabi--Yau threefold $X$. 
Due to this relation \eqref{relation}, we interchangeably use the terminology, K\"ahler parameters, chemical potentials, and parameters in the 5d theory. {We will also call $Q_{\beta}$ fugacity.} In section \ref{sec:Higgs.vertical} and \ref{sec:E8}, we use the partition function of the $T_3$ theory and the $T_6$ theory. These partition functions are computed by the refined topological vertex and then identified with the 5d Nekrasov partition functions of the theories with appropriate parameterisations.

\paragraph{5d superconformal index.}

The 5d superconformal index for a 5d theory (or equivalently the partition function on $S^1 \times S^4$) is defined as 
\be
I(\gamma_1,\gamma_2,m_i,u) = \text{Tr}\left[(-1)^F e^{-2(j_r+j_R)\gamma_1}e^{-2j_l\gamma_2}e^{-i \sum_i H_i m_i} u^k\right]\,, \label{5d.index}
\ee
where $j_r$ and $j_l$ are the Cartan generators of $SU(2)_r \times SU(2)_l \subset SO(5)$ with $j_r = \frac{j_1 + j_2}{2}$ and $j_l = \frac{j_1-j_2}{2}$, 
 $j_R$ is the Cartan generator of the $SU(2)_R$ R-symmetry group, $H_i$ are the Cartan generators for flavor symmetries and $k$
is the instanton number. $\gamma_1$ and $\gamma_2$ are related to the $\Omega$--deformation parameters by $\gamma_1 = \frac{i}{2}(\epsilon_1 + \epsilon_2)$ and $\gamma_2 = \frac{i}{2}(\epsilon_1 - \epsilon_2)$. Again, we use the same symbols for the Cartan generators and the eigenvalues of the states under the Cartans for simplicity. The explicit computation of the superconformal index can be performed using localisation techniques and the result is the product of contribution localised at the south pole of $S^4$ and the contribution localised at the north pole of $S^4$ \cite{Kim:2012gu} with integrations over holonomy variables corresponding to Coulomb branch moduli to extract gauge invariant operators.

\section{Tuning for coincident $(1,1)$ 5-branes}
\label{sec:Higgs.diagonal}

\begin{figure}[t]
\begin{center}
\includegraphics[width=60mm]{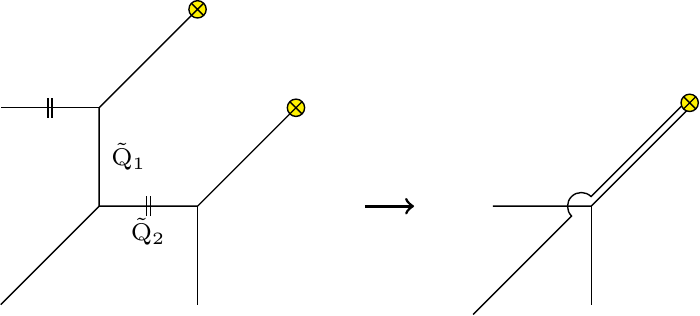}
\end{center}
\caption{The process of putting parallel diagonal external  5-branes together on one 7-brane.}
\label{fig:Higgs4}
\end{figure}

The Higgs branch of the $T_N$ theory opens up when we put parallel external 5-branes on a 7-brane. So far, we have discussed the tuning associated with putting the parallel vertical external 5-branes on one 7-brane \eqref{Higgs.vertical1} and \eqref{Higgs.vertical2} as well as the tuning associated with putting the parallel horizontal external 5-branes on one 7-brane \eqref{Higgs.horizontal1} and \eqref{Higgs.horizontal2}. We then find a similar tuning for putting parallel two diagonal external 5-branes on one single 7-brane as in figure \ref{fig:Higgs4}.
As with the case of putting two parallel vertical external 5-branes on one 7-brane, a pole in the superconformal index computation in this case is associated with an instanton fugacity. One can again change the preferred direction into the diagonal direction. Then we can sum up $\tilde{Q}_2$ as well as $\tilde{Q}_1$ and find a location of the poles. In fact, the tuning is essentially the same as the other ones associated with the parallel horizontal external legs or the parallel vertical external legs. We then propose that we can obtain the partition function of an infrared theory in the Higgs branch associated with the web in figure \ref{fig:Higgs4} by requiring 
\be
\tilde{Q}_1 = \tilde{Q}_2 = \left(\frac{q}{t}\right)^{\frac{1}{2}},
\label{Higgs.diagonal1}
\ee
or 
\be
\tilde{Q}_1 = \tilde{Q}_2 = \left(\frac{t}{q}\right)^{\frac{1}{2}}.
\label{Higgs.diagonal2}
\ee

We will again apply the prescription \eqref{Higgs.diagonal1} or \eqref{Higgs.diagonal2} to the partition functions of two Higgsed $T_3$ theories in order the exemplify the prescription. We have two types of the tuning associated with putting two parallel diagonal external 5-branes on one 7-brane. We will exemplify the prescription for each Higgs branch.

\subsection{Higgsed $T_3$ theory III}

\begin{figure}[t]
\begin{center}
\includegraphics[width=60mm]{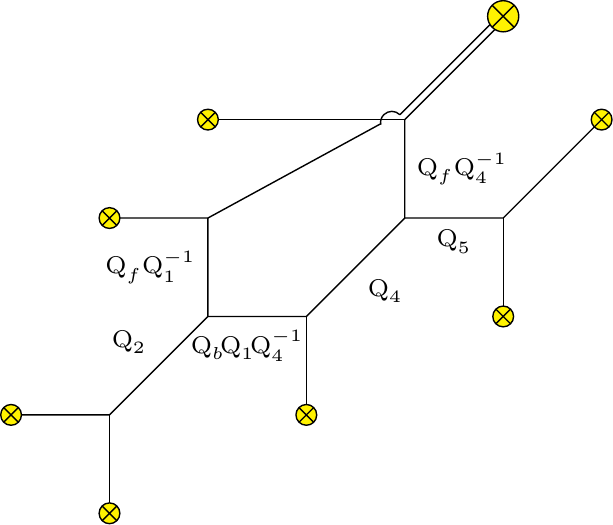}\qquad\qquad
\includegraphics[width=60mm]{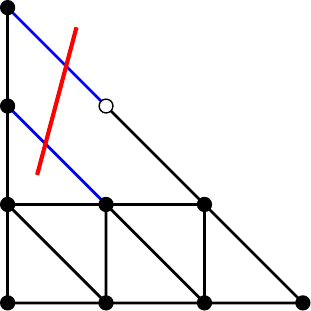}
\end{center}
\caption{Left: The web diagram of the first kind of the Higgsed $T_3$ theory associated with the coincident diagonal external 5-branes. Right: The corresponding dot diagram of the web diagram on the left. The red line shows the new external leg.}
\label{fig:T3Higgs3}
\end{figure}

We first consider putting two leftmost parallel diagonal external 5-branes on one 7-brane as in figure \ref{fig:T3Higgs3}. 
In order to obtain the partition function of the infrared theory in the Higgs branch arising from figure \ref{fig:T3Higgs3}, we adopt the tuning \eqref{Higgs.diagonal1} to K\"ahler parameters in figure \ref{fig:T3Higgs3} 
\be
Q_3 = \left(\frac{t}{q}\right)^{\frac{1}{2}}, \qquad Q_b = \left(\frac{t}{q}\right)^{\frac{1}{2}}. \label{Higgs3}
\ee
This means that we consider a pole located at 
\be
Q_3Q_be^{-2\gamma_1} = ue^{-\frac{i}{2}\left(m_1+m_2+m_3+m_4-m_5\right)}e^{-2\gamma_1} = 1. 
\ee
Therefore, the operator associated to the pole has charges $(j_r, j_l)=(0,0)$ and $j_R=1$. The operator has a vector of charges which form a weight of the Weyl spinor representation of $SO(10)$ with negative chirality and also carries an instanton number $1$. 

By inserting the conditions \eqref{Higgs3} to the partition function of the $T_3$ theory \eqref{T3}, one obtains 
\bea
Z_{T_{\mathcal{IR}}} = Z_0\cdot Z_{inst}\cdot Z_{dec}^{-1}
\eea
\bea
Z_0 &=& \prod_{i,j=1}^{\infty}\Big[\frac{\prod_{a=1,4}(1-e^{-i\lambda + im_a}q^{i-\frac{1}{2}}t^{j-\frac{1}{2}})(1-e^{-i\lambda - im_a}q^{i-\frac{1}{2}}t^{j-\frac{1}{2}})}{(1-q^{i}t^{j-1})^{\frac{1}{2}}(1-q^{i-1}t^{j})^{-\frac{1}{2}}(1-e^{-2i\lambda}q^{i}t^{j-1})}\nonumber\\
&&\times (1-e^{i\lambda + im_2}q^{i-\frac{1}{2}}t^{j-\frac{1}{2}})(1-e^{-i\lambda + im_2}q^{i-\frac{1}{2}}t^{j-\frac{1}{2}})\Big], \label{Higgs3-pert}
\eea
\bea
Z_{inst} &=& \sum_{\nu_2, \mu_5}\left(e^{\frac{i}{2}\lambda + \frac{i}{2}(m_1 + m_2 + m_4)}\left(\frac{t}{q}\right)^{\frac{3}{4}}\right)^{|\nu_2|} u_1^{|\mu_5|}\Big[\prod_{s\in\nu_2}\frac{\prod_{a=1,2}\left(2i\sin\frac{E_{2\emptyset}-m_a + i\gamma_1}{2}\right)(2i\sin\frac{E_{25}-m_4+i\gamma_1}{2})}{(2i)^2\sin\frac{E_{22}}{2}\sin\frac{E_{22}+2i\gamma_1}{2}(2i\sin\frac{E_{2\emptyset}- \lambda+2i\gamma_1}{2})}\nonumber\\
&&\prod_{s \in \mu_5}\frac{(2i\sin\frac{E_{52}+m_4+i\gamma_1}{2})(2i\sin\frac{E_{5\emptyset} - \lambda + m_4 + i\gamma_1}{2})}{(2i)^2\sin\frac{E_{55}}{2}\sin\frac{E_{55}+2i\gamma_1}{2}}\Big],  \label{Higgs3-inst}
\eea
\bea
Z^{-1}_{dec} &=& \prod_{i,j=1}^{\infty}\Big[(1-u_1e^{im_4}q^{i}t^{j-1})(1-q^{i-1}t^{j})(1-u_1e^{im_4}q^{i-1}t^{j})\nonumber\\
&&(1-u_1e^{-im_4}q^{i-1}t^{j})(1-e^{i\lambda + i(m_1+m_2 + m_4)}q^{i-\frac{3}{2}}t^{j+\frac{1}{2}})(1-u_1e^{i\lambda + i(m_1+m_2)}q^{i-\frac{3}{2}}t^{j+\frac{1}{2}})\Big], \nonumber \\ \label{Higgs3-decoupled}
\eea
Due to the first tuning of \eqref{Higgs3}, Young diagram summation of $\nu_1$ vanishes unless $\nu_1 = \emptyset$. To obtain \eqref{Higgs3-pert}--\eqref{Higgs3-decoupled}, we erased $m_3$ and $u_2$ by using \eqref{Higgs3}.

The instanton partition function \eqref{Higgs3-inst} can be again written by the products of the Plethystic exponentials
\bea
Z_{inst}  &=& \prod_{i,j=1}^{\infty}\Big[\frac{(1-Q_5q^{i-\frac{1}{2}}t^{j-\frac{1}{2}})(1-Q_bQ_1Q_4^{-1}q^{i-\frac{1}{2}}t^{j-\frac{1}{2}})(1-Q_bQ_1Q_5q^{i-\frac{1}{2}}t^{j-\frac{1}{2}})}{(1-u_1e^{im_4}q^{i}t^{j-1})(1-u_1e^{im_4}q^{i-1}t^{j})(1-u_1e^{-im_4}q^{i-1}t^{j})}\nonumber\\
&&\times\frac{(1-Q_bQ_1Q_2q^{i-\frac{1}{2}}t^{j-\frac{1}{2}})(1-Q_bQ_2Q_4^{-1}Q_fq^{i-\frac{1}{2}}t^{j-\frac{1}{2}})(1-Q_bQ_2Q_5Q_fq^{i-\frac{1}{2}}t^{j-\frac{1}{2}})}{(1-e^{i\lambda + i(m_1+m_2 + m_4)}q^{i-\frac{3}{2}}t^{j+\frac{1}{2}})(1-u_1e^{i\lambda + i(m_1+m_2)}q^{i-\frac{3}{2}}t^{j+\frac{1}{2}})(1-e^{-i\lambda+im_1}q^{i-\frac{1}{2}}t^{j-\frac{1}{2}})}\nonumber\\
&&\times\frac{(1-Q_bQ_4^{-1}Q_5Q_fq^{i-\frac{1}{2}}t^{j-\frac{1}{2}})(1-e^{-2i\lambda}q^{i}t^{j-1})}{(1-e^{-i\lambda+im_4}q^{i-\frac{1}{2}}t^{j-\frac{1}{2}})(1-e^{-i\lambda+im_2}q^{i-\frac{1}{2}}t^{j-\frac{1}{2}})}\Big] \label{Higgs3-PE}
\eea
The equality of \eqref{Higgs3-PE} can be checked in the same way as we have checked \eqref{eq:strange} in section \ref{sec:Higgs1}. We first write the equations on both sides of \eqref{Higgs3-PE} by the variables $Q_1, Q_2, Q_4^{-1}, Q_f$. Let us then focus on the order $\mathcal{O}(u_1^0)$. If we compute \eqref{Higgs3-inst} until the order $|\nu_2| = k$, the result is exact until the order $\mathcal{O}(Q_1^aQ_f^b)$ with $a + b = k$. Therefore, we can compare \eqref{Higgs3-inst} with \eqref{Higgs3-PE} until the order $\mathcal{O}(Q_1^aQ_f^b)$ with $a + b = k$. We have checked the equality until $k=3$. When $|\mu_5| = l$, we multiply \eqref{Higgs3-inst} by $Q_4^{-l}Q_f^{\frac{l}{2}}$ and then the result is exact until $\mathcal{O}(Q_1^aQ_f^b)$ with $a + b = k$ when we include the Young diagram summation of $\nu_2$ until $|\nu_2| = k$.  We have checked the equality \eqref{Higgs3-PE} until $(l, k) = (2, 2)$.

By combining \eqref{Higgs3-PE} with \eqref{Higgs3-pert}--\eqref{Higgs3-decoupled}, we finally obtain the partition function of the infrared theory of in the Higgs branch of the $T_3$ theory corresponding to figure \ref{fig:T3Higgs3}
\bea
Z_{T_{\mathcal{IR}}}  &=& \prod_{i,j=1}^{\infty}\Big[(1-e^{i(\nu_1+\nu_3^{\prime}-\mu)}q^{i-\frac{1}{2}}t^{j-\frac{1}{2}})(1-e^{i(\nu_2+\nu_2^{\prime}-\mu)}q^{i-\frac{1}{2}}t^{j-\frac{1}{2}})(1-e^{i(\nu_2 + \nu_3^{\prime}-\mu)}q^{i-\frac{1}{2}}t^{j-\frac{1}{2}})\nonumber\\
&&\times(1-e^{i(-\nu_1-\nu_1^{\prime}+\mu)}q^{i-\frac{1}{2}}t^{j-\frac{1}{2}})(1-e^{i(\nu_3 + \nu_2^{\prime} -\mu)}q^{i-\frac{1}{2}}t^{j-\frac{1}{2}})(1-e^{i(\nu_3 + \nu_3^{\prime}-\mu)}q^{i-\frac{1}{2}}t^{j-\frac{1}{2}})\nonumber\\
&&\times(1-e^{\nu_3+\nu_1^{\prime}-\mu}q^{i-\frac{1}{2}}t^{j-\frac{1}{2}})(1-e^{i(-\nu_2-\nu_1^{\prime}+\mu)}q^{i-\frac{1}{2}}t^{j-\frac{1}{2}})(1-e^{i(-\nu_1-\nu_2^{\prime}+\mu)}q^{i-\frac{1}{2}}t^{j-\frac{1}{2}})\Big] \nonumber\\
&&\times\Big[(1-q^{i-1}t^{j})^{\frac{3}{2}}(1-q^it^{j-1})^{-\frac{1}{2}}(1-e^{-3i\mu}q^{i-\frac{1}{2}}t^{j-\frac{1}{2}})\Big],\label{9free3}
\eea
where we rewrite the parameters by the chemical potentials associated with the unbroken global symmetry $SU(3) \times SU(3) \times U(1)$ in the Higgs branch. The generator of the unbroken global symmetry is 
\be
t_{SU(3) \times SU(3) \times U(1)} = -i\left(\mu_1(D_1 + D) + \mu_2D_2 + \mu_1^{\prime}D_3 + \mu_2^{\prime}D_4 + \mu(D + 2D_5 + D_6)\right),
\ee
and we further defined $\nu_i, \nu_i^{\prime}, (i=1, 2, 3)$ by \eqref{charge.rewrite1} and 
\bea
\nu_1^{\prime} = \mu_1^{\prime}, \qquad \nu_2^{\prime} = -\mu_1^{\prime} + \mu_2^{\prime}, \qquad \nu_3^{\prime} = -\mu_2^{\prime} \label{charge.rewrite3}
\eea
The explicit parameterisation is 
\bea
Q_1 &=& e^{i(-\nu_1 + \nu_2)}, \quad Q_2 = e^{i(-\nu_2 - \nu_1^{\prime} + \mu)}, \quad Q_4 = e^{i(-\nu_1 - \nu_2^{\prime} + \mu)}, \quad Q_5 = e^{i(\nu_1 + \nu_3^{\prime} - \mu)}, \nonumber\\
Q_f &=& e^{i(- 2\nu_2 + \nu_1^{\prime} - \mu)}.
\eea

The factors in the last line of \eqref{9free3} correspond to the singlet hypermultiplets in the Higgs branch. Those factors can be understood from the web diagram \ref{fig:T3Higgs3} as in the examples of section \ref{sec:Higgs1} and \ref{sec:Higgs2}. In particular, the very last factor in the last line of \eqref{9free3} may 
come from the contribution of strings between the new parallel external leg after the Higgsing. The new external leg is depicted in the red line in the dot diagram of figure \ref{fig:T3Higgs3}.

\subsection{Higgsing $T_3$ theory IV}

\begin{figure}[t]
\begin{center}
\includegraphics[width=60mm]{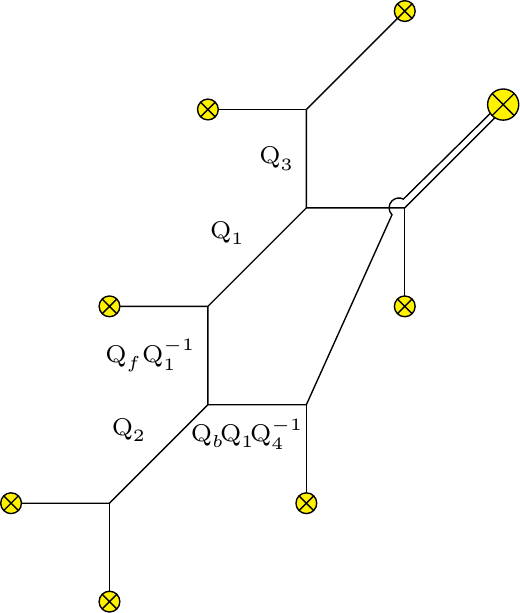}\qquad\qquad
\includegraphics[width=60mm]{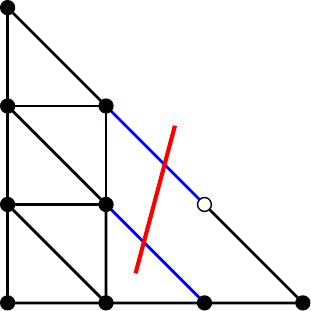}
\end{center}
\caption{Left: The web diagram of the second kind of the Higgsed $T_3$ theory associated with the coincident diagonal external 5-branes. Right: The corresponding dot diagram of the web diagram on the left. The red line shows the new external leg.}
\label{fig:T3Higgs4}
\end{figure}

We then consider the second type of tuning associated with putting the two rightmost parallel diagonal external 5-branes together on one 7-brane as in figure \ref{fig:T3Higgs4}.
For that, we adopt the tuning \eqref{Higgs.diagonal2}
\be
Q_5= \left(\frac{t}{q}\right)^{\frac{1}{2}}, \qquad Q_4^{-1}Q_f = \left(\frac{t}{q}\right)^{\frac{1}{2}}. \label{Higgs4}
\ee
This means that we consider a pole located at 
\be
Q_4^{-1}Q_5Q_fe^{-2\gamma_1} = e^{i(m_4-m_5)}e^{-2\gamma_1} = 1.
\ee
The pole is associated to the mesonic operator with $(j_r, j_l)=(0, 0), j_R=1$ and with charges forming a root of $SO(10)$.

With the conditions \eqref{Higgs4}, the partition function of \eqref{T3} becomes
\bea
Z_{T_\mathcal{IR}} &=&Z_0 \cdot Z_{inst}\cdot Z_{dec}^{-1}, \label{Higgs4-part}
\eea
\bea
Z_0 &=& \prod_{i,j=1}^{\infty}\Big[\frac{(1-e^{-i\lambda+im_1}q^{i-\frac{1}{2}}t^{j-\frac{1}{2}})(1-e^{-i\lambda-im_1}q^{i-\frac{1}{2}}t^{j-\frac{1}{2}})}{(1-q^{i}t^{j-1})^{\frac{1}{2}}(1-q^{i-1}t^{j})^{-\frac{1}{2}}(1-e^{-2i\lambda}q^{i-1}t^{j})}\nonumber\\
&&(1-e^{i\lambda + im_2}q^{i-\frac{1}{2}}t^{j-\frac{1}{2}})(1-e^{-i\lambda + im_2}q^{i-\frac{1}{2}}t^{j-\frac{1}{2}})(1-e^{i\lambda - im_3}q^{i-\frac{1}{2}}t^{j-\frac{1}{2}})(1-e^{-i\lambda-im_3}q^{i-\frac{1}{2}}t^{j-\frac{1}{2}})\Big], \nonumber\\
\label{Higgs4-pert} 
\eea
\bea
Z_{inst}&=&\sum_{\nu_1, \nu_2, \mu_5}u_2^{|\nu_1|+|\nu_2|}\left(e^{-i\lambda}\left(\frac{t}{q}\right)^{\frac{1}{2}}\right)^{|\mu_5|}\Big[\prod_{\alpha=1}^2\prod_{s \in \nu_{\alpha}}\frac{\left(\prod_{a=1}^32i\sin\frac{E_{\alpha\emptyset}-m_a+i\gamma_1}{2}\right)(2i\sin\frac{E_{\alpha 5} - \lambda  +2i\gamma_1}{2})}{\prod_{\beta=1}^2(2i)^2\sin\frac{E_{\alpha\beta}{2}}{2}\sin\frac{E_{\alpha\beta+2i\gamma_1}}{2}}\nonumber\\
&&\prod_{s\in\mu_5}\frac{\prod_{\alpha=1}^22i\sin\frac{E_{5\alpha}+\lambda}{2}}{(2i)^2\sin\frac{E_{55}}{2}\sin\frac{E_{55}+2i\gamma_1}{2}}\Big], \label{Higgs4-inst}
\eea
\bea
Z_{dec}^{-1}&=&\prod_{i,j=1}^{\infty}\Big[(1-q^{i-1}t^{j})(1-u_2e^{-\frac{i}{2}\lambda-\frac{i}{2}(m_1+m_2+m_3)}q^{i+\frac{1}{4}}t^{j-\frac{5}{4}})(1-u_2e^{-\frac{i}{2}\lambda-\frac{i}{2}(m_1+m_2+m_3)}q^{i-\frac{3}{4}}t^{j-\frac{1}{4}})\nonumber\\
&&(1-e^{-2i\lambda}q^{i-1}t^{j})(1-u_2e^{\frac{i}{2}\lambda+\frac{i}{2}(m_1+m_2+m_3)}q^{i-\frac{5}{4}}t^{j+\frac{1}{4}})(1-u_2e^{-\frac{3i}{2}\lambda+\frac{i}{2}(m_1+m_2+m_3)}q^{i-\frac{5}{4}}t^{j+\frac{1}{4}})\Big], \nonumber \\
\label{Higgs4-decoupled}
\eea
where we erased $u_1$ and $m_4$ to obtain \eqref{Higgs4-pert}--\eqref{Higgs4-decoupled}. As discussed in section \ref{sec:Higgs2}, not all the Young diagram summations with respect to $\mu_5$ contribute for a fixed order of $|\nu_1|=k$. The the contribution is non-zero if $\mu_{5, i} \leq \nu_{1, i}$ for all $i$.

The instanton partition function \eqref{Higgs4-inst} turns out to be the product of the Plethystic exponentials
\bea
Z_{inst} &=& \prod_{i,j=1}^{\infty}\Big[\frac{(1-Q_bQ_1Q_4^{-1}q^{i-\frac{1}{2}}t^{j-\frac{1}{2}})(1-Q_bQ_2Q_4^{-1}Q_fq^{i-\frac{1}{2}}t^{j-\frac{1}{2}})}{(1-u_2e^{-\frac{i}{2}\lambda-\frac{i}{2}(m_1+m_2+m_3)}q^{i+\frac{1}{4}}t^{j-\frac{5}{4}})(1-u_2e^{-\frac{i}{2}\lambda-\frac{i}{2}(m_1+m_2+m_3)}q^{i-\frac{3}{4}}t^{j-\frac{1}{4}})}\nonumber\\
&&\times\frac{(1-Q_bQ_1Q_2Q_3Q_4^{-1}Q_fq^{i-\frac{1}{2}}t^{j-\frac{1}{2}})(1-Q_bQ_3Q_4^{-1}Q_fq^{i-\frac{1}{2}}t^{j-\frac{1}{2}})}{(1-u_2e^{\frac{i}{2}\lambda+\frac{i}{2}(m_1+m_2+m_3)}q^{i-\frac{5}{4}}t^{j+\frac{1}{4}})(1-u_2e^{-\frac{3i}{2}\lambda+\frac{i}{2}(m_1+m_2+m_3)}q^{i-\frac{5}{4}}t^{j+\frac{1}{4}})}\Big]\label{Higgs4-PE}
\eea
We have checked the equality \eqref{Higgs4-PE} until $\mathcal{O}(u_2^2)$.

By combining the result \eqref{Higgs4-PE} with \eqref{Higgs4-pert}--\eqref{Higgs4-decoupled}, we obtain the partition function of the infrared theory in the Higgs branch of the $T_3$ theory 
\bea
Z_{T_{\mathcal{IR}}} &=& \prod_{i,j=1}^{\infty}\Big[(1-e^{i(\nu_2+\nu_3^{\prime}+\mu)}q^{i-\frac{1}{2}}t^{j-\frac{1}{2}})(1-e^{i(-\nu_2-\nu_2^{\prime}-\mu)}q^{i-\frac{1}{2}}t^{j-\frac{1}{2}})(1-e^{i(\nu_3 + \nu_2^{\prime}+\mu)}q^{i-\frac{1}{2}}t^{j-\frac{1}{2}})\nonumber\\
&&\times(1-e^{i(\nu_3+\nu_3^{\prime} + \mu)}q^{i-\frac{1}{2}}t^{j-\frac{1}{2}})(1-e^{i(-\nu_1 - \nu_3^{\prime}-\mu)}q^{i-\frac{1}{2}}t^{j-\frac{1}{2}})(1-e^{i(-\nu_1-\nu_2^{\prime} -\mu)}q^{i-\frac{1}{2}}t^{j-\frac{1}{2}})\nonumber\\
&&\times(1-e^{i(-\nu_3-\nu_1^{\prime}-\mu)}q^{i-\frac{1}{2}}t^{j-\frac{1}{2}})(1-e^{i(-\nu_2-\nu_1^{\prime}-\mu)}q^{i-\frac{1}{2}}t^{j-\frac{1}{2}})(1-e^{i(-\nu_1-\nu_1^{\prime} -\mu)}q^{i-\frac{1}{2}}t^{j-\frac{1}{2}})\Big]\nonumber\\
&&\times\Big[(1-q^{i-1}t^{j})^{\frac{3}{2}}(1-q^{i}t^{j-1})^{-\frac{1}{2}}(1-e^{-3i\mu}q^{i-\frac{1}{2}}t^{j-\frac{1}{2}})\Big],\label{9free4}
\eea
which can be explicitly seen as the partition function of nine free hypermultiplets up to singlet hypermultiplets. We have also parameterised the K\"ahler parameters by the chemical potentials associated with the unbroken flavour symmetry $SU(3)\times SU(3) \times U(1)$ as
\bea
Q_1 &=& e^{i(\nu_2 + \nu_3^{\prime} + \mu)}, \quad Q_2 = e^{i(\nu_3 + \nu_2^{\prime} + \mu)}, \quad Q_3 = e^{i(-\nu_1 - \nu_3^{\prime} - \mu)}, \quad Q_4 = e^{i(-\nu_2^{\prime} + \nu_3^{\prime})},\nonumber\\
Q_b&=&e^{\nu_1 + \nu_3^{\prime} - 2\mu}, \quad Q_f = e^{i(\nu_3 - \nu_2^{\prime})}. 
\eea
The generator of the unbroken global symmetry is 
\be
t_{SU(3) \times SU(3) \times U(1)} = -i\left(\mu_1 D_1 + \mu_2 D_2 + \mu_1^{\prime}D_3 + \mu_2^{\prime}(D_4 + D) + \mu(D_5 + 2D_6 + D)\right).
\ee
The relations between $\mu_i, \mu_i^{\prime}, (i=1, \cdots, 3)$ and $\nu_i, \nu_i^{\prime}, (i=1, \cdots, 3)$ are \eqref{charge.rewrite1} and \eqref{charge.rewrite3}.

The factors in the last big bracket of \eqref{9free4} are the contributions from the singlet hypermultiplets in the Higgs branch. Those factors again have the interpretation from the web diagram as discussed in \ref{sec:Higgs1} and \ref{sec:Higgs2}. In particular, the very last factor can be understood from the contribution of strings between the new parallel diagonal external leg. The new diagonal external leg after the tuning is depicted in the dot diagram of figure \ref{fig:T3Higgs4}. 

\section{Cartan generators of $SU(6) \times SU(3) \times SU(2)$}
\label{sec:Cartan}

We list up the Cartan generators which correspond to the $SU(6) \times SU(3) \times SU(2)$ in the Higgs vacuum of the $T_6$ theory corresponding to the web \ref{fig:higt6}. We will write each generator as $\sum_{i=1}^{25}a_iD_i$ and simply quote the coefficients $a_i$ for every generator. As discussed in section \ref{sec:parameterization}, the generators for $SU(6)$ can be determined as
\begin{eqnarray}
t_{SU(6)}^1 &=& \{0, 0, 0, 0, 0, 0, 0, 0, 1, 1, 1, 0, 0, 1, 0, 0, 0, 0, 0, 0, 0, 0, 0, 0, 0\}, \label{gen1}\\
t_{SU(6)}^2 &=& \{0, 0, 0, 0, 0, 0, 0, 0, 0, 0, 0, 0, 0, 0, 1, 1, 0, 0, 0, 0, 0, 0, 0, 0, 0\},\\
t_{SU(6)}^3 &=& \{0, 0, 0, 0, 0, 0, 0, 0, 0, 0, 0, 0, 0, 0, 0, 0, 0, 0, 0, 1, 0, 0, 0, 0, 0\},\\
t_{SU(6)}^4 &=& \{0, 0, 0, 0, 0, 0, 0, 0, 0, 0, 0, 0, 0, 0, 0, 0, 0, 0, 0, 0, 0, 0, 1, 0, 0\},\\
t_{SU(6)}^5 &=& \{0, 0, 0, 0, 0, 0, 0, 0, 0, 0, 0, 0, 0, 0, 0, 0, 0, 0, 0, 0, 0, 1, 0, 0, 1\},
\end{eqnarray}
Similarly, the generators for $SU(3)$ and $SU(2)$ are
\begin{eqnarray}
t_{SU(3)}^1 &=& \left\{0, 0, 0, 0, 0, \frac{1}{2}, \frac{1}{2}, \frac{1}{2}, 0, 0, 0, 1, 1, \frac{1}{2}, 0, 0, \frac{1}{2}, \frac{1}{2}, 0, 0, 0, 0, 0, 0, 0\right\},\\
t_{SU(3)}^2 &=&\left \{0, 0, 0, 0, 0, 0, 0, 0, 0, 0, 0, 0, 0, 0, 0, 0,  \frac{1}{2},  \frac{1}{2}, 0, 0, 1,  \frac{1}{2}, 0,  \frac{1}{2}, 0\right\},
\end{eqnarray}
and
\begin{equation}
t_{SU(2)} =\left\{\frac{1}{3}, \frac{2}{3}, 1, \frac{2}{3}, \frac{1}{3}, 0, \frac{1}{3}, \frac{2}{3}, \frac{2}{3}, \frac{1}{3}, 0, 0, \frac{1}{3}, \frac{1}{3}, \frac{1}{3}, 0, 0, \frac{1}{3}, 0, 0, 0, 0, 0, 0, 0\right\}
\end{equation}
respectively. The gauge generator is 
\begin{equation}
t_{\text{gauge}} = \{0, 0, 0, 0, 0, 0, 0, 0, 0, 0, 0, 0, 0, 0, 0, 0, 0, 0, 1, 0, 0, 0, 0, 0, 0\}.\label{gen.gauge}
\end{equation}
We then define parameters associated with the generators \eqref{gen1}--\eqref{gen.gauge} as
\begin{equation}
t =-i\left(\mu_1t_{SU(6)}^1 + \mu_2t_{SU(6)}^2 + \mu_3t_{SU(6)}^3 + \mu_4t_{SU(6)}^4 + \mu_5t_{SU(6)}^5 + \mu_1^{\prime}t_{SU(3)}^1 + \mu_2^{\prime}t_{SU(3)}^2 + \tilde{\mu}t_{SU(2)} + \lambda t_{\text{gauge}}\right). \label{generator}
\end{equation}

By using the definition of the masses and the tentative instanton fugacity \eqref{Sp1.para2}, we find their relation with the chemical potentials for particles in the canonical simple roots of $SU(6)$ 
\begin{eqnarray}
2\mu_1 - \mu_2 &=& m_2 - m_4,\\ \label{E8root1}
-\mu_1 + 2\mu_2 - \mu_3 &=& -m_2 - m_3,\\
-\mu_2 + 2\mu_3 - \mu_4 &=&  m_1 - \tilde{u},\\
-\mu_3 + 2\mu_4 - \mu_5 &=& -m_6 + m_7,\\
-\mu_4 + 2\mu_5 &=& -m_5 + m_6.
\end{eqnarray}
Similarly, the chemical potentials for particles in the canonical simple roots of $SU(3)$ are
\begin{eqnarray}
2\mu_1^{\prime} - \mu_2^{\prime} &=& -m_3 - m_5 - m_6 - \tilde{u},\\
-\mu_1^{\prime} + 2\mu_2^{\prime} &=& -m_2 - m_4 + m_7 - \tilde{u}.
\end{eqnarray}
The chemical potential for a particle in the canonical simple root of $SU(2)$ is 
\begin{equation}
2\tilde{\mu} = m_1 - m_2 - m_4 - m_5  - m_6 - \tilde{u}.  \label{E8root2}
\end{equation}
Eq.~\eqref{E8root1}--\eqref{E8root2} yield \eqref{SU(2)} in section \ref{sec:parameterization}. 


\end{document}